\documentclass[aps,pra,onecolumn,superscriptaddress]{revtex4}

\usepackage{revquantum}
\usepackage{siunitx}
\newcommand{\figurefolder}{../fig}

\newcommand{\NV}{NV$^{-}$}

\newcommand{\MLE}{\text{MLE}}
\newcommand{\BCE}{\text{BCE}}
\newcommand{\MSE}{\text{MSE}}
\newcommand{\Bayes}{\text{Bayes}}
\newcommand{\Bnorm}{\mathcal{N}}
\newcommand{\Setup}{{\text{S}}}

\newcommand{\sgz}{{\text{g},0}}
\newcommand{\sgp}{{\text{g},+1}}
\newcommand{\sgm}{{\text{g},-1}}
\newcommand{\sez}{{\text{e},0}}
\newcommand{\sep}{{\text{e},+1}}
\newcommand{\sem}{{\text{e},-1}}
\newcommand{\sss}{{\text{s}}}

\renewcommand{\figurefolder}{./fig}
\begin{document}

%=============================================================================
% BEGIN TITLE BLOCK
%=============================================================================

\title{Statistical Inference with Quantum Measurements:\\
Methodologies for Nitrogen Vacancy Centers in Diamond}

\author{Ian Hincks}
\email{ihincks@uwaterloo.ca}
\affilUWAMath
\affilIQC

\author{Christopher Granade}
\affilUSydPhys
\affilEQuSUSyd

\author{David G. Cory}
\affilUWChem
\affilIQC
\affilCIFAR
\affilPI

\date{\today}

\begin{abstract}
    The analysis of photon count data from the standard nitrogen vacancy 
    (NV) measurement process is 
    treated as a statistical inference problem.
    This has applications toward gaining better and more rigorous
    error bars for tasks such as parameter estimation (eg. magnetometry), 
    tomography, and randomized benchmarking.
    We start by providing a summary of the standard phenomenological model of 
    the NV optical process in terms of Lindblad jump operators.
    This model is used to derive random variables describing emitted photons 
    during measurement, to which finite visibility, dark counts, and 
    imperfect state preparation are added.
    NV spin-state measurement is then stated as an abstract statistical inference 
    problem consisting of an underlying biased coin obstructed by 
    three Poisson rates. 
    Relevant frequentist and Bayesian estimators are provided, discussed, and 
    quantitatively compared.
    We show numerically that the risk of the maximum likelihood estimator is 
    well approximated by the Cram\'er-Rao bound, for which we provide a simple formula.
    Of the estimators, we in particular promote the Bayes estimator, owing to its slightly
    better risk performance, and straight-forward error propagation 
    into more complex experiments.
    This is illustrated on experimental data, where Quantum Hamiltonian 
    Learning is performed and cross-validated in a fully Bayesian setting, 
    and compared to a more traditional weighted least squares fit.
\end{abstract}

%=============================================================================
% END TITLE BLOCK
%=============================================================================

\maketitle

%=============================================================================
\section{Introduction}
\label{sec:intro}
%=============================================================================

Developing tools for the characterization of quantum systems is an increasingly 
important problem.
As more performance is demanded out of quantum devices, more knowledge about 
these quantum devices is also required.
This applies as much to large-scale multi-qubit quantum information processors as it does to 
small single-qubit quantum sensors.
Importantly, this knowledge must include not only estimates of relevant quantities,
but also careful estimates of the uncertainty of these quantities.
Indeed, if the application is, for example, metrology, then the very nature of the 
problem demands that one should be as 
confident in one's ability to produce meaningful error bars as one's ability to produce 
the estimate itself.
Or, if one is using knowledge about a quantum system to design control sequences 
(such as unitary gates), then it is important to know how much system parameters 
are expected to vary through space and time.
If the estimate of this parameter distribution is too tight, 
the control sequence will not meet specifications, and if 
it is too broad, the control sequence will not have optimal 
efficiency~\cite{rabitz_quantum_2004,pravia_robust_2003,hocker_characterization_2014,henry_fidelity_2007}.

These considerations imply that data from quantum experiments should be analyzed 
on firm statistical footing.
This, in particular, requires a detailed model that computes the likelihoods of 
experimental outcomes given a specified set of system parameters or hyperparameters.
This is not to say that we need perfect statistical models, but rather, that models 
and methods should be well enough defined so that rigorous questions can be
asked and answered unambiguously.
Most of the widespread characterization protocols in quantum information have been 
described by statistical models.
Quantum mechanics is ultimately a statistical theory and so this is usually a natural 
thing to do.
State and process tomography have been studied extensively as matrix-valued 
inference~\cite{huszar_adaptive_2012, blume-kohout_robust_2012, granade_practical_2016}, 
randomized benchmarking (RB)
and derivative protocols are inherently statistical~\cite{magesan_characterizing_2012,epstein_investigating_2014,granade_accelerated_2015,wallman_randomized_2014}, 
and Hamiltonian parameter learning is often considered from a machine 
learning perspective~\cite{wiebe_quantum_2014,granade_robust_2012,kosut_optimal_2004,wang_experimental_2017}.
These statistical models of characterization, however, usually stop short of 
platform dependent considerations; for example, the RB protocol does not tell you how to 
interpret the noisy voltage measurement of a superconducting qubit.
Such divisions are drawn to achieve cross-quantum-platform generality.

Specific platforms, however, really do need to think carefully about how to append their own 
messy details onto these well-established models.
Incorrect models may lead to estimates biased in unknown ways, or, just as bad, 
to error-bars which are inaccurate given the actual data.
This leads us to the purpose of the present paper:
a detailed and explicit model of the measurement process of the
negatively charged
Nitrogen Vacancy (\NV) center in diamond, along with an analysis 
of relevant estimators and methodologies.

\NV~centers have been studied extensively due to a number of 
remarkable physical properties~\cite{doherty_theory_2012}.
These include long coherence times at room temperature~\cite{balasubramanian_ultralong_2009},
the ability to address and readout a single defect in 
isolation~\cite{gruber_scanning_1997,jelezko_read-out_2004}, the
ability to initialize to a (nearly) pure state on demand~\cite{harrison_optical_2004}, and the ability 
to selectively interact with nearby 
nuclei~\cite{jelezko_observation_2004,dutt_quantum_2007,felton_hyperfine_2009,jiang_repetitive_2009,neumann_single-shot_2010,dreau_high-resolution_2012}.
Moreover, the \NV~center's sensitivity to external macroscopic properties like
magnetic fields, electric fields, and temperature, in combination with its nanoscopic 
spacial profile, have shown it to be highly suitable as a quantum 
sensor~\cite{dolde_electric-field_2011,acosta_temperature_2010,rondin_magnetometry_2014}.

As such, it is of value to carefully examine precisely what one 
learns from the \NV~center when a measurement is made.
If the application is quantum sensing, then it is crucial to be able 
to accurately report error bars of the quantity of interest.
Or, if an application requires high-fidelity control, it is important 
to be able to quantify one's knowledge of the system parameters so 
that control sequences can be designed to be robust against uncertainty.
If these same control sequences are being tested in a tomography
or benchmarking protocol, then sensible estimates of
figures of merit also presuppose a thorough understanding of the 
measurement data.

These demands all reduce to statistical inference, 
which is the process of inferring values of interest from a model,
given a dataset of measurements.
However, in order to trust an inference one must first be 
able to trust the sufficiency of the model.
Therefore, we begin by taking the accepted physical description 
of the \NV~measurement process and use it to derive a 
statistical model describing the expected distributions 
of experimental measurements, including annoyances such as 
limited visibility, dark counts, imperfect state preparation, and
reference drift.

This paper is organized as follows.
\autoref{sec:sys-description} introduces a phenomenological model of the 
optical dynamics of an \NV~center in terms of Lindblad jump operators.
\autoref{sec:meas-dynamics} provides a model for the measurement process,
including photon loss and dark counts.
\autoref{sec:state-initialization} describes the state initialization 
process and derives the pseudo-pure state and its contrast-reduced references.
\autoref{sec:drifting-refs} introduces a stochastic model of the 
reference drift process.
\autoref{sec:stats-of-meas} uses the derived description of the measurement 
process to state it as a statistical inference problem -- readers who are 
less interested in the physical derivation of the model may choose to begin with this 
section; though it is the culmination of the preceding sections, it has been written 
to be comprehensible without them.
\autoref{sec:estimators} provides a couple of estimators for this inference 
problem, and compares them.
\autoref{sec:qhl-example} gives an example of the model being used
with experimental data for Quantum Hamiltonian Learning \cite{wiebe_quantum_2014}.
All raw data and code necessary to reproduce the results of this paper 
are openly available online \cite{code_and_data}.

%=============================================================================
\section{Optical Dynamics of the \NV~Center in Diamond}
\label{sec:sys-description}
%=============================================================================

Here we briefly summarize the optical dynamics of the \NV center in diamond, 
setting up our notation for future sections.
More complete descriptions can be found elsewhere, see, for example, the 
review article by \citet{doherty_nitrogen-vacancy_2013}~and references 
therein.

The \NV~defect is an impurity in the diamond lattice consisting 
of a nitrogen atom adjacent to an empty lattice site, replacing 
two carbon atoms that would normally be in those positions. 
This defect, when negatively charged, has six bound electrons whose 
spacial wavefunction extends on the order of a dozen lattice sites before 
becoming negligible compared to nuclear dipolar 
interactions~\cite{smeltzer_13_2011,gali_textitab_2008,dreau_high-resolution_2012}.
These six electrons form 
an effective spin-1 electron in the electronic ground 
state~\cite{doherty_theory_2012}.
It is standard practice to pick two out of these three spin energy levels
to define a qubit.

Importantly, the \NV~center also has optical properties; it can absorb and emit 
photons.
The energy difference between the optical ground state and the 
first excited state is $\SI{637}{nm}$ (red).
%At room temperature, however, a spontaneously emitted photon is usually
%accompanied with phonons of varying energy, so that the emission 
%spectrum of an \NV center is spread out from about $600$nm to $800$nm,
%with a small zero-phonon spike at $637$nm~\cite{gruber_scanning_1997}.
It is experimentally convenient to optically excite the defect from the ground state 
to the first excited state using light with a wavelength outside 
of the emission spectrum, for example $\SI{532}{nm}$ (green).
This is possible due to the presence of higher energy 
states above the first excited state which very quickly decay to 
the first excited state, while preserving spin populations.
Separating red and green allows a confocal microscope 
to be set up so that optical cycling of a single 
isolated \NV center can be studied~\cite{gruber_scanning_1997}:
incident green light is delivered to a small region inside of a diamond,
roughly a cubic micron in volume, and red light is extracted from 
the same region.
Assuming the use of a diamond whose impurities are sparse enough, this region
can be chosen to contain a single \NV~center.

The dynamics of the \NV~center are usually described for the spin system alone,
with the optical degrees of freedom assumed to be in the ground state.
However, since we are interested in the measurement process, we 
must include both degrees of freedom.
We describe the dynamics using a 
seven level system: three levels for the optical ground state spin system,
three levels  for an optical excited state spin system, and one level for 
an optical inter-system-crossing (ISC).
It is known that more levels exist~\cite{doherty_nitrogen-vacancy_2013}, 
but adding them to this discussion 
will not change our model of the measurement process.

We decompose our Hilbert space as the direct sum
\begin{align}
    \Hilbert=\Hilbert_\text{ground}\oplus\Hilbert_\text{excited}\oplus\Hilbert_\text{isc},
\end{align}
where $\dim(\Hilbert_\text{ground})=\dim(\Hilbert_\text{excited})=3$ and 
$\dim(\Hilbert_\text{isc})=1$.
We define a basis for $\Hilbert$ as 
\begin{align}
    \ket{\sgp}=\begin{pmatrix} 1\\0\\0\\0\\0\\0\\0 \end{pmatrix},~
    \ket{\sgz}=\begin{pmatrix} 0\\1\\0\\0\\0\\0\\0 \end{pmatrix},~
    \ket{\sgm}=\begin{pmatrix} 0\\0\\1\\0\\0\\0\\0 \end{pmatrix},~
    \ket{\sep}=\begin{pmatrix} 0\\0\\0\\1\\0\\0\\0 \end{pmatrix},~
    \ket{\sez}=\begin{pmatrix} 0\\0\\0\\0\\1\\0\\0 \end{pmatrix},~
    \ket{\sem}=\begin{pmatrix} 0\\0\\0\\0\\0\\1\\0 \end{pmatrix},~
    \ket{\sss}=\begin{pmatrix} 0\\0\\0\\0\\0\\0\\1 \end{pmatrix}
\end{align}
where $\{+1,0,-1\}$ are spin labels corresponding to the eigenvalues
of $\Sz=\diag(1,0,-1)$, and $(\text{g},\text{e},\text{s})$ refer to optical ground, excited and singlet,
respectively.
In this way we can write, for example, the spin-1 $z$ operator in the 
optical ground state as $\Sz \oplus 0 \oplus 0=\ket{\sgp}\bra{\sgp}-\ket{\sgm}\bra{\sgm}$.

Given an external magnetic field applied
along the z-axis with strength $\omega_e$ (in angular frequency units), 
the Hamiltonian of the system is given  by 
\begin{equation}
    H=H_\text{ground}\oplus0\oplus0 + 0\oplus H_\text{excited}\oplus0
\end{equation}
with 
\begin{align}
    H_\text{ground}&=\Delta_g\Sz^2+\omega_e\Sz \nonumber \\
    H_\text{excited}&=\Delta_e\Sz^2+\omega_e\Sz
\end{align}

The terms $\Delta_g\approx \SI{2.87}{GHz}$ and $\Delta_e\approx \SI{1.4}{GHz}$ 
denote the zero-field splittings.
As the name implies, they are energies that enter into the Hamiltonian 
without the application of an external field; they result from the couplings
between the electrons constituting the \NV~center.

Absorption of a photon takes the system from $\Hilbert_\text{ground}$ 
to $\Hilbert_\text{excited}$, and vice versa for spontaneous photon
emission.
These processes are known to be spin-conserving.
Although coherent optical control is possible,
we restrict our attention 
to the more commonly used dissipative regime.
Therefore, we describe excitation and spontaneous emission
using spin-conserving Lindblad operators,
\begin{align}
    L_1 &= \sqrt{\gamma_{eg}}\left( \ket{\sgp}\bra{\sep} + \ket{\sgz}\bra{\sez} + \ket{\sgm}\bra{\sem} \right) \\
    L_2 &= \sqrt{k\cdot\gamma_{eg}}\left( \ket{\sep}\bra{\sgp} + \ket{\sez}\bra{\sgz} + \ket{\sem}\bra{\sgm} \right)
\end{align}
where $\gamma_{eg}$ is the rate of spontaneous emission, about $\SI{77}{MHz}$.
Without the additional dynamics described below, this would imply an average 
excited state lifetime of about $1/\gamma_{ea}=\SI{13}{ns}$.
The dimensionless parameter $k$ corresponds to the 
laser power and in our experiments is typically on the order of unity when 
the laser is on.
It can be set to $0$ in periods where the laser is off.

Spin selective measurement is possible because of an additional decay path that is 
not spin-conserving and emits photons with a wavelength outside of 
the \SIrange{600}{800}{nm} emission spectrum.
This route preferentially allows the excited $\ket{\sem}$ 
and $\ket{\sep}$ states to decay to the ground states through
the ISC to the state $\ket{\sss}$.
It can be modeled using the Lindblad dissipaters
\begin{align}
    L_3 &= \sqrt{\gamma_{es}/2} \ket{\sss}\bra{\sep} \\
    L_4 &= \sqrt{\gamma_{es}/2} \ket{\sss}\bra{\sem} \\
    L_5 &= \sqrt{\gamma_{sg}/3} \ket{\sgp}\bra{\sss} \\
    L_6 &= \sqrt{\gamma_{sg}/3} \ket{\sgz}\bra{\sss} \\
    L_7 &= \sqrt{\gamma_{sg}/3} \ket{\sgm}\bra{\sss}.
\end{align}
The first two move support on the excited $\ket{\sem}$ 
and $\ket{\sep}$ states to the ISC state.
The last three spread ISC population approximately evenly (and incoherently) 
over the ground space. 
This may seem counterintuitive given that optical excitation of sufficient 
duration results in high spin state polarization; the resolution is that 
the selective decay from $\ket{\sep},\ket{\sem}$ to $\ket{\sss}$ is
what actually drives polarization, as measured by \citet{robledo_spin_2011}.
The rate of the spin-selective 
decay is roughly $\gamma_{es}=\SI{30}{MHz}$, and comparing this 
to $\gamma_{ea}$, shows that 
the excited $\pm1$ states take the ISC path roughly 1/3$^\text{rd}$ of 
the time.
The lifetime of the singlet is quite long, with a decay
rate of roughly 
$\gamma_{sg}=\SI{3}{MHz}$; this is the time scale that will end up
dominating the optimal measurement time.

Small non-spin conserving transitions are estimated to have a rate
of about $\gamma_{01}=\SI{1}{MHz}$~\cite{doherty_nitrogen-vacancy_2013}.
They can be modelled as the Lindblad operators
\begin{align}
    L_8 &= \sqrt{\gamma_{01}/4}\ket{\sgp}\bra{\sez} \nonumber \\
    L_9 &= \sqrt{\gamma_{01}/4}\ket{\sgm}\bra{\sez} \nonumber \\
    L_{10} &= \sqrt{\gamma_{01}/4}\ket{\sgz}\bra{\sep} \nonumber \\
    L_{11} &= \sqrt{\gamma_{01}/4}\ket{\sgz}\bra{\sem}.
    \label{eq:spin-non-cons-lindblad}
\end{align}

\begin{figure}
    \begin{center}
        \includegraphics[width=\textwidth]{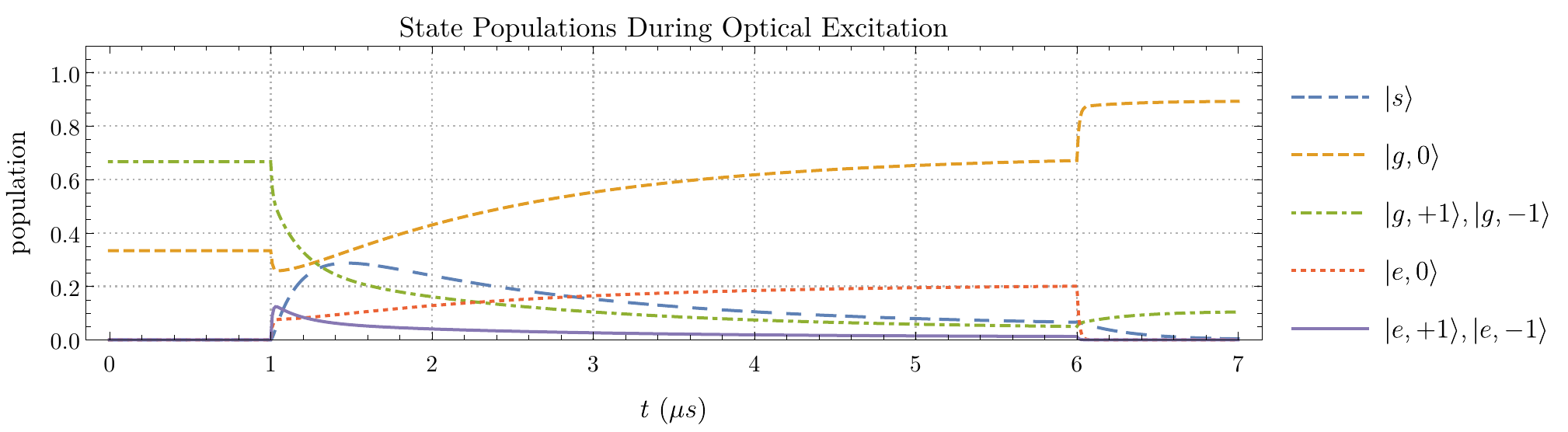}
    \end{center}
    \caption{
        The quantum state is initialized to fully mixed state in the optical
        ground spin-1 manifold, which is approximately the 
        Boltzmann distribution at room temperature and low magnetic field.
        Therefore at $t=0$, the combined $\ket{\sgp}$, $\ket{\sgm}$ manifold
        has twice as much population as $\ket{\sgz}$.
        A laser pulse of duration $5\mu$s is applied starting at $1\mu$s.
        The populations of the subspaces spanned by the 
        pure states in the legend are tracked through time.
        The values used are $k=0.3$, $\gamma_{eg}=\SI{77}{MHz}$,
        $\gamma_{es}=\SI{30}{MHz}$, $\gamma_{sg}=\SI{3}{MHz}$, and
        $\gamma_{01}=\SI{0.5}{MHz}$.
        At $6\mu$s when the laser pulse is turned off, the excited
        states quickly decay to the ground states, and the singlet 
        state slowly leaks back to the ground state; normally we wait
        around 5 or 10 times the characteristic decay time before
        resuming activity.
        Note that full polarization will never be reached due to the 
        nonzero mixing rate $\gamma_{01}$.
    }
    \label{fig:state-populations}
\end{figure}

Solving the Lindblad master equation,
\begin{align}
    \frac{\dd \rho(t)}{\dd t}
        = -\ii[H,\rho(t)]
            +\sum_{i=1}^{11} \left( 
                L_k\rho(t)L_i^\dagger 
                - (\rho(t)L_i^\dagger L_i+L_i^\dagger L_i\rho(t))/2 
            \right),
    \label{eq:lindblad-master}
\end{align}
with some initial state $\rho(0)$ allows us to track the populations 
and coherences of the quantum system through time.
If we start with spin-state coherences, they will quickly die off due 
to the mismatch in zero-field splittings between the optical ground 
and excited states.
Since we have no coherence generating terms in our internal Hamiltonian, describing
the optical dynamics in a fully quantum setting is overkill.
If we were to simultaneously turn on a resonant microwave field and the
green laser, the simplifications we will make in \autoref{sec:rate-eqn} 
would not apply.

%=============================================================================
\section{Measurement Dynamics}
\label{sec:meas-dynamics}
%=============================================================================

It is possible to gain information about the spin state of the NV system by
simultaneously illuminating it and counting the photons it 
emits in the process.
This works because the ISC is 
spin selective.
States initially with support on the subspace $\vectorspan(\ket{\sgm},\ket{\sgp})$, 
once excited, have a decay path which does not 
emit a detectable photon.
Therefore, states initially with support in this space, on
average, appear dimmer.
We now formalize this.

%=============================================================================
\subsection{Measurement Description}
\label{sec:meas-description}

A single measurement of an \NV~center consists of turning the laser on 
for some amount of time (on the order of $\SI{500}{ns}$) and counting the 
spontaneously emitted photons in this duration.
We assume in our model that no information about arrival time or spectral properties 
is recorded.

The probability of a quantum system spontaneously emitting a detectable photon
within a short duration $dt$ at time $t$ is given by the product of the 
rate of spontaneous emission, the length of the time window, and 
probability of being in the excited triplet manifold~\cite{suter_physics_1997}:
\begin{equation}
    \prob{\text{photon};[t,t+dt]}=\gamma_{ea}\Tr\left[P_e\cdot\rho(t)\right] dt,
\end{equation}
where $P_e=\ket{\sep}\bra{\sep}+\ket{\sez}\bra{\sez}+\ket{\sem}\bra{\sem}$ 
is the projector onto the excited subspace $\Hilbert_\text{excited}$ and 
$\rho(t)\in\Density(\Hilbert)$ is the state of the system at time $t$.
This defines an inhomogeneous Poisson process, where the rate of 
events is time-dependent, given by
\begin{equation}
    \mu(t)=\gamma_{ea}\Tr\left[P_e\cdot\rho(t)\right].
    \label{eq:inhom-rate}
\end{equation}
This is a generalized version
of the more common homogeneous Poisson process, where the event rate $\lambda$ 
is constant, and the probability of $k$ events in the time duration $t$
is given by $\prob{k}=\e^{-\lambda t}(\lambda t)^k/k!$.
An inhomogeneous Poisson process has a similar formula given by
\begin{equation}
    \prob{n_e|\rho(0)}=
    \frac{\left(\int_0^{\Delta t}\mu(t)dt\right)^{n_e}}{n_e!}\e^{-\int_0^{\Delta t}\mu(t)dt}
    \label{eq:prob-emit}
\end{equation}
where $n_e$ is the number of emitted photons during the interval $[0,\Delta t]$.
The expected number of emitted photons is then
\begin{equation}
    \expect[n_e|\rho(0)]=\int_0^{\Delta t}\mu(t)dt.
    \label{eq:expected-emit}
\end{equation}
With our typical parameters, 
the expected number of emitted photons is on the order of a dozen.
Optimal measurement times $\Delta t$ will be discussed in \autoref{sec:cond-fisher-info}.

Notice that we are conditioning on what we call 
\textit{the pre-measurement state}, $\rho(0)$.
If the parameters in the dynamics are known and fixed, then so too is $\rho(t)$
for $0\leq t\leq \Delta t$ given the pre-measurement state.
It follows that the inhomogeneous rate function $\mu(t)$ is
conditioned upon the pre-measurement state, $\mu(t)=\mu(t|\rho(0))$, but we often
omit this for notational simplicity, writing simply $\mu(t)$.

Given the parameters of the Hamiltonian and Lindblad dissipaters, the pre-measurement
state, and an integration time, we now have a concrete method of 
calculating the expected number of photons emitted.

%=============================================================================
\subsection{The Rate Equation Simplification}
\label{sec:rate-eqn}

Due to the absence of coherence, a full open quantum system 
simulation of $\rho(t)$ is unnecessary to compute $\mu(t)$.
Instead, we can reduce our dynamics to a rate equation picture 
without further approximation.

To this end, we define the probabilities
\begin{subequations}
    \begin{align}
        p_{g0}(t)&=\Tr\left[\ketbra{\sgz}\rho(t)\right] \\
        p_{g1}(t)&=\Tr\left[(\ketbra{\sgm}+\ketbra{\sgp})\rho(t)\right] \\
        p_{e0}(t)&=\Tr\left[\ketbra{\sez}\rho(t)\right] \\
        p_{e1}(t)&=\Tr\left[(\ketbra{\sem}+\ketbra{\sep})\rho(t)\right] \\
        p_{s}(t)&=\Tr\left[\ketbra{\sss}\rho(t)\right] ,
    \end{align}
    \label{eq:prob-components}
\end{subequations}
as well as the vector $\vec{p}(t)=(p_{g0}(t),p_{g1}(t),p_{e0}(t),p_{e1}(t),p_s(t))^\T\in[0,1]^5$.
Here, $\rho(t)$ is the solution to the Lindblad master equation~\autoref{eq:lindblad-master}.
Notice that the components of this vector sum to unity, $\sum_{i=1}^5p_i(t)=1$
because the projectors used in the definitions of its components sum to the identity.

Combining the master equation with the definitions from 
Equation~\autoref{eq:prob-components}, we can compute the time evolution of each 
component of $\vec{p}(t)$.
For example, we have
\begin{align}
    \dot{p}_{g0}(t) 
        &= \Tr[\ketbra{\sgz}\dot{\rho}(t)] \nonumber \\
        &= \Tr\left[\ketbra{\sgz}\left(
            -i[H,\rho(t)]+\sum_{k}\left( L_k\rho(t)L_k^\dagger - (\rho(t)L_k^\dagger L_k+L_k^\dagger L_k\rho(t))/2 \right)		
        \right)\right] \nonumber \\
        &= -k\gamma_{eg}p_{g0}(t) + \gamma_{eg}p_{e0}(t) + \frac{\gamma_{sg}}{3}p_{s}(t)
\end{align}
where we have skipped a few lines of algebra.
In this way we end up with a set of coupled linear differential equations 
involving rates from the Lindblad operators which can be described by the matrix DE
\begin{equation}
    \begin{pmatrix}
        \dot{p}_{g0}(t) \\ \dot{p}_{g1}(t) \\ \dot{p}_{e0}(t) \\ \dot{p}_{e1}(t) \\ \dot{p}_s(t)
    \end{pmatrix}
    =
    \begin{pmatrix}
        -k\gamma_{eg} & 0 & \gamma_{eg} & \gamma_{01}/2 & \gamma_{sg}/3 \\
        0 & -k\gamma_{eg} & \gamma_{01}/2 & \gamma_{eg} & 2\gamma_{sg}/3 \\
        k\gamma_{eg} & 0 & -\gamma_{eg}-\gamma_{01}/2 & 0 & 0 \\
        0 & k\gamma_{eg} & 0 & -\gamma_{eg}-\gamma_{es}-\gamma_{01}/2 & 0 \\
        0 & 0 & 0 & \gamma_{es} & -\gamma_{sg}
    \end{pmatrix}
    \cdot
    \begin{pmatrix}
        p_{g0}(t) \\ p_{g1}(t) \\ p_{e0}(t) \\ p_{e1}(t) \\ p_s(t)
    \end{pmatrix},
\end{equation}
which we write in condensed notation as
\begin{equation}
    \dot{\vec{p}}(t)=R\cdot\vec{p}(t).
    \label{eq:rate-eqn}
\end{equation}
Notice that the columns of $R$ sum to $0$ which ensures that 
$\vec{p}(t)$ remains a probability vector as it evolves.
This condensation of the Lindblad master equation into a 
rate equation of probabilities is possible because, in our 
special case, the Hamiltonian commutes with the projectors 
and the Lindblad dissipaters have no dynamics within these 
subspaces.
This assumption would break if the NV were placed in a magnetic 
field with off-axis field components comparable to the zero-field
splittings, or if near-resonance microwave fields were turned 
on during the laser pulse.

The solution of Equation~\autoref{eq:rate-eqn} is $\vec{p}(t)=\e^{tR}\vec{p}(0)$. 
Thus the inhomogeneous Poisson rate from Equation~\autoref{eq:inhom-rate}, 
$\mu(t)=\gamma_{ea}\Tr\left[P_e\cdot\rho(t)\right]$, can be simplified to 
$\mu(t)=\gamma_{ea}(p_{e0}(t)+p_{e1}(t))$, or in terms of the initial state,
\begin{align}
    \mu(t)&=\gamma_{ea}\vec{m}\cdot\e^{tR}\vec{p}(0) \nonumber \\
    \quad\text{ with }\quad
    \vec{p}(0) & =(\Tr[\ketbra{\sgz}\rho(0)],\Tr\left[(\ketbra{\sgm}+\ketbra{\sgp})\rho(0)\right],0,0,0)^\T
\end{align}
where $\vec{m}=(0,0,1,1,0)^\T$ is the projector onto the excited space,
and where we have assumed that the pre-measurement state
$\rho(0)$ has support only on $\Hilbert_\text{ground}$.
This new expression is much more tractable as it involves just a $5\times 5$ 
matrix exponential, instead of a $49\times 49$ matrix exponential in 
superoperator space.
It also makes it simpler to derive the following relationship:
\begin{align}
    \mu(t|\rho(0))
    &= \gamma_{ea}\vec{m} \cdot \e^{tR}\vec{p}(0) \nonumber \\
    &= \gamma_{ea}\vec{m} \cdot \e^{tR}(\Tr[\ketbra{\sgz}\rho(0)],\Tr\left[(\ketbra{\sgm}+\ketbra{\sgp})\rho(0)\right],0,0,0)^\T \nonumber \\
    &= \Tr[\ketbra{\sgz}\rho(0)]\gamma_{ea}\vec{m} \cdot \e^{tR}(1,0,0,0,0)^\T \nonumber \\
        &\quad\quad + \Tr\left[(\ketbra{\sgm}+\ketbra{\sgp})\rho(0)\right]\gamma_{ea}\vec{m} \cdot \e^{tR}(0,1,0,0,0)^\T  \nonumber \\
    &= \Tr[\ketbra{\sgz}\rho(0)]\mu(t,\ketbra{\sgz}) + \Tr\left[(\ketbra{\sgm}+\ketbra{\sgp})\rho(0)\right]\mu(t,\ketbra{\sgp}) \nonumber \\
    &= p\cdot\mu(t,\ketbra{\sgz}) + (1-p)\cdot\mu(t,\ketbra{\sgp})
\end{align}
where $p=\Tr[\ketbra{\sgz}\rho(0)]$, again assuming $\rho(0)$ 
has support only on $\Hilbert_\text{ground}$.
Note that the choice of using $\ket{\sgp}$ rather than $\ket{\sgm}$ (or 
any superposition of both) was arbitrary.
Therefore the rate of photon emission 
during measurement given the 
pre-measurement state $\rho(0)$ is the convex combination of the the 
photon emission rates of the states $\ket{\sgz}$ and $\ket{\sgp}$,
where the convex combination parameter is the overlap of $\rho(0)$ 
with $\ket{\sgz}$.
It is this relationship that ultimately justifies the notion of 
taking reference measurements to calibrate the signal measurement.

An immediate corollary of this is that Equation~\autoref{eq:expected-emit} 
can be simplified to 
\begin{equation}
    \expect[n_e|\rho(0)]=p\expect[n_e|\ketbra{\sgz}] + (1-p)\expect[n_e|\ketbra{\sgp}]
    \label{eq:expect-emit-simp}
\end{equation}
which says that the expected number of emitted photons during 
measurement for the pre-measurement state $\rho(0)$ is the convex
combination of the expected number of photons for the pre-measurement 
states $\ket{\sgz}$ and $\ket{\sgp}$.
We already know that $n_e$ is always a Poisson distribution for any 
pre-measurement state (\autoref{sec:meas-description}), 
and that a Poisson  distribution is characterized 
completely in terms of its expected value, therefore once $\mu_0:=\expect[n_e|\ketbra{\sgz}]$,
$\mu_1:=\expect[n_e|\ketbra{\sgp}]$, and $p=\Tr[\ketbra{\sgz}\rho(0)]$ are known,
everything about $\prob{n_e|\rho(0)}$ is also known:
\begin{equation}
    \prob{n_e|\rho(0)}=
    \frac{\left(p\mu_0+(1-p)\mu_1\right)^{n_e}}{n_e!}\e^{-(p\mu_0+(1-p)\mu_1)}
    \label{eq:prob-emit-simp}
\end{equation}
which is a more tractable version of Equation~\autoref{eq:prob-emit}.

%=============================================================================
\subsection{Measurement Visibility and Noise}
\label{sec:meas-vis-and-noise}

There are two mechanisms that will affect our photon counting:
photons can get lost along the way to the detector, or photons can be
detected that did not originate from the NV.
In \autoref{sec:meas-description} we derived the probability of the 
NV emitting some number of photons during a measurement, $\prob{n_e}$, and in 
\autoref{sec:rate-eqn} we provided a simpler formula for the 
same quantity.

In this section we introduce two new variables, $\Gamma$ and $\eta$.
Let $\Gamma$ be the rate of dark counts per unit time, due both to noise in the photon 
counter itself and to stray photons.
Similarly, let $\eta$ be the probability that a photon emitted by the NV center will  
be collected by the detector.
This parameter is largely determined by the quality of the confocal
microscope and the solid angle of emitted photons in view.

It is useful 
to define $n_{dc}$ as the random variable representing those detected 
photons which were dark counts, and $n_{td}$ as the random variable 
representing those detected photons 
which originated from the NV, dubbed `true detections'. 
We have the relationship
\begin{equation}
    n_d = n_{dc} + n_{td}
\end{equation}
where $n_d$ is the random variable representing all detected photons 
during a single measurement window.
Note that $n_{dc}$ and $n_{td}$ are independent random variables, and also,
that they are not in practice distinguishable.

Assuming the rate of dark counts is constantly equal to $\Gamma$ over 
the a measurement duration $\Delta t$ we simply have 
\begin{equation}
    n_{dc}\sim\Poisson{\Gamma \Delta t}.
\end{equation}
The true detections are slightly more complicated. 
Each photon emitted by the NV has an (independent) probability $\eta$ 
of arriving at the detector and being detected.
We therefore have the conditional distribution
\begin{equation}
    n_{td}|n_{e} \sim \Binom{n_e}{\eta}
\end{equation}
where we are conditioning on a particular number of photons being emitted
by the NV, $n_e$.
Recall that $n_e$ is Poisson distributed with a mean which we 
label $\mu=\expect{n_e}:=\expect[n_e|\rho(0)]$ for now.
We can therefore use the law of total probability to compute
\begin{align}
    \prob{n_{td}=n}
        &= \sum_{m=0}^{\infty} \prob{n_{td}=n|n_e=m}\prob{n_e=m} \nonumber \\
        &= \sum_{m=0}^{n-1}0\cdot\prob{n_e=m}+\sum_{m=n}^{\infty} \binom{m}{n}\eta^{n}(1-\eta)^{m-n}\frac{\mu^me^{-\mu}}{m!} \nonumber \\
        &= \frac{\eta^n(1-\eta)^{-n}e^{-\mu}}{n!}\sum_{m=n}^{\infty} \frac{(1-\eta)^m \mu^m}{(m-n)!} \nonumber \\
        &= \frac{(\eta\mu)^ne^{-\eta\mu}}{n!},
\end{align}
which shows that $n_{td}$ is also Poisson distributed, with a mean 
given by $\eta\mu$.
Since the sum of two independent Poisson variables is also Poisson, 
we conclude that the random varibale of interest, $n_d=n_{td}+n_{td}$,
is Poisson distributed,
\begin{align}
    n_d &\sim \Poisson{\Gamma\Delta t + \eta \expect[n_e|\rho(0)]}.
\end{align}
This, in combination with \autoref{eq:expect-emit-simp}, gives
\begin{align}
    n_d &\sim \Poisson{\Gamma\Delta t + \eta(p\mu_0 +(1-p)\mu_1)} \nonumber \\
        &= \Poisson{p(\Gamma\Delta t +\eta\mu_0) +(1-p)(\Gamma\Delta t + \eta\mu_1)}
    \label{eq:prob-nd-general}
\end{align}
where $\mu_0:=\expect[n_e|\ketbra{\sgz}]$,
$\mu_1:=\expect[n_e|\ketbra{\sgp}]$, and $p=\Tr[\ketbra{\sgz}\rho(0)]$.

Therefore, just as $\mu_0$ and $\mu_1$ served as the maximum and minimum reference
counts for the expected emitted number of photons, $\expect[n_e]$, we see that in the 
case of measurement visibility and noise, the numbers $(\Gamma\Delta t +\eta\mu_0)$ 
and $(\Gamma\Delta t +\eta\mu_1)$ serve as the new references for $\expect[n_d]$.

%=============================================================================
\section{State Initialization}
\label{sec:state-initialization}
%=============================================================================

Simply illuminating the NV center for a period of time with 
a laser, and then waiting for the system to settle, results 
in a highly polarized spin state.
This can be shown by analysing the steady state solution to 
the rate equation.
Importantly, the equilibrium state is unique, so that no matter 
what the initial conditions are to the initialization procedure, the 
final state is the same.

%=============================================================================
\subsection{The Steady State Solution to the Rate Equation}
\label{sec:steady-state}

The rate equation derived in \autoref{sec:rate-eqn}, while much simpler 
than the Lindblad master equation it was derived from, is still 
not analytically solvable unless certain terms like $\gamma_{01}$ 
are assumed to be zero.
In order to estimate, for example, the time required to initialize the
NV center, it will be helpful to consider the steady state solutions
to the rate equation.

The rate matrix $R$ from Equation~\autoref{eq:rate-eqn} has only 
non-positive eigenvalues, and must have a non-trivial null space.
To see this, examine the solution to the rate equation,
$\vec{p}(t)=e^{tR}\vec{p}(0)$.
If $R$ had positive or complex eigenvectors, the probability 
vector $\vec{p}(t)$ would either blow up or gain complex 
entries, neither of which are allowed under a Lindblad 
master equation.
Moreover, if all of the eigenvectors were strictly negative,
$\vec{p}(t)$ would asymptote to $0$, and would therefore 
violate conservation of probability, hence at least one 
of the eigenvectors must be $0$.

The null space completely specifies the steady state solution.
Indeed, suppose that we have an initial set of populations
$\vec{p}(0)$ and decompose this into an eigenbasis 
$\vec{v}_1,...,\vec{v}_5$ for $R$, where it holds that 
$\vec{v}_1,..,\vec{v}_k$ all have eigenvalue zero, and 
$\vec{v}_{k+1},..,\vec{v}_5$ have strictly negative eigenvalues
$\lambda_{k+1},...,\lambda_5$.
This gives $\vec{p}(0)=\sum_{i=1}^5 a_i \vec{v}_i$ for some
real coefficients $a_1,...,a_5$.
The evolution is now described by
\begin{align}
    \vec{p}(t)
        &= \sum_{i=1}^5 a_i e^{Rt}\vec{v}_i 
        = \sum_{i=1}^5 a_i e^{\lambda_i t}\vec{v}_i 
        = \sum_{i=1}^k a_i \vec{v}_i +\sum_{i=k+1}^5 a_i e^{\lambda_i t} \vec{v}_i  
        \xrightarrow{t\rightarrow\infty} \sum_{i=1}^k a_i \vec{v}_i,
\end{align}
which is to say that only the population that was originally 
within the null subspace can remain in the steady state.
This analysis also shows that the non-zero eigenvalue with the smallest 
absolute value largely determines the rate of decay --- populations in 
this subspace take the longest to die.

As a technical aside, note that the rate matrix $R$ is not normal, 
so that its eigenspaces are not orthogonal.
This means that the population of the null space cannot be 
naively computed by projecting onto that subspace; a full linear 
inversion must be performed.

When the laser is off, we have $k=0$ and the null space is 
two dimensional, spanned by $(1,0,0,0,0)^\T$ and $(0,1,0,0,0)^\T$,
which correspond to the populations of the 
optical ground state -- this is not surprising.
The three remaining eigenvalues are $-(\gamma_{eg}+\gamma_{01}/2)$,
$-(\gamma_{eg}+\gamma_{es}+\gamma_{01}/2)$, and $-\gamma_{sg}$.
The first two die off quickly --
these correspond to population escaping from the optically 
excited states.
The last rate, $\gamma_{sg}$, is the most important, having an 
eigenvector $(1/3,2/3,0,0,1)^\T$, corresponds to population 
slowly seeping out of the singlet state and entering the three optical
ground state levels.
State initialization is complete once the singlet state is 
sufficiently depleted.
The process is exponential, so one need only wait some 
small multiple of $1/\gamma_{sg}$ before the population 
of $\ket{\sss}$ is vanishingly small.

When the laser is on, so that $k>0$, the eigenstructure of the 
rate matrix is not as tractable, but we can still make progress.
Firstly, it can be shown  that the null space is now only one dimensional. 
We denote the probability vector spanning the null subspace as 
$\vec{p}_{ss}=(p_{g0}^{ss},p_{g1}^{ss},p_{e0}^{ss},p_{e1}^{ss},p_{s}^{ss})^\T$.
This is extremely important because it validates the \NV 
initialization procedure: no matter what the quantum state was 
before the laser is turned on, if you let it equilibrize 
under optical illumination, it will reach a unique steady-state.

\begin{figure}
    \begin{center}
        \includegraphics[width=0.6\textwidth]{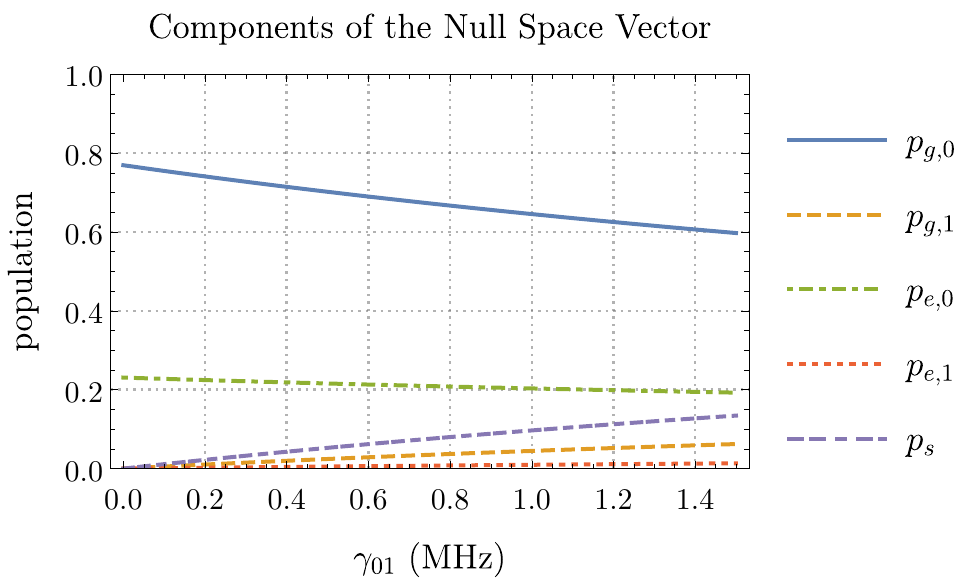}
    \end{center}
    \caption{
        The components of $\vec{p}_{ss}$,
        the vector spanning the one dimensional null space of the rate matrix
        $R$, as a function of the spin-flip rate $\gamma_{01}$.
        Population of this vector outside of the $\ket{\sez}$, $\ket{\sgz}$ subspace 
        results in imperfect polarization.
        The other rate values used are $k=0.3$, $\gamma_{eg}=\SI{77}{MHz}$,
        $\gamma_{es}=\SI{30}{MHz}$, and $\gamma_{sg}=\SI{3}{MHz}$.
    }
    \label{fig:null-space-vector}
\end{figure}

Under the approximation that $\gamma_{01}=0$, we have
$\vec{p}_{ss}=(1/(1+k),0,k/(1+k),0,0)^\T$ which has support 
only on the $m_S=0$ spin state.
We see that if $k=1$ so that the rate of spontaneous emission is equal to the 
rate of optical excitation, the ground and excited $m_S=0$ 
populations equilibrize to the same value of $1/2$.
The null eigenspace is much more complicated when $\gamma_{01}>0$.
Though an analytic expression can be derived, it is a large 
unhelpful mess of divisions and multiplications of the various rates
which we have been unable to simplify.
To gain some insight, in \autoref{fig:null-space-vector} we 
plot the components of $\vec{p}_{ss}$ as a function of $\gamma_{01}$.
As $\gamma_{01}$ increases, two properties are apparent:
the steady-state population of the $m_S=\pm 1$ spin states 
increases, as expected, but also, the steady-state singlet 
state population increases dramatically. 
The latter effect occurs because the singlet state's relatively long lifetime 
makes it a good storage location, and now due to $\gamma_{01}$, there 
is always some population to feed it.
Indeed, expanding $\vec{p}_{ss}$ in a $\gamma_{01}$ power series about 
$0$, the steady state population of the singlet is 
$p_s^{ss}=\frac{3k\gamma_{01}}{2(k+1)\gamma_{sg}}+\mathop{O}(\gamma_{01}^2)$.

%=============================================================================
\subsection{Imperfect Preparation: The Pseudo-pure State}
\label{sec:imperfect-prep}

Due to small but important spin-non-conserving terms in the master equation 
during continuous optical excitation (Equation~\autoref{eq:spin-non-cons-lindblad}), 
the steady state density matrix 
must have non-zero support on the states $\ket{\sgm}$,
$\ket{\sgp}$, $\ket{\sem}$, and $\ket{\sep}$.
This can be seen in \autoref{fig:null-space-vector}.
This means that the initialization process does not asymptote to the pure state
$\ket{\sgz}$, but rather to a mixed state $\rho_0$ which is mostly pure.
Since our optical model does not distinguish between $m_s=+1$ and $m_s=-1$, 
they must have equal population in the pumping steady-state,
and therefore $\rho_0$ can be written as a pseudo-pure state \cite{cory_ensemble_1997},
\begin{equation}
    \rho_0 = q \ketbra{\sgz} + (1-q)\frac{\I}{3},
\end{equation}
where the purity parameter $q$ depends in
a non-trivial way on all of the parameters of the rate equation, but
generally decreases as the rates of spin-non-conserving processes increase.
It is pseudo-pure in the sense that any unitary (and in general, unital) 
process only acts on the first term,
\begin{equation}
    U\rho_0U^\dagger = qU\ketbra{\sgz}U^\dagger + (1-q)\frac{\I}{3}
\end{equation}
so that its lack of complete purity, in practice, serves only to limit 
the contrast of measurement.

To see this, we begin by computing the expected number photons emitted during the 
measurement of our preparation procedure $\rho_0$ using 
Equation~\autoref{eq:expect-emit-simp},
\begin{align}
    \mu_0'
        &:=\expect[n_e|\rho_0] 
        = \frac{1+2q}{3}\mu_0 
        + \frac{2-2q}{3}\mu_1.
\end{align}
Next, if prior to measuring $\rho_0$ we perform an operation 
$\ket{\sgz}\mapsto\ket{\sgp}$, which may be implemented as 
an adiabatic inversion with a microwave pulse, we have a 
pre-measurement state
\begin{equation}
    \rho_1 = q\ketbra{\sgp} + (1-q)\frac{\I}{3}
\end{equation}
which when measured emits an expected number of photons
\begin{align}
    \mu_1'
        &:=\expect[n_e|\rho_1] 
        = \frac{1-q}{3}\mu_0 
        + \frac{2+q}{3}\mu_1.
\end{align}
Previously we had considered $\mu_0$ and $\mu_1$ as the quantities that
define the reference measurements in the absence of noise.
However, given that $\rho_0$ is the 
best achievable initial state using standard techniques, 
it is in practice more convenient to 
use $\mu_0'$ and $\mu_1'$.
Indeed, if we prepare the state $\rho_0$ and perform any unital operation 
resulting in a pre-measurement state
\begin{equation}
    \rho=q\rho_\psi + (1-q)\frac{\I}{3},
    \label{eq:arb-allowed-state}
\end{equation}
then some simple algebra shows that
\begin{align}
    \expect[n_e|\rho]
        &= p\mu_0' + (1-p)\mu_1'
\end{align}
where $p=\Tr[\rho_\psi\ketbra{\sgz}]$.

Finally, if we take into consideration finite visibility $\eta$ and 
dark count rate $\Gamma$ as discussed in \autoref{sec:meas-vis-and-noise},
we may define
\begin{align}
    \alpha &\defeq \expect[n_d|\rho_0] = \Gamma\Delta t + \eta\mu_0' \nonumber \\
    \beta &\defeq \expect[n_d|\rho_1] = \Gamma\Delta t + \eta\mu_1' \nonumber \\
    \gamma &\defeq \expect[n_d|\rho].
\end{align}
to arrive at
\begin{align}
    \expect[n_d|\rho] &= \gamma = p\alpha + (1-p)\beta \nonumber \\
    n_d|\rho &\sim \Poisson{p\alpha + (1-p)\beta}
\end{align}
which is analagous to Equation~\autoref{eq:prob-nd-general} but for our pseudo-pure 
state preparation.
The quantity $\gamma$, which we will call the \textit{signal}, is by 
definition conditioned on the pre-measurement 
state $\rho$, and will henceforward generically refer to the expected 
number of detected photons given the pre-measurement state of interest or, 
equivalently, the experiment of interest which when performed on the 
preparation state $\rho_0$ yields $\rho$.
We will call the quantities $\alpha$ and $\beta$ our \textit{references} 
because they bound the expected values of detected photons for arbitrary 
states of the form in Equation~\autoref{eq:arb-allowed-state}.
More specifically, $\alpha$ is the \textit{bright reference} and 
$\beta$ is the \textit{dark reference} since $\alpha>\beta$.

%=============================================================================
\section{Drifting References}
\label{sec:drifting-refs}
%=============================================================================

One of the main complications of experimental NV measurement is drifting of the
references $\alpha$ and $\beta$ in time.
Though there are many mechanisms which can cause this, for us, the most prevalent 
is due to relative movement between the NV center and the 
focal spot of the confocal microscope.
This region of focus has a 3D Gaussian profile, and 
as temperatures or other properties of the lab change in time, 
movement of the center of this region off of the point-sized NV center causes 
both a drop-off of delivered laser power, $k$, and 
collection efficiency, $\eta$.
Other mechanisms may include fluctuation of the laser power, and 
quality of the confocal microscope's alignment, which will affect $\eta$, $k$,
and $\Gamma$.
If left unchecked, the drift would eventually cause the NV under study 
to no longer be in the focal region at all.
To avoid this, a \textit{tracking} procedure is periodically run, whose 
purpose is to recenter the NV with the focal region by taking a series 
spatial of images and using feedback to realign.

%=============================================================================
\subsection{Experiment Ordering}
\label{sec:exp-ordering}

\begin{figure}
    \begin{center}
        \includegraphics[width=\textwidth]{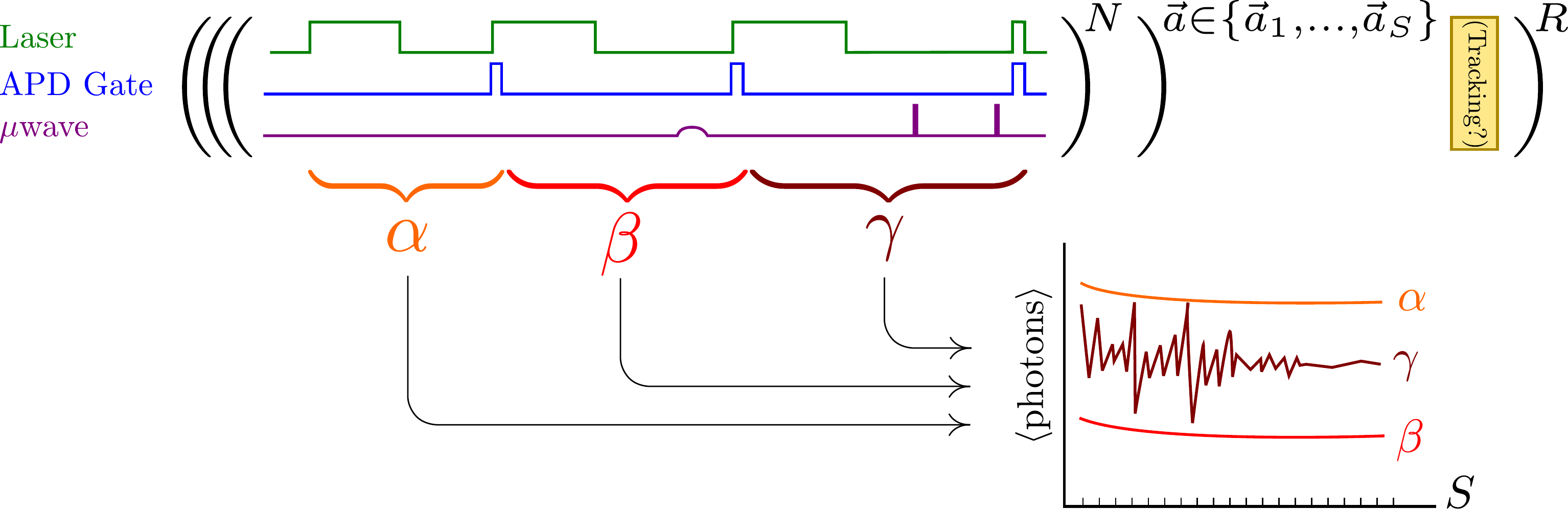}
    \end{center}
    \caption{
        The considered experimental ordering.
        Another popular ordering transposes the inner two levels.
        Given a particular parameter configuration of the experiment, 
        $\vec{a}\in\{\vec{a}_1,...,\vec{a}_S\}$ (in the above example, 
        the parameter is the distance between the two last microwave pulses),
        $N$ repetitions are performed of both the experiment, $\gamma$, 
        and the references, $\alpha$ and $\beta$.
        The bright reference $\alpha$ is measured by initializing with a
        laser pulse, waiting for metastable optical states to decay, and 
        taking a measurement by opening the APD counting gate while the laser is on.
        The dark reference $\beta$ is similar, except an inversion pulse is applied
        prior to measurement.
        The pulse sequence prior to the reference measurement $\gamma$ depends on the
        current parameter $\vec{a}$.
        Each time $N$ repetitions have been made of all $S$ parameter configurations, 
        the system decides whether to track or not, and this is all 
        repeated $R$ times.
        A sketch of the resulting data is shown, averaged over both $N$ and $R$.
    }
    \label{fig:experiment-sequence}
\end{figure}

Experiments typically have a set of parameters that are varied.
For the sake of concreteness, we choose the specific experiment and repetition 
ordering described in \autoref{fig:experiment-sequence}; 
variations of this ordering may require a modified (though similar) analysis to that which follows.
We denote the list of experiment parameter configurations as $\vec{a}_1,...,\vec{a}_S$.
For example, in the case of a Ramsey experiment for magnetometry 
\cite{taylor_high-sensitivity_2008}, the distance between two
$\frac{\pi}{2}$ pulses is varied, so that $\vec{a}_s=(t_s)$ where 
$t_s$ is the pulse spacing.
If additionally the phase of the second Ramsey pulse is varied in 
proportion to the spacing, we have instead $\vec{a}_s=(t_s,\omega t_s)$ 
for some angular frequency $\omega$.
A parameter configuration $\vec{a}\in\{\vec{a}_1,...,\vec{a}_S\}$ is fixed and 
a large number, $N$, of back-to-back \textit{repetitions} with this parameter configuration
is performed before moving on to the next.
The choice of $N$ is motivated, for instance, by the time required for experimental
control hardware to switch between choices of configuration $\vec{a}$.
Each time all parameter configurations have been dealt with, the tracking procedure 
is run, and then the entire procedure is repeated $R$ times, which we call
\textit{averages}.

A signal measurement $\gamma$ depends only on the current value of $\alpha$ 
and $\beta$, the current parameter configuration $\vec{a}$, and any noise 
operations acting on $\vec{a}$.
We denote the true value of $\alpha$, $\beta$, and $\gamma$ at the 
$n^\text{th}$ repetition of the $s^\text{th}$ 
experiment $\vec{a}_s$ in the $r^\text{th}$ average as 
$\alpha_{n,s,r}$, $\beta_{n,s,r}$, and $\gamma_{n,s,r}$, for
$1\leq n\leq N$, $1\leq s\leq S$, and $1\leq r\leq R$.
It holds that  $\gamma_{n,s,r}=p_s\alpha_{n,s,r}+(1-p_s)\beta_{n,s,r}$
where we have used our assumption of the independence of $p$ from the
drift processes to label $p$ by only $s$.
This yields the random variables
\begin{align}
    X_{n,s,r} &\sim \Poisson{\alpha_{n,s,r}} \nonumber \\
    Y_{n,s,r} &\sim \Poisson{\alpha_{n,s,r}} \nonumber \\
    Z_{n,s,r} &\sim \Poisson{\alpha_{n,s,r}}
\end{align}
with the corresponding variates $(x_{n,s,r},y_{n,s,r},z_{n,s,r})_{n,s,r=1}^{N,S,R}$.

%=============================================================================
\subsection{Combining Experiments}
\label{sec:combining-experiments}

Due to low visibility, the repetition number $N$ in 
\autoref{fig:experiment-sequence} is often quite high.
It is usually cumbersome to store the results of the experiment for each 
individual measurement.
We therefore assume that the data is summed over the $N$ repetitions.
Because of the additive property of the Poisson distribution,
if we define $X_{s,r}\defeq \sum_{n=1}^NX_{n,s,r}$,
$Y_{s,r}\defeq \sum_{n=1}^NY_{n,s,r}$, and 
$Z_{s,r}\defeq \sum_{n=1}^NZ_{n,s,r}$
we get
\begin{align}
    X_{s,r} &\sim \Poisson{\alpha_{s,r}} \nonumber \\
    Y_{s,r} &\sim \Poisson{\alpha_{s,r}} \nonumber \\
    Z_{s,r} &\sim \Poisson{\alpha_{s,r}}
\end{align}
where $\alpha_{s,r}\defeq \sum_{n=1}^N\gamma_{n,s,r}$,
$\beta_{s,r}\defeq \sum_{n=1}^N\gamma_{n,s,r}$,
and $\gamma_{s,r}\defeq \sum_{n=1}^N\gamma_{n,s,r}$.

In a slight abuse of notation, in \autoref{sec:stats-of-meas} and on, when
 we talk about unscripted $\alpha$, $\beta$, and $\gamma$, we
are referring to $\alpha_{s,r}$, $\beta_{s,r}$, and $\gamma_{s,r}$ 
for a particular index $(s,r)$
where some $N\geq 1$ is implicitly understood.
Furthermore, in this context $\delta$ will refer to $\sum_{n=1}^N \Gamma_{n,s,r}\Delta t$,
that is, the total contribution to the dark counts.
Similarly, $\nu$ will refer to $\sum_{n=1}^N \eta_{n,s,r}(\mu_0')_{n,s,r}$ and 
$\kappa$ to $\frac{1}{\nu}\sum_{n=1}^N \eta_{n,s,r}(\mu_1')_{n,s,r}$.
The fraction $0<\kappa< 1$ represents how much dimmer the reference state $\rho_1$ 
is as compared to $\rho_0$.
This results in the relationship
\begin{align}
    \alpha &= \delta + \nu \nonumber \\
    \beta &= \delta + \kappa\nu.
    \label{eq:drift-params-breakdown}
\end{align}

%=============================================================================
\subsection{Correlations of Model Parameters}
\label{sec:param-correlations}

The three quantities $\mu_0'$, $\mu_1'$, and $\eta$ all depend on the quality 
of the coupling between the confocal microscope and the NV defect; as the 
relative displacement between the defect and the center of focus drifts, 
$\eta$ decreases.
This could be due to, for example, temperature changes in the lab
which expand or contract the components on the optical table.
However, $\mu_0'$ and $\mu_1'$ also change with decreased optical coupling 
because the preparation state $\rho_0$ depends on the laser power 
which is coupling dependent.
Therefore we expect correlations between $\mu_0'$, $\mu_1'$, and $\eta$.

Nominally $\Gamma\Delta t$ should be independent of each of the quantities 
$\eta$, $\mu_0'$, $\mu_1'$, and $p$.
However, this may break if the power of the laser varies in time and a significant 
portion of the dark counts are caused by unwanted reflections of, or excitations 
due to the laser; we may end up with correlations between $\Gamma\Delta t$ and 
each of $\eta$, $\mu_0'$, and $\mu_1'$.

Finally, and perhaps most importantly, the quantity $p$ will normally be independent of 
the quantities $\eta$, $\mu_0'$, $\mu_1'$, and $\Gamma \Delta t$.
However, it is also possible for this to fail.
For example, if the process that takes the preparation state $\rho_0$ 
to the pre-measurement state $\rho_0$ is
non-unital, we will not end up with exactly the form $q\rho_\phi+(1-q)\I/3$ 
(as seen in \autoref{sec:imperfect-prep}) leading 
to errors when inferring $p$ from $p\alpha + (1-p)\beta$.
Non-unitality could occur due to $T_1$ relaxation, or leakage 
of laser light when it is supposed to be off.

%=============================================================================
\subsection{A Stochastic Model of Drift}
\label{sec:stoch-model}

The references $\alpha$ and $\beta$ are best viewed as 
stochastic processes with autocorrelations in time.
As discussed above, they will also be correlated 
with each other.
They will undergo a discontinuous jump every time a tracking operation 
is performed.

As derived in \autoref{sec:imperfect-prep}, conditioned on set of parameters, 
including $k$, $\Gamma$, $\Delta t$, $\eta$, $\gamma_{es}$, $\gamma_{sg}$, 
$\gamma_{eg}$, and $\gamma_{01}$ the references are given by
$\alpha= \Gamma\Delta t + \eta \mu_0'$ and
$\beta= \Gamma\Delta t + \eta \mu_1$
at a particular instance in time.
Given the number variables and unknowns that likely go into the drift process itself 
(which will in turn affect $k$, $\Gamma$, and $\eta$) on top of the already 
complex conditional model stated above, writing down an analytic model 
for the stochastic processes $\alpha$ and $\beta$ would be difficult, if not 
impossible.

We therefore restrict our attention to simpler effective models which still well 
describe expected and observed behaviour.
We assume that the stochastic process $(\alpha,\beta)$ is a Gaussian process.
This is a weak assumption especially given that in later sections we will only 
make use of the first two moments.

We may relate this continuous stochastic process to the discrete 
variables defined in \autoref{sec:exp-ordering} with a standard
discretization as follows.
Consider the $r^\text{th}$ average of the experiment and 
a particular realization of the stochastic process $(\alpha(t),\beta(t))$ 
during this average.
Then we have that
$\alpha_{n,s,r}=\alpha(t_{n,s})$ and $\beta_{n,s,r}=\beta(t_{n,s})$
where $t_{n,s}$ is the time of the $n^\text{th}$ repetition of the $s^\text{th}$
experiment relative to the start of the $r^\text{th}$ average.
This gives, for example, the discrete Cox process 
$X_{n,s,r}\sim\Poisson{\alpha(t_{n,s})}$.

We take the time value $t=0$ to mean the time directly after performing 
a tracking operation.
The tracking operation has the effect of drawing the initial values 
$\alpha(0)$ and $\beta(0)$ from a fixed normal distribution 
\begin{equation}
    (\alpha(0),\beta(0)) \sim \Normal{(\alpha_0,\beta_0)}{\Sigma_0}
\end{equation}
where the variances in $\Sigma_0$ are determined by the noise and error in the 
tracking procedure, and the mean is a property of the confocal microscope's
quality and the NVs optical properties.
Assuming a Gaussian stochastic process leads to the distribution
\begin{equation}
    (\alpha(t),\beta(t)) \sim \Normal{(\alpha_t,\beta_t)}{\Sigma_t}
\end{equation}
at time $t\geq 0$.
%Note that for a Gaussian process, marginalizing the initial conditions
%over a Gaussian distribution at $t=0$ can be made equivalent to setting 
%some specific initial condition at some $t<0$.

As a concrete example, consider the stochastic model
\begin{align}
    \alpha_1 &\sim \Normal{\alpha_0}{\sigma_\alpha} \nonumber \\
    \nu &\sim \OU{0,\sigma_\nu,\theta_\nu,\alpha_1-\Gamma\Delta t} \nonumber \\
    \kappa &\sim \OU{\kappa_0,\sigma_\kappa,\theta_\kappa,0} \nonumber \\
    (\alpha,\beta) &= (\Gamma\Delta t + \nu,\Gamma\Delta t+\kappa\cdot\nu).
    \label{eq:stoch-model}
\end{align}
Here, $\Gamma\Delta t$ are the expected dark counts in a single measurement, 
which we have assumed to be deterministic and constant for simplicity.
However, the expected number of photons due to the bright state $\rho_0$, 
denoted an $\nu$, is an 
Ornstein--Uhlenbeck process with long-time mean $0$, volatility
$\sigma_\nu$, mean reversion speed $\theta_\nu$, and initial value 
$\alpha_1-\Gamma\Delta t$, where the initial value is marginalized 
over $\Normal{\alpha_0}{\sigma_\alpha}$ representing 
imperfections in the tracking process.
This implies that, on average, $\alpha(0)=\alpha_0$ and 
$\alpha(\infty)=\Gamma\Delta t$ with average decay rate $\mu_\nu$, where `shakiness' 
in getting there is determined by $\sigma_\nu$.
In order to correlate $\alpha$ with $\beta$, and also to help enforce the 
physical constraint $\beta\leq\alpha$, we relate the two components 
with another Ornstein--Uhlenbeck process $\kappa$.

Note that this model does not guarantee, for example, that $\alpha(t) >0$
or that $0\leq \kappa(t) \leq 1$.
We can only make these scenarios improbable by choosing low volatilities.
This is the compromise of having such a simple model in terms of 
the well known Ornstein-Uhlenbeck Gaussian process.

\begin{figure}
    \begin{center}
        \includegraphics[width=0.5\textwidth]{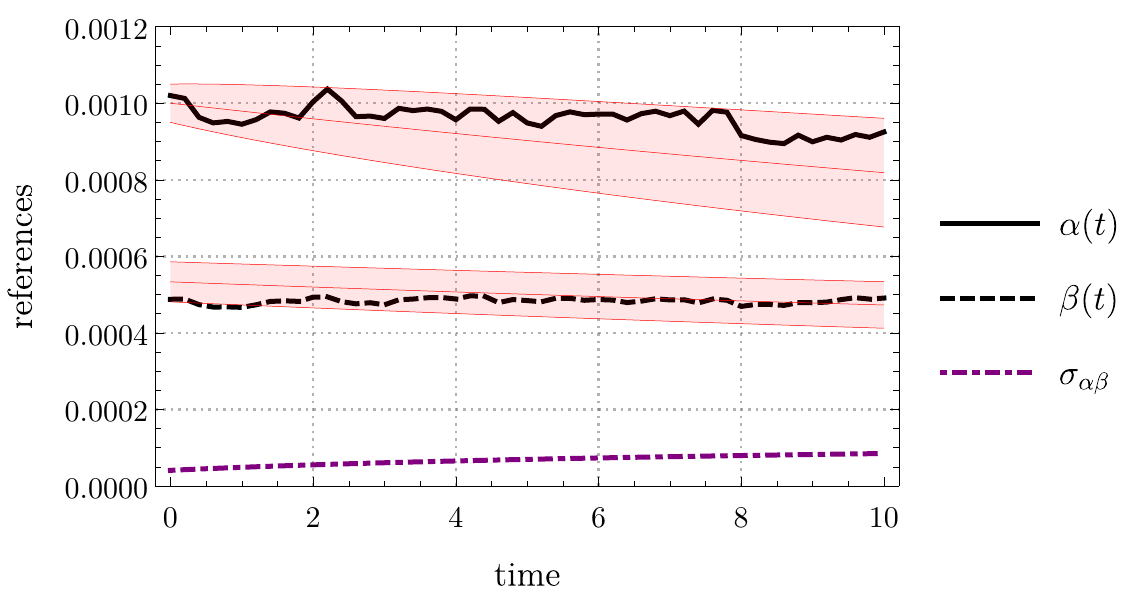}
    \end{center}
    \caption{Simulation of a severe case of drift.	
    A random instance of the process $(\alpha(t),\beta(t))$ defined in Equation~\autoref{eq:stoch-model} is shown 
    on top of their first moments with a shaded single standard deviation. 
    The dot-dashed purple line shows the square root of covariance.
    The model parameters used are $\alpha_0=10^{-3}$, $\sigma_\alpha=5\times 10^{-5}$, $\sigma_\nu=5\times 10^{-5}$, $\theta_\nu=0.03$, $\Gamma\Delta t=3\times 10^{-4}$, $\kappa_0=1/3$, $\sigma_\kappa=0.01$, and $\theta_\kappa=0.01$.
The time units are arbitrary; scaling the x-axis is equivalent to scaling $\theta_\nu$ and $\theta_\kappa$.}
    \label{fig:stoch-variate}
\end{figure}

Solving for the moments of this stochastic process we arrive at the expressions
\begin{align}
    \expect[\alpha(t)] &= \Gamma\Delta t + (\alpha_0-\Gamma\Delta t)\e^{-t\theta_\nu} \nonumber \\
    \expect[\beta(t)] &= \Gamma\Delta t + \kappa_0(\alpha_0-\Gamma\Delta t)\e^{-t\theta_\nu} \nonumber \\
    \Var[\alpha(t)] &= \sigma_\alpha^2\e^{-2 t \theta_\nu}
        +\sigma_\nu^2\frac{(1-\e^{-2 t \theta_\nu})}{2\theta_\nu} \nonumber \\
    \Var[\beta(t)] &= \kappa_0^2\sigma_\alpha^2\e^{-2 t \theta_\nu}
        + \sigma_\kappa^2\frac{((\alpha_0-\Gamma\Delta t)^2 + \sigma_\alpha^2) \e^{-2t\theta_\nu}}{2\theta_\kappa}
        + \sigma_\nu^2\frac{(2\theta_\kappa\kappa_0^2+1)(1-\e^{-2 t \theta_\nu})}{4\theta_\kappa\theta_\nu} \nonumber \\
    \Cov[\alpha(t),\beta(t)] &= \kappa_0\sigma_\alpha^2\e^{-2 t \theta_\nu}
        + \frac{\kappa_0\sigma_\nu^2(1-\e^{-2t\theta_\nu})}{2\theta_\nu}.
    \label{eq:stoch-moments}
\end{align}
These calculations can be found in \autoref{app:stoch-moments-code}.
In \autoref{fig:stoch-variate}, these moments are plotted along with a single 
random trajectory of the stochastic process defined above.

The purpose of this section has been to demonstrate that it is possible to 
meaningfully model the drift process as a stochastic process, which 
leads to reference count variances and covariances at each time 
step.
These variances will correspond to the moments of the hyperparameter 
distribution for the references described in \autoref{sec:stat-model}.

%=============================================================================
\section{Statistical Models of Measurement}
\label{sec:stats-of-meas}
%=============================================================================

In this section we state the measurement of an NV center as a 
statistical inference problem.
This statement is a direct result of the derivations of the previous sections,
however, it is intentionally written so that previous sections may 
be ignored by those presently uninterested in the physical derivation of 
the model.

%=============================================================================
\subsection{The Statistical Model of Measurements}
\label{sec:stat-model}

Consider the inner-product space $\Hilbert=\Complex^3$ with the canonical
basis $\ket{1}=(1,0,0)^\T$, $\ket{0}=(0,1,0)^\T$, and $\ket{-1}=(0,0,1)^\T$.
We call a state of the system prior to the measurement procedure the 
pre-measurement state.
Suppose that the pre-measurement state of interest is given by the 
density matrix $\rho\in\Density(\Complex^3)$.
Define $p=\Tr[\rho P_0]$ where $P_0$ is the projector onto $\ket{0}$.
In the case of a strong quantum measurement we would have access to 
random variables drawn from the distribution $\Bernoulli{p}$.
Instead, however, we have access to the random triplet 
$(X,Y,Z)|\alpha,\beta$ defined by
\begin{align}
    X &\sim \Poisson{\alpha} \nonumber \\
    Y &\sim \Poisson{\beta} \nonumber \\
    Z &\sim \Poisson{p\alpha+(1-p)\beta} \equiv \Poisson{\gamma} 
\end{align}
where $\alpha$ is the expected number of photons collected in $N$ 
independent measurements of the pre-measurement state $\ket{0}$, 
$\beta$ is the expected number of photons collected in $N$ 
independent measurements of the pre-measurement state $\ket{1}$
\footnote{For those astute readers who have read the previous sections:
we are abusing language slightly to keep the statement of the 
model compact; $\alpha$ and $\beta$ actually refer to the expected photons
collected due to the pseudo-pure states $\rho_0$ and $\rho_1$.}, 
and $\gamma$ is the expected number of photons collected in $N$ 
independent measurements of the pre-measurement state $\rho$.
It is always true that $0\leq p\leq 1$ and $0\leq \beta \leq \gamma\leq \alpha$.

The references $\alpha$ and $\beta$ are in turn random variables 
drawn from the distribution
\begin{align}
    \matrixtwobyone{\alpha}{\beta}
        \sim \Normal{\matrixtwobyone{\bar\alpha}{\bar\beta}}{\matrixtwobytwo{\sigma_\alpha^2}{\sigma_{\alpha\beta}}{\sigma_{\alpha\beta}}{\sigma_\beta^2}}
\end{align}
with $\sigma_{\alpha\beta}>0$.
The normality of this distribution is typically irrelevant, and is 
stated as such just to be concrete. 
We are usually only interested in its first two moments.
The `true' distribution is almost certainly quite complicated, and arises 
from the stochastic processes described in \autoref{sec:drifting-refs}.
In later sections this multinormal distribution will be replaced with a 
product gamma distribution, or a mixture of product gamma distributions,
as discussed in \autoref{app:conjugate-priors}.
To be clear, note that when a variate $(x,y,z)$ is sampled from this distribution,
all three variates are conditional on the same values of $\alpha$ and $\beta$.

This creates a hierarchical model with nuisance hyperparameters $\bar\alpha$, 
$\bar\beta$, $\sigma_\alpha$, $\sigma_\beta$, and $\sigma_{\alpha,\beta}$.
We will see that this second layer is rarely useful, so that the conditional 
model $(X,Y,Z)|\alpha,\beta$ should usually be used in practice.
However, it is important to remember that the second layer exists, because it 
makes it clear that multiple identical samples cannot 
be taken from the conditional model.
Indeed, the values of $\alpha$ and 
$\beta$ will change each time the set of $N$ measurements are made.

%=============================================================================
\subsection{Moment Calculations}
\label{sec:model-moments}

We first work out the first two moments of the random variables $X$,
$Y$, and $Z$, both conditional and unconditional on the hyperparameters 
$\alpha$ and $\beta$.

The conditional moments are trivially given by
\begin{align}
    \expect[X|\alpha] &= \Var[X|\alpha] = \alpha \nonumber \\
    \expect[Y|\beta] &= \Var[Y|\beta] = \beta \nonumber \\
    \expect[Z|p,\alpha,\beta] &= \Var[Z|p,\alpha,\beta] = p\alpha + (1-p)\beta \nonumber \\
    \text{and } \Cov[X,Y|\alpha,\beta] &= 0
    \label{eq:cond-model-means}
\end{align}
using basic properties of the Poisson distribution.
The law of total expectation can be used to compute
\begin{align}
    \expect[X] &= \expect_{\alpha,\beta}[\expect[X|\alpha]] = \bar\alpha \nonumber \\
    \expect[Y] &= \expect_{\alpha,\beta}[\expect[Y|\beta]] = \bar\beta \nonumber \\
    \text{and }  \expect[Z] &= \expect_{\alpha,\beta}[\expect[Y|p,\alpha,\beta]] = p\bar\alpha + (1-p)\bar\beta,
    \label{eq:model-means}
\end{align}
showing that the variance of $\alpha$ and $\beta$ do not affect the mean.
Similarly, the law of total variance gives
\begin{align}
    \Var[X] &= \expect_{\alpha,\beta}[\Var[X|\alpha]] + \Var_{\alpha,\beta}[\expect[X|\alpha]] 
            = \bar\alpha + \sigma_\alpha^2 \nonumber \\
    \Var[Y] &= \expect_{\alpha,\beta}[\Var[Y|\beta]] + \Var_{\alpha,\beta}[\expect[Y|\beta]] 
            = \bar\beta + \sigma_\beta^2 \nonumber \\
    \Var[Z] &= \expect_{\alpha,\beta}[\Var[Z|p,\alpha,\beta]] + \Var_{\alpha,\beta}[\expect[Z|p,\alpha,\beta]] 
            = p\bar\alpha + (1-p)\bar\beta + p^2\sigma_\alpha^2 + (1-p)^2\sigma_\beta^2 + p(1-p)\sigma_{\alpha,\beta}
\end{align}
which shows that the variances have two parts, one due to the 
usual finite sampling error of a Poisson variable, and one 
due to the underlying fluctuation of the Poisson parameters.

%=============================================================================
\subsection{Three Advantages of the Conditional Model}

If the measurement model is not stated as concretely as it was 
in \autoref{sec:stat-model}, there can be a slight subtlety 
in the interpretation of random variates.
Misunderstanding this point could result in reporting incorrect 
error bars or confidence intervals/credible regions.
In an attempt to be as clear as possible, we illustrate with an example.

Suppose Yves and Zoey together collect $R$ variates of the 
random variable $(X,Y,Z)$, sampled 
identically and independently, giving the results 
$(x_1,y_1,z_1),(x_2,y_2,z_2),...,(x_R,y_R,z_R)$.
They go back to their separate offices and try to analyse the data.
Yves calculates the sample mean and variance of $x_1,x_2,...,x_R$
as $x_\text{samp}=\frac{1}{R}\sum_{r=1}^R x_r$ and 
$\sigma_{x,\text{samp}}^2=\frac{1}{R-1}\sum_{r=1}^R(x_r-x_\text{samp})^2$,
respectively.
He gets $x_\text{samp}=200$ and $\sigma_{x,\text{samp}}=20$.
Looking at Equation~\autoref{eq:model-means}, and using the standard error of 
the mean, this informs his rough belief that 
$\bar\alpha\approx x_\text{samp}\pm \sigma_{x,\text{samp}}/\sqrt{R}=200\pm 20/\sqrt{R}$.

Zoey recalls that each variate $x_r$ was obtained by summing $N$ independent 
measurements.
She wonders why they bothered batching the results into $R$ sets, and why 
they didn't just take $N\times R$ measurements to begin with.
She therefore does a sum to get the new quantity $x_\text{sum}=\sum_{r=1}^R x_r$.
She knows that for each $1\leq r\leq R$, $x_r$ was drawn from the distribution 
$\Poisson{\alpha_r}$ for some specific but unknown value of $\alpha_r$.
Therefore, knowing the summative property of Poisson distributions, she 
correctly deduces that $x_\text{sum}$ was sampled from $\Poisson{\sum_{r=1}^R \alpha_r}$.
The Poisson distribution's standard deviation is the square-root of its mean,
she therefore adopts the rough belief that 
$\sum_{r=1}^R \alpha_r \approx x_\text{sum}\pm\sqrt{x_\text{sum}}$, and therefore that 
$\frac{1}{R}\sum_{r=1}^R \alpha_r \approx x_\text{sum}/R \pm \sqrt{x_\text{sum}/R})=200\pm 14/\sqrt{R}$.

Zoey and Yves get back together and compare their results,
and wonder why Zoey is more confident than Yves about 
the quantity she has estimated, when, naively, it seems that they
have estimated the same thing with the same data.

The discrepancy comes down to the fact that they are estimating parameters 
using different models.
Yves is estimating the hyperparameter $\bar\alpha$ from the hierarchical 
model in \autoref{sec:stat-model}, justified by the moment calculations 
in \autoref{sec:model-moments}.
Zoey is foregoing the hyperparameter layer of the model and making a direct 
inference about the sum of the \textit{particular} references 
$\alpha_1,\alpha_1,...,\alpha_R$ they happened to draw in their measurements. 
Neither is wrong, they are simply estimating different but related 
quantities.

We saw in \autoref{sec:model-moments} that 
$\Var[X|\alpha] = \bar\alpha$
and 
$\Var[X] = \bar\alpha + \sigma_\alpha^2$.
Yves was assuming he made $R$ independent measurements of 
$X$ and Zoey was assuming she made a single measurement of 
$X|\alpha$ with a combined $\alpha=\sum_{r=1}^R \alpha_r$.
Therefore the difference between their error bars is due 
to $\sigma_\alpha$.

The end goal, of course, is not to estimate the reference $\alpha$ 
or its mean, but to estimate quantities related to the 
quantum state like $p=\Tr[\rho P_0]$.
However, the better one's accuracy in estimating the references, the better one's accuracy in $p$.
Therefore, given the above discussion, it is apparent that there 
is no advantage to drawing multiple samples of $(X,Y,Z)$ when it 
is possible to increase the number of measurements $N$ instead, often 
by adding samples together as Zoey did.
There are, however, special circumstances where drawing multiple 
samples is desirable.
This occurs in cases where $p$ is not fixed shot to shot, but is instead
drawn from a distribution on each shot.

There is a second and less obvious advantage to Zoey's method.
Yves' model assumes that the multinormal distribution on $(\alpha,\beta)$ 
is a good approximation to the moments of the stochastic process governing
$\alpha$ and $\beta$ discussed in \autoref{sec:stoch-model}.
This will usually be good enough.
But if, for example, there are daily
temperature patterns in the lab which significantly affect optical alignment, 
this could cause the normal approximation to be inaccurate.
Zoey's model, however, makes no assumptions about the nature of the drift.

Finally, the third advantage of the conditional model is that it is 
simpler and therefore more tractable.
Just solving for the MLE of the hierarchical model analytically would be 
difficult or impossible; it is painful enough for the conditional model.

%=============================================================================
\subsection{Basic Inference Problem}
\label{sec:basic-inference-problem}

We state the basic NV measurement inference problem for both the conditional
and hierarchical models.

For the conditional model, recall that we are, although without much loss 
of generality, limited to drawing a single sample $(x,y,z)$ from 
$(X,Y,Z)|\alpha,\beta$.
The inference problem is, given the likelihood function
\begin{align}
    \Lhood(	p,\alpha,\beta|x,y,z	) 
        &= \prob{X=x,Y=y,Z=z|p,\alpha,\beta} \nonumber \\
        &= \pdf{Pois}{x;\alpha} \cdot \pdf{Pois}{y;\beta} \cdot \pdf{Pois}{z;p\alpha+(1-p)\beta} \nonumber \\
        &= \frac{\alpha^x\e^{-\alpha}}{x!}
            \cdot \frac{\beta^y\e^{-\beta}}{y!}
            \cdot \frac{(p\alpha+(1-p)\beta)^z\e^{-(p\alpha+(1-p)\beta)}}{z!},
    \label{eq:cond-likelihood}
\end{align}
to infer the value of $p$.
Note that $\alpha$ and $\beta$ are nuisance parameters.

In the case of the hierarchical model, if we take $R$ iid samples 
$(x_r,y_r,z_r)$ from $(X,Y,Z)$ we have the likelihood function
\begin{align}
    \Lhood(
        p,\bar\alpha,&\bar\beta,\sigma_\alpha,\sigma_\beta,\sigma_{\alpha,\beta}|
        (x_1,y_1,z_1),(x_2,y_2,z_2),...,(x_R,y_R,z_R)
    ) = \nonumber \\
        & \prod_{r=1}^R  \int_{-\infty}^\infty\int_{-\infty}^\infty
            \pdf{Pois}{x_r;\alpha} \cdot \pdf{Pois}{y_r;\beta} \cdot \pdf{Pois}{z_r;\gamma}
            \cdot \pdf{Norm}{\alpha,\beta;\inlinetwobyone{\bar\alpha}{\bar\beta},\inlinetwobytwo{\sigma_\alpha^2}{\sigma_{\alpha\beta}}{\sigma_{\alpha\beta}}{\sigma_\beta^2}}\dd\alpha\,\dd\beta
\end{align}
where $\bar\alpha$, $\bar\beta$, $\sigma_\alpha$, 
$\sigma_\beta$, and $\sigma_{\alpha,\beta}$ 
are nuisance parameters, and we are still trying to infer the 
value of $p$.
This integral is generally intractable.

%=============================================================================
\subsection{Generalized Inference Problems}
\label{sec:generalized-inference-problems}

In the previous subsection, the inference problem was stated such that 
the survival probability $p=\Tr[\rho P_0]$ was the quantity of interest.
We may of course modify this if some other quantity is preferred.

For example, suppose we are interested in state tomography.
We define the ideal unitary operators $\{U_1,...,U_9\}\in\Unitary(\Hilbert)$
in such a way that $\{U_n \ketbra{0} U_n^\dagger\}_{n=1}^9$ is a basis 
for $\Linear(\Hilbert)$.
Our scheme is to prepare $\rho$, implement the gate $U_n$ for some $n$, 
and measure the resulting state, so that $p_n=\Tr[U_n\rho U_n^\dagger P_0]$
with corresponding random variables
\begin{align}
    Z_n &\sim \Poisson{p_n\alpha+(1-p_n)\beta} \equiv \Poisson{\gamma}.
\end{align}
Then our likelihood function becomes
\begin{align}
    \Lhood(	\rho,\alpha,\beta|x,y,z_1,z_2,...,z_9)
        = \pdf{Pois}{x;\alpha} \cdot \pdf{Pois}{y;\beta} \cdot \prod_{n=1}^9\pdf{Pois}{z_n;p_n\alpha+(1-p_n)\beta}
\end{align}
and we are interested in inferring $\rho\in\Density(\Hilbert)$.
This will be similar for the hierarchical model.
We have taken one set of reference measurements for all $Z_1,...,Z_9$.
We could also choose to take one for each, resulting in the data 
$(x_1,y_1,z_1),...,(x_9,y_9,z_9)$ with a similar likelihood function
of the form
$\Lhood(\rho,\alpha_1,\beta_1,...,\alpha_9,\beta_9|x,y,z_1,z_2,...,z_9)$. 
These sorts of details come down to the particulars of the experimental implementation.

It is clear that any measurement inference problem can be stated in a similar way,
such as process tomography, and Hamiltonian parameter inference, as will be 
seen in \autoref{sec:qhl-example}.

%=============================================================================
\subsection{Fisher Information and the Cram\'er--Rao Bound}
\label{sec:cond-fisher-info}

The average curvature of the likelihood function provides a measure of how
informative data are.
Generally, a highly curved (unimodal) likelihood function implies a 
tight region of support, so that data will tell you a lot about the 
parameters of interest.
This is formalized by Fisher information and the Cram\'er--Rao bound
\cite{cover_elements_2006}.
Given a likelihood function $\Lhood(\vec{\theta}|\vec{d})$ of parameters
$\vec{\theta}$ given data $\vec{d}$, the Fisher information is the 
average curvature of $\Lhood$, more specifically,
it is the negative expected Hessian matrix of the log-likelihood, 
$I(\vec{\theta})_{i,j}=-\expect_{\vec{d}}[\frac{\partial^2 \log L(\vec{\theta}|\vec{d})}{\partial \theta_i \partial \theta_j}|\vec{\theta}]$.
The Cram\'er--Rao bound asserts that no unbiased estimator of $\vec{\theta}$ 
can outperform this intrinsic curvature.
Namely, if $\hat{\theta}$ is any unbiased estimator, which takes data 
$\vec{d}$ and outputs estimates of the true value of $\vec{\theta}$, 
then its covariance is lower-bounded by the inverse Fisher information
matrix,
\begin{align}
    \Cov[\hat{\theta}]
        &\geq I(p,\alpha,\beta)^{-1}.
    \label{eq:cramer-rao-unbiased}
\end{align}
As good estimators will often make this inequality nearly tight,
the inverse Fisher information sets a benchmark for estimators to aim at.
There is a generalized inequality for biased estimators that will be used 
in \autoref{sec:mle}.

\begin{figure}
    \begin{center}
        \includegraphics[width=0.5\textwidth]{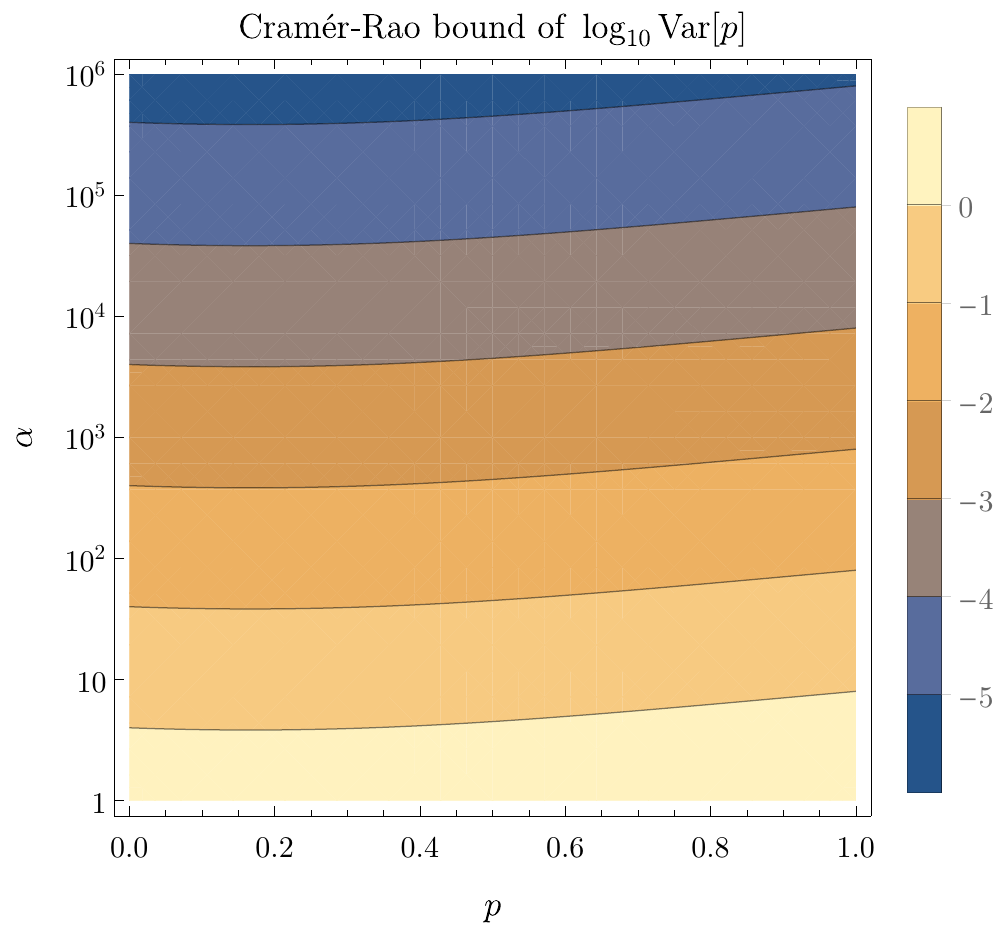}
    \end{center}
    \caption{
        With $\beta$ fixed as $\alpha/2$, the Cram\'er-Rao 
        bound of $\hat{p}(x,y,z)$ 
        is plotted as a function of $\alpha$ and $p$.	
        We see that values of $p$ closer to $1$ are slightly
        more difficult to estimate.
    }
    \label{fig:cramer-rao}
\end{figure}

The Fisher information matrix of the conditional model $(X,Y,Z)|\alpha,\beta$ 
can be computed exactly as
\begin{align}
    I(p,\alpha,\beta) &=
        \begin{pmatrix}
            \frac{(\alpha -\beta )^2}{p (\alpha -\beta )+\beta } & 
            \frac{p (\alpha -\beta )}{p (\alpha -\beta )+\beta } & 
            \frac{\alpha }{\beta +\alpha  p-\beta  p}-1 \\
            \frac{p (\alpha -\beta )}{p (\alpha -\beta )+\beta } & 
            \frac{p^2}{p \alpha -p \beta +\beta }+\frac{1}{\alpha } & 
            -\frac{(p-1) p}{p (\alpha -\beta )+\beta } \\
            \frac{\alpha }{p \alpha -p \beta +\beta }-1 & 
            -\frac{(p-1) p}{p (\alpha -\beta )+\beta } & 
            \frac{p \alpha +(p-2) (p-1) \beta }{\beta  (p (\alpha -\beta )+\beta )} \\
        \end{pmatrix}
\end{align}
with an inverse matrix given by
\begin{align}
    I(p,\alpha,\beta)^{-1}
        &=
        \begin{pmatrix}
            \frac{p (p+1) \alpha +(p-2) (p-1) \beta }{(\alpha -\beta )^2} & 
            \frac{p \alpha }{\beta -\alpha } & 
            \frac{(p-1) \beta }{\alpha -\beta } \\
            \frac{p \alpha }{\beta -\alpha } & 
            \alpha  & 
            0 \\
            \frac{(p-1) \beta }{\alpha -\beta } & 
            0 & 
            \beta
        \end{pmatrix}.
\end{align}
See \autoref{app:likelihood-cumulants} for the calculation.
The Cram\'er--Rao bound for the top left entry states that
\begin{align}
    \Var[\hat{p}(x,y,z)] 
        &\geq 
        \frac{p (p+1) \alpha +(p-2) (p-1) \beta }{(\alpha -\beta )^2},
    \label{eq:cramer-rao-bound}
\end{align}
for any unbiased estimator $\hat{p}$ given the data triple
of photon counts $(x,y,z)$.
This bound is plotted in \autoref{fig:cramer-rao} for the slice
$\beta=\alpha /2$.
In \autoref{sec:estimator-risk} we will see that this bound is very
close to the average error incurred by a few important estimators, including the 
widely used maximum likelihood estimator, and in many disparate 
but relevant parameter regimes.
This means that this inequality is perhaps better thought of as an approximation,
except in exceptionally low contrast regimes.

We can use this result to derive a formula that tells us roughly how much data we will 
need to collect in order to lower the error bars on $p$ to a specified level.
We approximate that $\hat{p}\sim\Normal{p}{(I(p,\alpha,\beta)^{-1})^{-1/2}_{1,1}}$
with $\alpha$ and $\beta$ known (whereas we will generally only have estimates of them).
This gives the $100(1-\zeta)\%$ confidence interval 
\begin{align}
    p\pm \Delta p=p\pm c_{\zeta/2}\frac{\sqrt{p(p+1)\alpha+(p-2)(p-1)\beta}}{\alpha-\beta}
    \label{eq:fisher-confidence}
\end{align}
for $\hat{p}$ where $c_{\zeta/2}=\sqrt{2}\operatorname{erf}^{-1}(1-\zeta)$
with $\operatorname{erf}(x)=\pi^{-1/2}\int^{x}_{-x}e^{-t^2}\dd t$ the error function.
Supposing a reference contrast of $C=\frac{\alpha-\beta}{\alpha+\beta}$, we need
\begin{align}
    \alpha \approx \frac{c_{\zeta/2}^2}{2\Delta p^2}(1+1/C)^2
    \label{eq:fisher-confidence-requirement}
\end{align}
to make $\hat{p}\approx p\pm \Delta p$ a $100(1-\zeta)\%$ confidence interval
\footnote{We use this definition for contrast 
simply because it shows up naturally in \autoref{eq:mle-bias}; $\frac{\alpha-\beta}{\alpha}$ 
is arguably a better choice.}.
To derive this formula we have assumed the worst case, $p=1$.
These calculations show that, for example, if we desire a 95\% confidence
interval of $\pm 0.01$ for $p$, then we need to do at least enough experiments 
$N$ so that $\alpha$ is on the order of $\sim 170,000$.

\begin{figure}
    \begin{center}
        \includegraphics[width=0.95\textwidth]{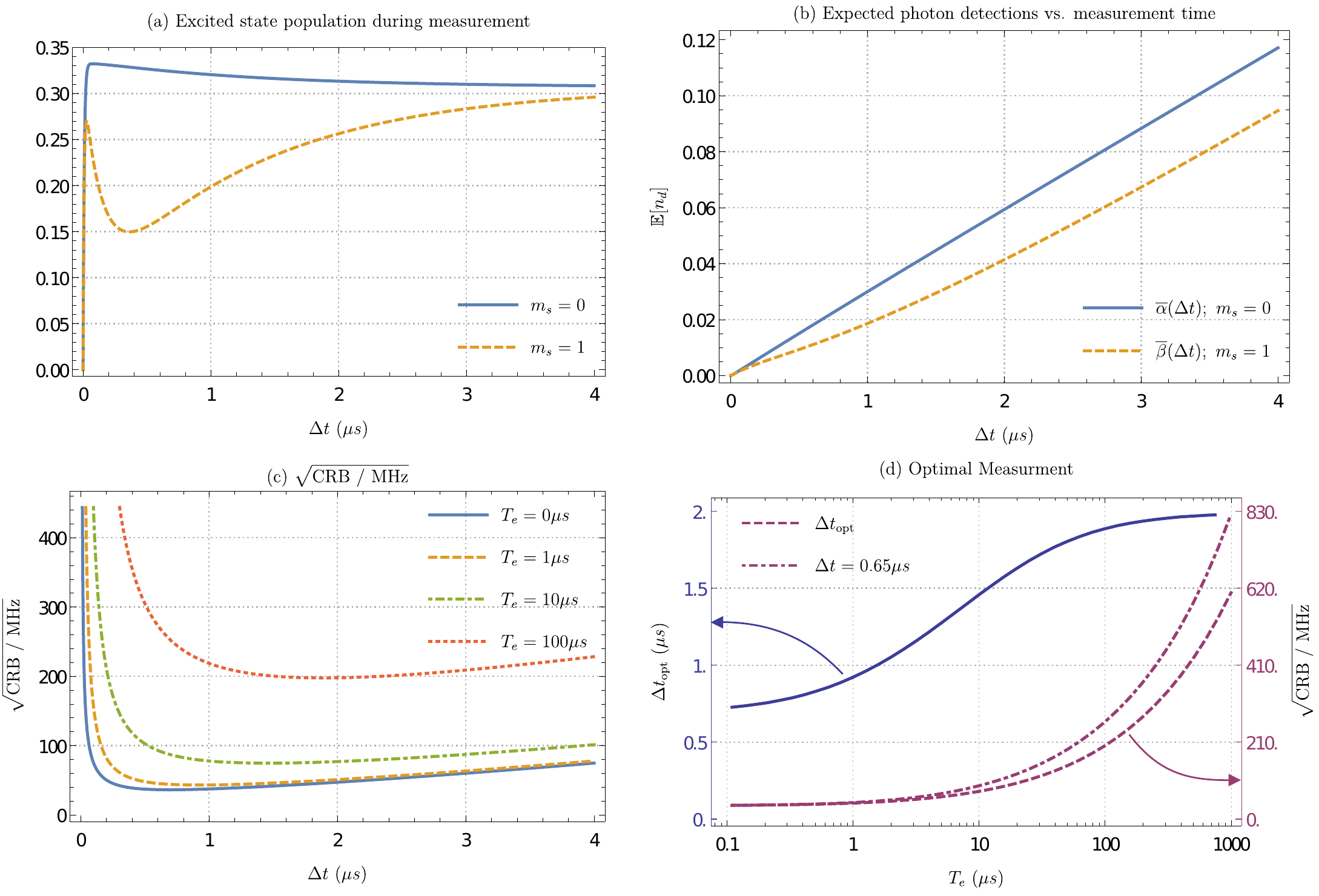}
    \end{center}
    \caption{
        Simulated example of optimizing measurement time for a low visibility
        experiment.
        The same analysis holds for high visibility experiments.
        (a) The population of the optical excited state is plotted for 
        two initial states, (b) which results in distinguishable numbers
        of detected photons given that we average enough repetitions.
        They are labeled $\overline{\alpha}(\Delta t)$ and  $\overline{\beta}(\Delta t)$
        in the main body and asymptote to the same slope since they both
        end up in the same steady state of the master equation.
        (c) These curves can be used to estimate the standard deviation
        of $p$ normalized to square-root runtime for various experiment lengths.
        For example, given $\sqrt{\text{CRB/MHz}}=400/\sqrt{\text{MHz}}$,
        a total run time of $\SI{100}{s}=\SI{e8}{us}$ will approximately 
        reduce the uncertainty of $\Delta p$ to $0.04$.
        (d) As a function of $T_e$, optimal measurement window length 
        $\Delta t_\text{opt}$ is shown (left axis) along with the corresponding
        $\sqrt{\text{CRB/MHz}}$ values for both the optimal measurement time,
        and a fixed measurement time of $\SI{0.65}{us}$ (right axis).
        It is seen that in this regime the payoff of using the optimal measurement
        time is rather slim.
    }
    \label{fig:opt-measurement}
\end{figure}

This heuristic holds for any optical transition rates and choice of measurement 
time $\Delta t$ in the measurement protocol detailed in \autoref{sec:meas-dynamics}.
However, certain choices of measurement time are better than others.
We can use the Cram\'er--Rao bound to estimate the optimal such time,
namely, we wish to choose the measurement time $\Delta t$ that maximizes the temporal 
information density of $p$.
%This assumes that $p$ is the quantity of interest, when usually, some other 
%quantity, like the strength of the static magnetic field along the 
%principle axis, will be of greater interest.
%However, since $p$ is present as an intermediate latent variable in any NV
%inference problem, minimizing its variance will always be a good thing to do.
Begin by supposing that the total runtime of a fixed experiment, including taking
bright and dark reference counts, is $T=N(T_e+3\Delta t)$,
where $N$ is the number of repetitions, and $T_e$ is the amount of time per repetition
not spent counting photons (initialization, wait periods, pulse sequences, etc.).
We must multiply $\Delta t$ by 3 to account for all three 
of the signal, bright reference, and dark reference counting windows.
Writing $\alpha=N\overline{\alpha}$ and $\beta=N\overline{\beta}$, with $\overline{\alpha}$
and $\overline{\beta}$ the average per-shot reference values, 
again at the worst case $p=1$, gives
$\Delta p^2=\frac{2\alpha}{(\alpha-\beta)^2}$, or rearranging, gives
\begin{align}
    \Delta p \sqrt{T}
        &= \frac{\sqrt{2(T_e+3\Delta t)\overline{\alpha}(\Delta t)}}
            {\overline{\alpha}(\Delta t)-\overline{\beta}(\Delta t)}.
\end{align}
We have written $\overline\alpha=\overline{\alpha}(\Delta t)$ 
and $\overline\beta=\overline{\beta}(\Delta t)$ to emphasize their 
implicit dependence on $\Delta t$.
These two functions are easily estimated experimentally by sweeping the 
length of the measurement window.
Then for any given $T_e$, the the quantity $\Delta p \sqrt{T}$ can be
be minimized, visualized in \autoref{fig:opt-measurement}.
As the experiment time $T_e$ grows, it becomes increasingly worthwhile 
to lengthen the duty cycle of measurement.
This formula and its units are analogous to widely used magnetometry sensitivity 
formulas; see \citet{taylor_high-sensitivity_2008} or
\citet{hirose_continuous_2012} for two examples out of many.

Of note is the steep increase of $\Delta p \sqrt{T}$ as $\Delta t\rightarrow 0$.
While the slope is relatively gentle as $\Delta t$ gets larger, causing little harm 
even if $\Delta t$ is twice as big as the optimal value for a given $T_e$, there is a large penalty
for choosing a measurement of $\Delta t$ which is too short.
Long refocusing sequences like CPMG are especially at risk of falling into this trap.

%=============================================================================
\section{Simple Estimators}
\label{sec:estimators}
%=============================================================================

The previous section discussed inference models of NV measurement 
in some detail.
In this section, we introduce two estimators for the basic inference 
problem of the conditional model defined in \autoref{sec:basic-inference-problem}.
We also compare their relative strengths and weaknesses.

%=============================================================================
\subsection{Maximum Likelihood Estimator}
\label{sec:mle}

The most obvious estimator turns out to be quite a good one, and the one 
that has been used almost universally in practice.
Suppose $(x,y,z)$ is a variate of $(X,Y,Z)|\alpha,\beta$.
Equation~\autoref{eq:cond-model-means} shows that $x$, $y$, and $z$ 
are unbiased estimates of $\alpha$, $\beta$, and $p\alpha+(1-p)\beta$,
respectively.
We invert the equation $p\alpha+(1-p)\beta$ for $p$ substituting in our 
estimates above to get the estimator
\begin{align}
    \hat{p}_\MLE = \frac{z-y}{x-y}
    \label{eq:mle}
\end{align}
for $p$.
Although appearing quite simple, it is difficult to work with 
this estimator analytically; it is the ratio of two 
correlated Skellam distributions which does not have many nice properties.
However, it can be shown with Lagrange multipliers that this is the maximum 
likelihood estimator (MLE) of the model, that is,
$\left(\frac{z-y}{x-y},x,y\right)$ is the (unique) maximum of the 
function $\Lhood(p,\alpha,\beta|x,y,z)$ on the domain $0\leq p\leq 1$,
$0< \beta \leq \alpha$, for any $x,y,z\geq 0$.
See \autoref{app:mle-derivation} for details.

There is always a finite probability that $x=y$, in which case this estimator will
divide by zero.
With sufficient magnitude of and contrast between $\alpha$ and $\beta$ this 
is highly unlikely.
It still poses a problem if we wish to prove anything about it, for example,
if we wish to find its expectation value.
To avoid this situation we define the slightly modified estimator
\begin{align}
    \hat{p}_{\MLE,\epsilon} = \frac{z-y}{x-y+\epsilon}
\end{align}
for some non-integer value of $\epsilon$.
One might consider using this estimator instead of the MLE 
if $x=y$ has a significant probability.
After quite a bit of work, shown in \autoref{app:mle-bias}, 
it is seen that the bias of $\hat{p}_{\MLE,\epsilon}$ is 
non-zero, a linear function of $p$, and exactly given by the integral
\begin{align}
    \Bias[\hat{p}_{\MLE,\epsilon}]
    &= \re\int_0^\pi
            \ii (-1)^{-\epsilon } e^{-\alpha(1+e^{\ii\phi})-\beta(1+e^{-\ii\phi})}
            ((\beta+(\alpha-\beta)p) e^{\ii\phi\epsilon}+\beta e^{\ii\phi(\epsilon-1)})
            \dd \phi -p.
\end{align}
Taking the limit as $\epsilon$ approaches zero gives an expression for the bias of
the original estimator,
\begin{align}
    \Bias[\hat{p}_{\MLE}]
        &= \lim_{\epsilon\rightarrow 0} \Bias[\hat{p}_{\MLE,\epsilon}] \nonumber \\
        &= \int_0^\pi
        e^{-(\alpha+\beta)(1+\cos\phi)}\left[
            (\gamma+\beta\cos\phi)\sin((\alpha-\beta)\sin\phi)+(\beta\sin\phi)\cos((\alpha-\beta)\sin\phi)
        \right] \dd\phi-p.
\end{align}
This integral can be solved numerically to find the exact bias.
If $\alpha+\beta\gg 1$, we can derive an asymptotic approximation to this integral,
\begin{align}
    \Bias[\hat{p}_\MLE]
        &\approx
        \left(
            p - \frac{\beta}{\alpha+\beta}
        \right)
        \frac{\alpha+\beta}{(\alpha-\beta)^2} + \order{(\alpha+\beta)^{-2}},
        \label{eq:mle-bias}
\end{align}
which through numerics can be shown to be valid for 
$\alpha+\beta\gtrsim 300$.
Note that the bias vanishes at $p=\frac{\beta}{\alpha+\beta}$,
and that if the contrast 
$C=\frac{\alpha-\beta}{\alpha+\beta}$ is fixed, then the worst case bias 
scales as $\frac{\alpha+\beta}{(\alpha-\beta)^2}=\order{(\alpha+\beta)^{-1}}$.

Although estimators with no bias are generally preferred, 
we see that this estimator has the more important property
of being consistent, meaning that $\Bias[\hat{p}_\MLE]\rightarrow 0$ as 
$\alpha\rightarrow \infty$ with fixed contrast.
We can make an even stronger statement by 
using the Cram\'er--Rao bound for biased estimators, which is
a generalization of \autoref{eq:cramer-rao-unbiased}
known as the van Trees inequality or the Bayesian Cram\'er--Rao bound
\cite{van_trees_detection_1968}, stating that
\begin{align}
    \Cov[\hat{\theta}]
        &\geq J_{\hat{\theta}}(p,\alpha,\beta)I(p,\alpha,\beta)^{-1}J_{\hat{\theta}}^\T(p,\alpha,\beta)
\end{align}
where $J_{\hat{\theta}}(p,\alpha,\beta)$ is the expectation value of
the Jacobian matrix of the possibly biased estimator $\hat{\theta}$.
Using the approximation from \autoref{eq:mle-bias}, this gives us the inequality
\begin{align}
    \Var[\hat{p}_\MLE] 
        &\geq \frac{p (p+1) \alpha +(p-2) (p-1) \beta }{(\alpha -\beta )^2}
             + \order{(\alpha+\beta)^{-2}}
\end{align}
for the maximum likelihood estimator, again assuming that the contrast is
fixed as $\alpha$ and $\beta$ increase.
This is approximately the same bound as the unbiased Cram\'er--Rao bound 
discussed earlier. Indeed, this is generic, as the van Trees inequality approaches
the Cram\'er--Rao bound for large data sets, such that incorporating prior 
information can be thought of as an important correction for finite data sets
\cite{opper_online_1998}.

%=============================================================================
\subsection{Bayes Estimator}
\label{sec:bayes-estimator}

If we assume our prior knowledge of the parameters $(p,\alpha,\beta)$ is 
encoded in the probability distribution $\pi(p,\alpha,\beta)$, then
assuming our model is correct, Bayes' theorem will tell us how to best
update our beliefs about the parameters after we have measured the variate 
$(x,y,z)$:
\begin{align}
    \pi^*(p,\alpha,\beta)
        \equiv \prob{p,\alpha,\beta|x,y,z}
        &= \frac{
            \prob{x,y,z|p,\alpha,\beta}\pi(p,\alpha,\beta)}{
            \int\prob{x,y,z|p,\alpha,\beta}\pi(p,\alpha,\beta)\dd p\,\dd \alpha\,\dd \beta
            } 
        = \frac{L(p,\alpha,\beta|x,y,z)\pi(p,\alpha,\beta)}{\Bnorm}	.	
\end{align}
Here, $L$ is the likelihood function from Equation~\autoref{eq:cond-likelihood} 
and $\Bnorm$ is a normalization constant.

If we assume a separable prior $\pi(p,\alpha,\beta)=\pi(p)\pi(\alpha,\beta)$,
and it would be strange not to, then Bayes' theorem can be applied sequentially.
This relies on the conditional independence of $X$, $Y$, and $Z$.
We can first update our prior distribution using the datum $(x,y)$ to get
\begin{align}
    \prob{p,\alpha,\beta|x,y}
        &= \frac{
            \prob{x,y|p,\alpha,\beta}\pi(p,\alpha,\beta)}{
            \int\prob{x,y|p,\alpha,\beta}\pi(p,\alpha,\beta)\dd p\,\dd \alpha\,\dd \beta
            } 
        = \frac{
            \prob{x|\alpha}\prob{y|\beta}\pi(\alpha,\beta)}{
            \int \prob{x|\alpha}\prob{y|\beta}\pi(\alpha,\beta)\dd \alpha\,\dd \beta
            }\pi(p) 
        \equiv \pi^*(\alpha,\beta)\pi(p)
\end{align}
and subsequently
\begin{align}
    \prob{p,\alpha,\beta|z}
        &= \frac{
            \prob{z|p,\alpha,\beta}\pi^*(\alpha,\beta)\pi(p)}{
            \int\prob{z|p,\alpha,\beta}\pi^*(\alpha,\beta)\pi(p)\dd p\,\dd \alpha\,\dd \beta
            } 
        = \pi^*(p,\alpha,\beta).
\end{align}
This sequential break down is useful because a conjugate prior can be found for the 
likelihood $\prob{x,y|p,\alpha,\beta}$ in a couple of useful cases,
meaning the posterior $\pi^*(\alpha,\beta)$ can be computed exactly.
Formulas for two different conjugate priors for the references are
given in \autoref{app:conjugate-priors}.
The full likelihood $L$, however, almost certainly does not have a conjugate prior.

Given the posterior distribution $\pi^*$, there are many choices for the estimator.
The most common is the mean square error (MSE) Bayes estimator 
which simply takes the expectation value of $\pi^*$.
For the particular parameter $p$, we have
\begin{align}
    \hat{p}_\Bayes
        &= \expect[p|x,y,z]
        = \int_0^1 \int_0^\infty \int_0^\infty p \pi^*(p,\alpha,\beta) \dd p\,\dd\alpha\,\dd\beta.
\end{align}
This is known to be the estimator which minimizes the 
expected MSE of the estimate over all possible estimators, 
\begin{align}
    (\hat{p}_\Bayes ,\hat{\alpha}_\Bayes ,\hat{\beta}_\Bayes )
        &= \operatorname{argmin}_{\hat{\theta}} \expect[(\hat\theta(x,y,z)-(p,\alpha,\beta))^2].
\end{align}

%=============================================================================
\subsection{Comparing Estimators: Risk}
\label{sec:estimator-risk}

\floatsetup{subcapbesideposition=top}
\begin{figure}
    \sidesubfloat[]{\includegraphics[width=0.4\textwidth]{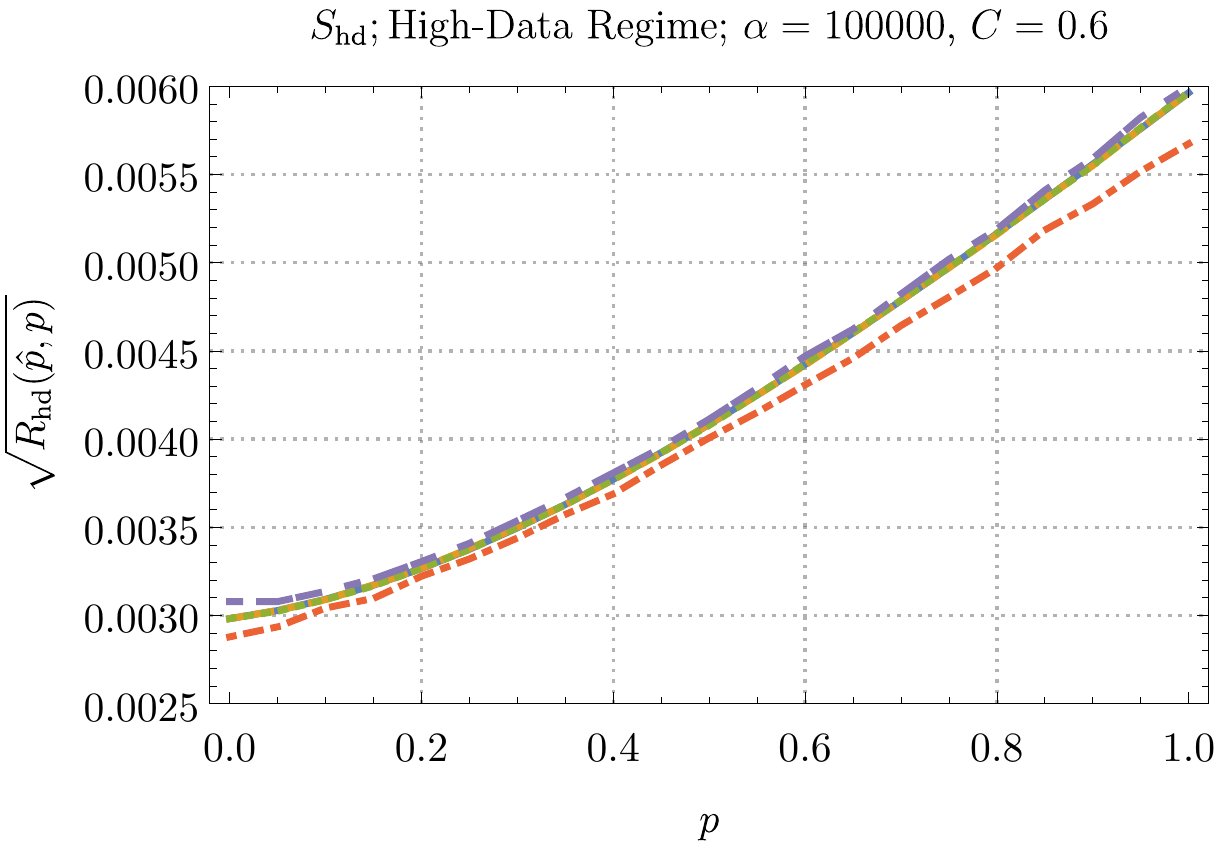}} \quad
    \sidesubfloat[c]{\includegraphics[width=0.4\textwidth]{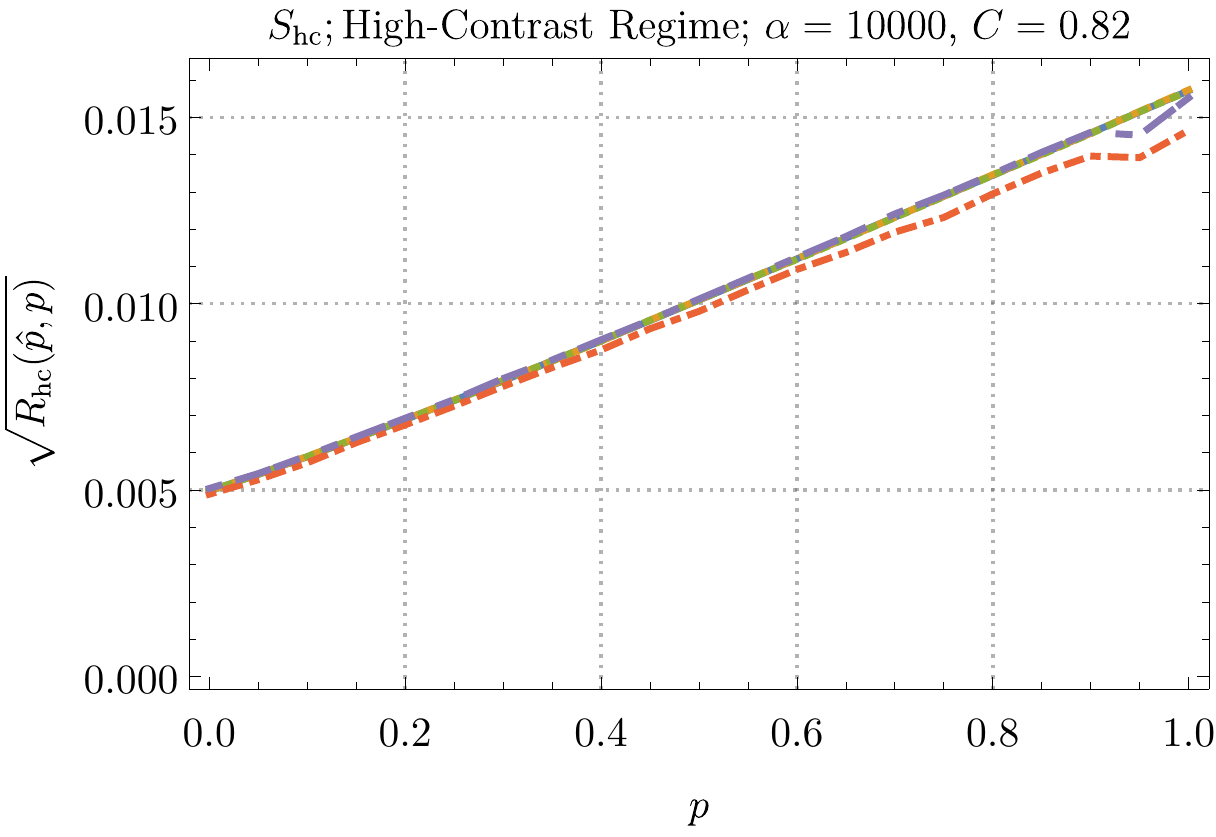}
        \raisebox{0.7\height}{\includegraphics[width=0.11\textwidth]{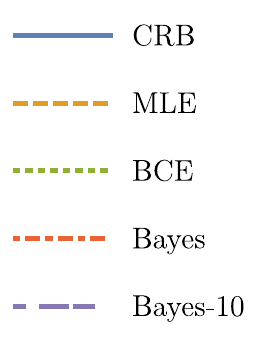}}	} \\
    \sidesubfloat[]{\includegraphics[width=0.4\textwidth]{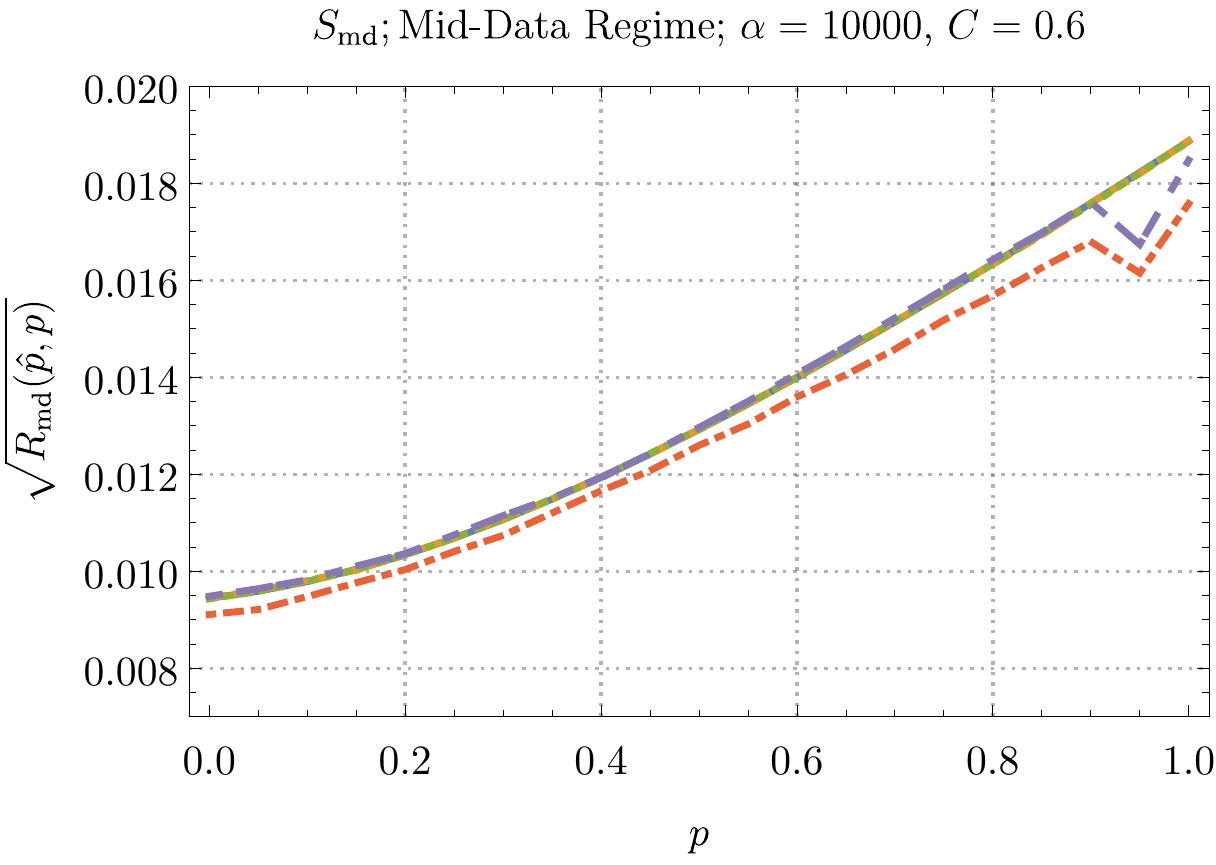}} \quad
    \sidesubfloat[]{\includegraphics[width=0.4\textwidth]{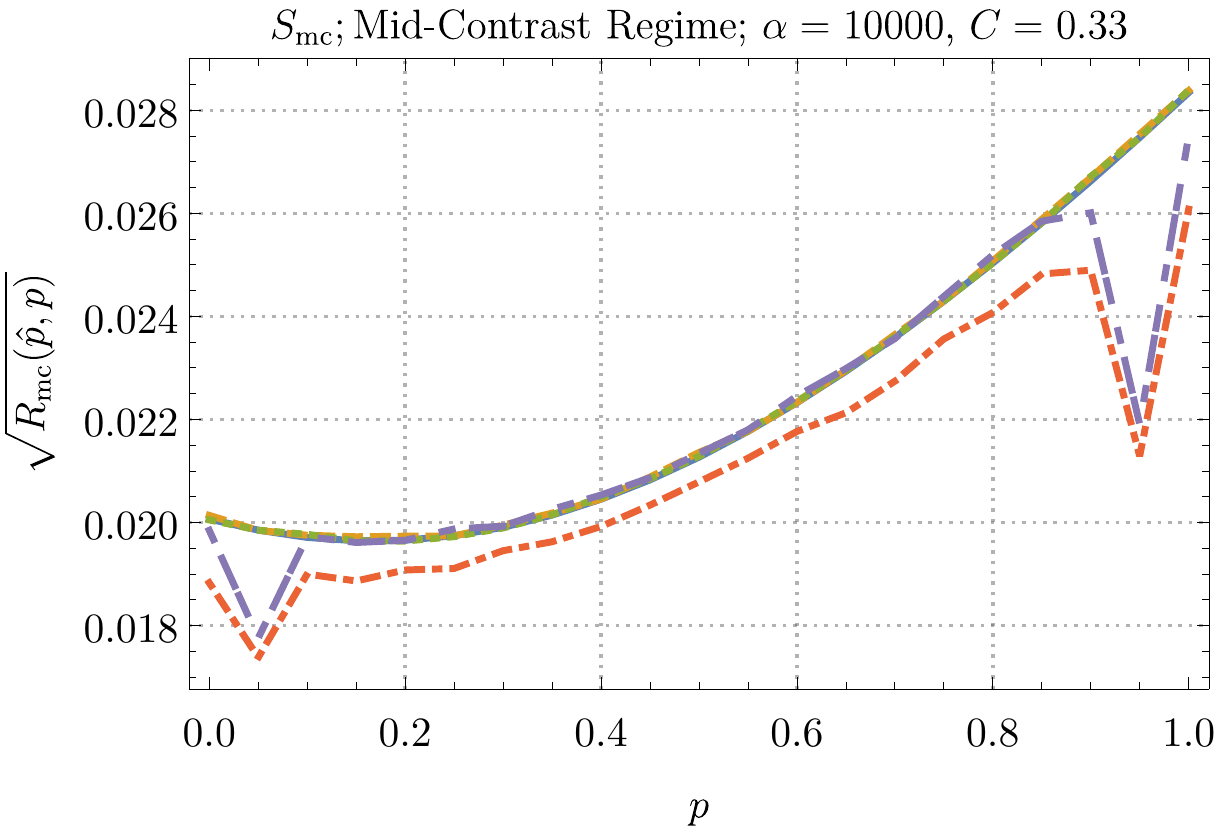}}\hfill\null \\
    \sidesubfloat[]{\includegraphics[width=0.4\textwidth]{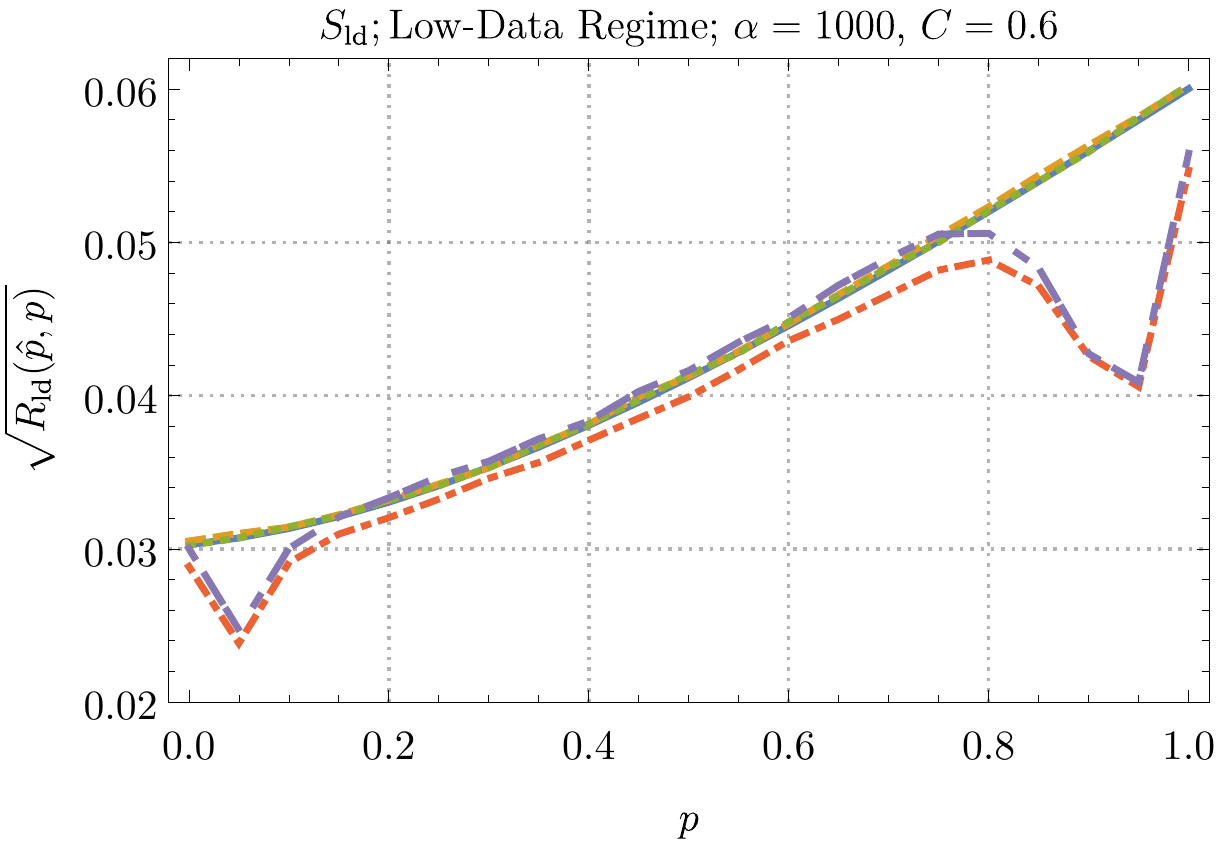}} \quad
    \sidesubfloat[]{\includegraphics[width=0.4\textwidth]{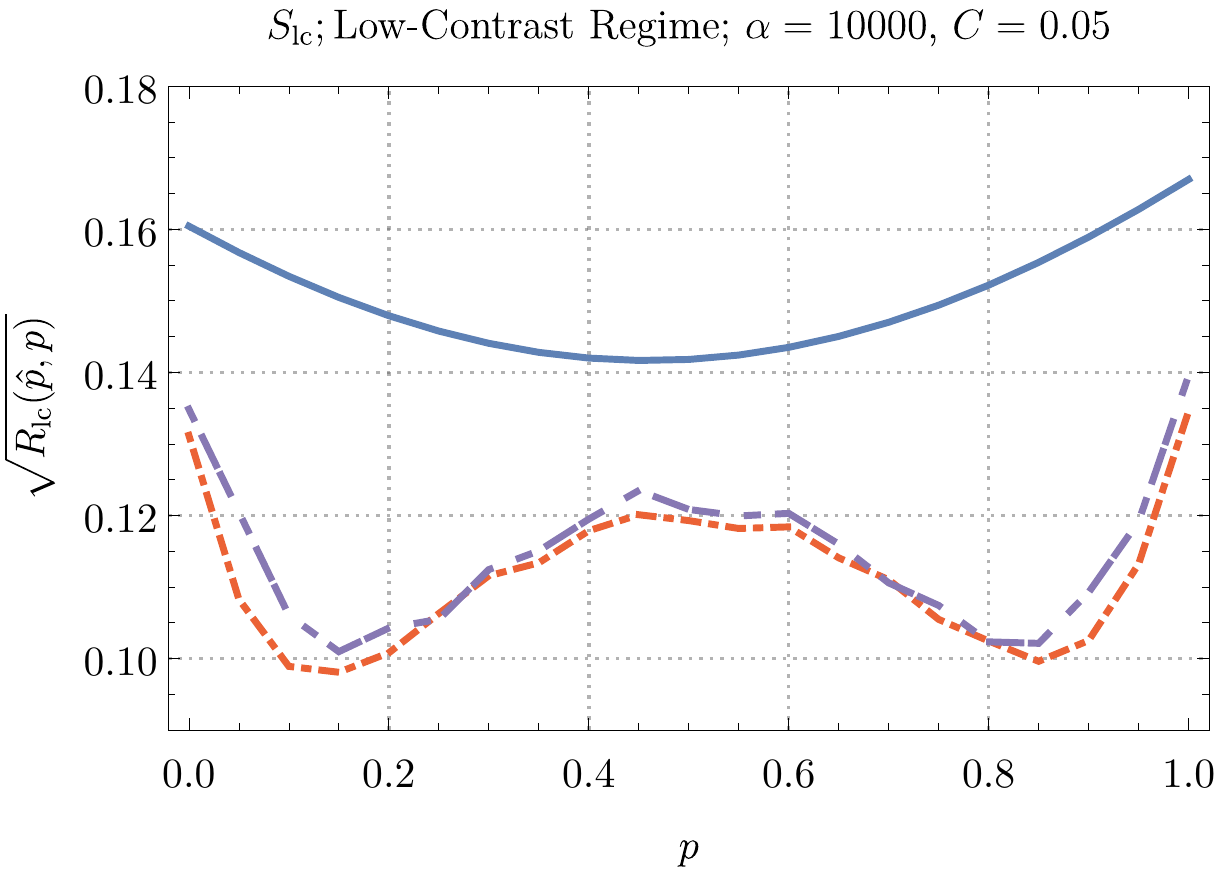}}\hfill\null
    \caption{The MSE risk for several estimators of $p$, labeled in the
    legend, is plotted for six different regimes of experimental setup, (a)-(f).
    The square root has been taken so that the units of the $y$-axes have the same 
    units as $p$.
    The estimators under study 
    are the maximum likelihood estimator, $\hat{p}_{\MLE}$,
    the bias corrected estimator (see \autoref{app:bias-corr-estimator}), 
    $\hat{p}_\BCE$, and the Bayes estimator,
    $\hat{p}_{\Bayes}$, with two different priors.
    These priors are denoted by ``Bayes'' and ``Bayes-10'',
    with the latter being a more conservative prior corresponding
    to a ten-fold increase in the assumed
    covariance, as explained in the main body.
    Sharp peaks for the Bayes estimators are artefacts of the coarse sampling
    along the $x$-axis; risk was evaluated at $p$ ranging from $0$ to $1$ in 
    steps of $0.05$.
    The risks of $\hat{p}_{\MLE}$ and $\hat{p}_{\BCE}$ are much bigger than 
    $1$ for the low-contrast regime due to the common occurence of $y>x$, 
    and are therefore not plotted.}
    \label{fig:mse-risk}
\end{figure}

We wish to compare the quality of the above estimators.
Since one estimator is frequentist and one is Bayesian, we 
choose to accept the idea of a `true value' of $p$.
Given an estimate $\hat p(x,y,z)$ of $p$ which depends on the 
data $(x,y,z)$ drawn from $(X,Y,Z)|\alpha,\beta$, there are many ways of 
quantifying the distance between the estimate and the true value.
We consider the mean-squared-error (MSE) loss function
\begin{align}
    \Loss_\MSE(\hat p(x,y,z), p)=(\hat p(x,y,z) - p)^2.
\end{align}

It is reasonable to assume that the operators of a given experimental setup 
will have a rough idea of what to expect as their reference counts.
Or, given the results of a set of experiments, all of 
the reference counts can be pooled to empirically construct a distribution 
of reference counts.
As such, we assume the existence of a probability distribution 
$P_\Setup(\alpha,\beta)=\Pr(\alpha,\beta|\text{experimental setup } \Setup)$
which characterizes a particular setup called $\Setup$ 
(assuming a fixed number of shots, $N$, per experiment).

Given an estimator $\hat{p}$, we 
can define its associated risk with respect to the true 
values $(p,\alpha,\beta)$ as the average value of the loss,
\begin{align}
    \Risk (\hat p, p, \alpha, \beta)
        &= \expect_{x,y,z|p,\alpha,\beta} [\Loss_\MSE(\hat p,p)] 
        = \sum_{x,y,z=0}^\infty \Loss_\MSE(\hat p(x,y,z),p)\Pr(x,y,z|\alpha,\beta,p).
\end{align}
With a particular setup $\Setup$ we can then quantify its overall risk,
\begin{align}
    \Risk_{\Setup} (\hat p, p)
        &= \expect_{\alpha,\beta} [\Risk(\hat p,p,\alpha,\beta)] 
        = \int \Risk(\hat{p},p,\alpha,\beta) P_\Setup(\alpha,\beta) \dd\alpha\,\dd\beta,
    \label{eq:risk-of-setup}
\end{align}
by marginalizing over our knowledge about it.
If we were to additionally marginalize over a distribution of $p$, this would be 
the Bayes risk, and an estimator minimizing this quantity would be a
Bayes estimator.
Therefore, the expression above can be seen as a hybrid between Bayes 
risk and frequentist risk, where we marginalize over $\alpha$ and $\beta$,
but not $p$.

We will treat the prior of a Bayesian estimator as an implicit
property of the estimator.
We can then compare, for example, $\hat{p}_\Bayes$ with itself 
under different priors.
The prior on $\alpha$ and $\beta$ does not need to have any relationship with 
$P_\Setup(\alpha,\beta)$.
A large part of choosing a prior has to do with assessing one's level of 
paranoia, and one may be 
more paranoid about generating a fair risk comparison than about giving 
the estimator an over-informed prior, or vice versa.
However, setting them equal to each other will often be the most
sensible thing to do.

We study a few regimes of experimental setups.
We consider a high-data regime, $\Setup_\text{hd}$, where $\overline\alpha= 100,000$,
a mid-data regime, $\Setup_\text{md}$, where $\overline\alpha= 10,000$, and a 
low-data regime, $\Setup_\text{ld}$ where $\overline\alpha= 1,000$.
In these three cases the contrast is the same, $C=\frac{\overline\alpha-\overline\beta}{\overline\alpha+\overline\beta}=0.6$.
We additionally consider mid-data regimes of varying contrast, 
where $\overline\alpha= 10,000$
for $C=0.05,0.33,0.82$.
This defines the respective low, medium, and high contrast setups 
$\Setup_{lc}$, $\Setup_{mc}$, and $\Setup_{hc}$. 
The above setup descriptions only supply the mean values of their 
respective distributions $P_\Setup$,  that is, 
$\expect_{P_\Setup}[(\alpha,\beta)]=(\overline\alpha,\overline\beta)$.
To keep things simple, we take these distributions to all be binormal 
with super-Poisson standard deviations $\sigma_\alpha=2 \sqrt{\overline{\alpha}}$ and
$\sigma_\beta=2 \sqrt{\overline{\beta}}$, and covariances 
defined by $\sigma_{\alpha,\beta}=1.5 \beta$.

In \autoref{fig:mse-risk}, $\sqrt{\Risk_{\Setup}(\hat p, p)}$ is plotted for each of the setups 
described above, and for each of the estimators $\hat{p}_\MLE$, $\hat{p}_\BCE$, and 
$\hat{p}_\Bayes$. 
The square root was taken so that both axes have the same units.
Two different priors are used for the Bayes estimator on each setup,
both are product gamma distributions, discussed further 
in \autoref{app:drift-prior-uncor}.
The first, `Bayes', uses the same mean value and diagonal
covariance elements as $P_\Setup$.
The second, `Bayes-10', uses the same mean value as $P_\Setup$,
but standard deviations which are ten times larger than $P_\Setup$, corresponding to a rather uninformative prior.
The more sophisticated prior discussed in \autoref{app:drift-prior-cor}, which allows 
for correlations between $\alpha$ and $\beta$, should in theory be strictly better 
than the ones used in these calculations, but were found to be too computationally 
expensive for naive implementations of risk computation.
For all setups and estimators, risk is computed by Monte Carlo sampling; 
for each value of $p$,
many pairs $(\alpha,\beta)$ are 
sampled from $P_\Setup(\alpha,\beta)$, for each pair many variates ($x$,$y$,$z$) 
are drawn from the likelihood distribution, and the loss $L_\MSE$ is computed for each.
The average of these loss values for this value of $p$ forms an 
estimate of $\Risk_\Setup(\hat{p},p)$.

These plots show that the Cram\'er--Rao bound \autoref{eq:cramer-rao-bound}
is an excellent estimate of the risk of the \MLE~in most regimes.
Further, under our loss function, the Bayes estimator never has more risk than the 
\MLE, and has superior performance especially near the boundaries of $[0,1]$,
even for the rather uninformative Bayes-10 prior. 

%=============================================================================
\section{Example: Hamiltonian Learning with Bayesian Inference}
\label{sec:qhl-example}
%=============================================================================

One of the primary advantages of using a Bayesian approach to NV measurement 
is that it can be used as an overlay model on other estimation problems.
This results in seamless propagation of error bars to the final
quantities of interest.
In frequentist settings, it is usually a pain to justifiably propagate error bars 
near the boundaries of an interval like $[0,1]$ because the usual 
normality approximations are dubious.
To illustrate this Bayesian approach, in this section we provide a thorough 
example of Quantum Hamiltonian Learning (QHL) \cite{wiebe_quantum_2014}
using experimental data.
In a recent and related experimental work, QHL was shown to be a 
powerful method of characterizing quantum systems~\cite{wang_experimental_2017}.
Raw data and reproducible code for this paper can be found in our online  repository~\cite{code_and_data}.

Other advantages of the Bayesian inference algorithm we employ, though not used 
by us here, include the ability to adaptively choose the next experiment to maximize
information gain, and the ability to treat reference drift as a time dependent
stochastic process \cite{granade_qinfer_2017}.

%=============================================================================
\subsection{Hamiltonian Model}
\label{sec:qhl-hamiltonian-model}

We assume that our spin-1 Hamiltonian (in the optical ground state) is of the form
\begin{align}
    H_{m_\Iop} = \delta\Delta \Sz^2 + (\omega_e + m_\Iop A_N)\Sz + \Omega(t) \Sx
    \label{eq:hamiltonian}
\end{align}
where we are in a frame rotating near the ground state ZFS, 2.87GHz, and 
have invoked the secular and rotating wave approximations.
Here, $\delta\Delta$ is the mismatch between the applied microwave frequency 
and the ZFS, $\omega_e$ is the projection of the static magnetic field 
onto the $z$-axis, $A_N$ is the hyperfine splitting due to 14-Nitrogen, 
$m_\Iop$ is the spin number of the nitrogen atom,
and $\Omega$ is the microwave nutation strength.
All of the parameter units are angular frequency, $2\pi\cdot$MHz, except $m_\Iop$ which is unitless.
Our goal is to learn these parameters, as well as the $T_2$ decay time.
In particular, we would like good error bars on $\omega_e$
since this is the quantity of interest in magnetometry,
considering other parameters as nuisances.

At room temperature, the nitrogen atom is equally likely to be in each of 
its three energy states $\ket{m_\Iop\in\{-1,0,+1\}}$.
Since the nitrogen $T_1$ is much longer than a single experiment, we assume 
that for each experiment, the state of the nitrogen is fixed.
Hence in the Hamiltonian above, we simply treat the axial hyperfine coupling 
between the \NV and the nitrogen as a small $m_\Iop$ dependent shift in 
the static magnetic field.

With an initial state $\rho_i=\ketbra{\sgz}$, at time $t$ the density matrix
is described by $\rho(t)=\frac{1}{3}\sum_{m_\Iop} \mathcal{S}(t)_{m_\Iop}[\rho_i]$
where $S_{m_\Iop}(t)$ is the evolution superoperator under the Hamiltonian $H_\Iop$,
along with a single dephasing Lindblad term $L=\sqrt{T_2}\Sz$.
The superoperators $\mathcal{S}_{m_\Iop}$ can be computed by exponentiating the 
supergenerator derived from the Lindblad master equation.
We will only consider constant or piecewise constant values of $\Omega(t)$, which 
simplifies simulation (finite rise-times are ignored).
This convex combination approach is valid because we will be summing over many 
trials of the same experiment, and hence will see an equal mixture of all 
three nitrogen states on average.

%=============================================================================
\subsection{Experiment Choices and Data}
\label{sec:qhl-choices}

The time dependence of the nutation envelope $\Omega$ is controlled by 
the experimentalist.
To learn the parameters of this system, we choose to do two types of experiments.
The first is the Rabi experiment, where $\Omega$ is finite and constant 
for a period $t_r$ 
and the state is subsequently measured.
Rabi experiments are primarily sensitive to nutation frequency.
The second is the Ramsey experiment, where there is a wait period with $\Omega=0$ 
of length $t_w$ between 
two identical unitary gates which are created by turning $\Omega$ on at full 
power for a duration $t_p$.
Ramsey experiments are sensitive to fields along the $z$-axis.
These two experiments are sensitive to 
roughly orthogonal regions of parameter space.

Experiments were performed on a microscope with relatively poor optical
characteristics; for a single repetition ($N=1$) we had 
an average number of detected bright reference photons $\alpha\approx 0.006$,
and an average number of detected dark reference photos $\beta\approx 0.004$,
giving a contrast value of $0.2$.
The static magnetic field acting on the \NV center was just the ambient 
stray field in the laboratory; some combination of Earth's field, building 
characteristics, and nearby electronics.
We took $R=400$ averages of $N=30000$ repetitions for both the Rabi
and Ramsey experiments.
Rabi flops were sampled at 100 linearly spaced points between $t_r=8$ns and $t_r=800$ns.
The Ramsey wait times were sampled at 200 linear spaced points between
$t_w=0.01\mu$s and $t_w=2\mu$s, with a pulse time $t_p=44$ns.
Raw (summed) data is plotted in \autoref{app:qhl-figures}.

\floatsetup{subcapbesideposition=top}
\begin{figure}
    \sidesubfloat[]{\includegraphics[width=0.45\textwidth]{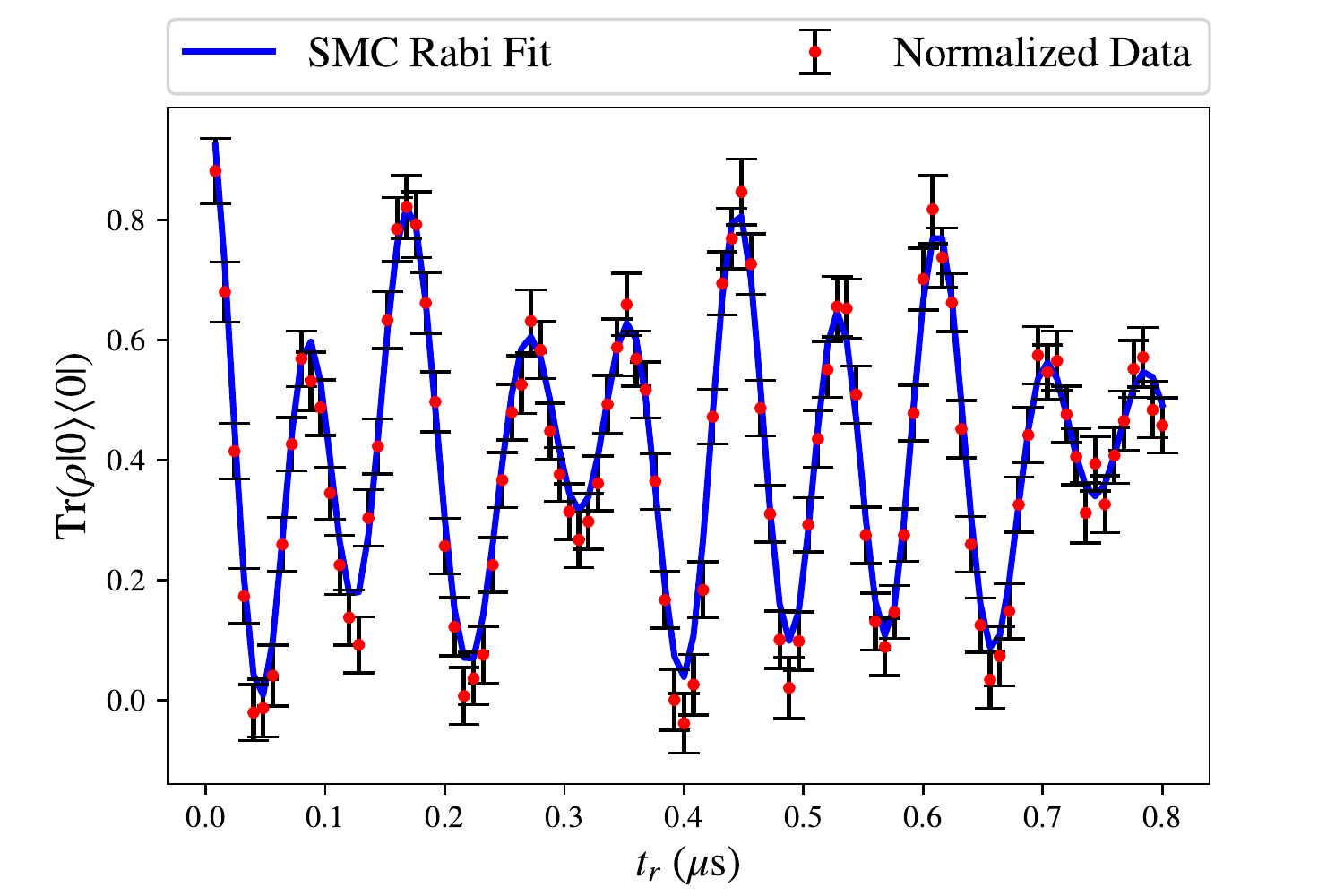}} \quad
    \sidesubfloat[]{\includegraphics[width=0.45\textwidth]{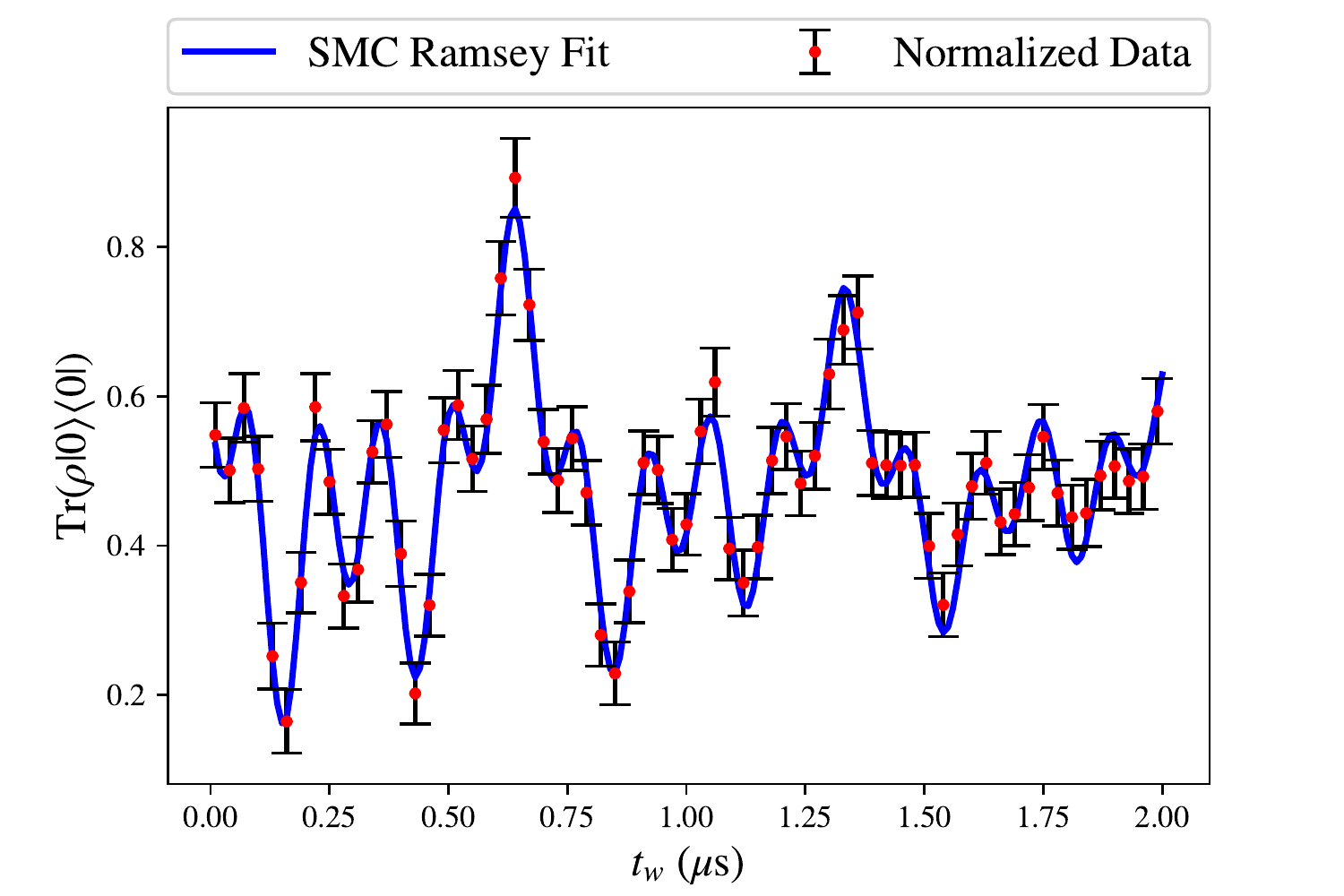}} \\
    \sidesubfloat[]{\includegraphics[width=0.45\textwidth]{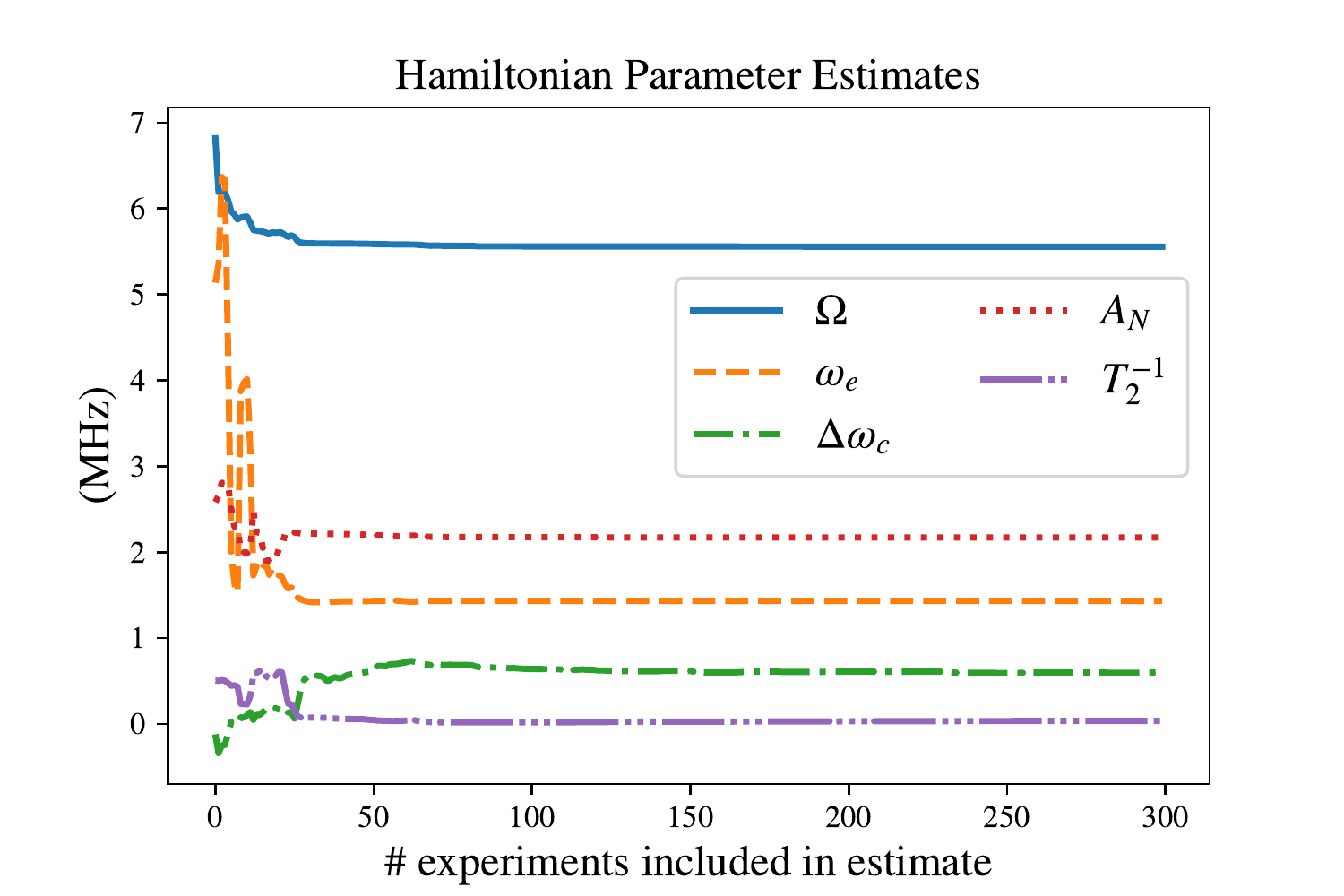}} \quad
    \sidesubfloat[]{\includegraphics[width=0.45\textwidth]{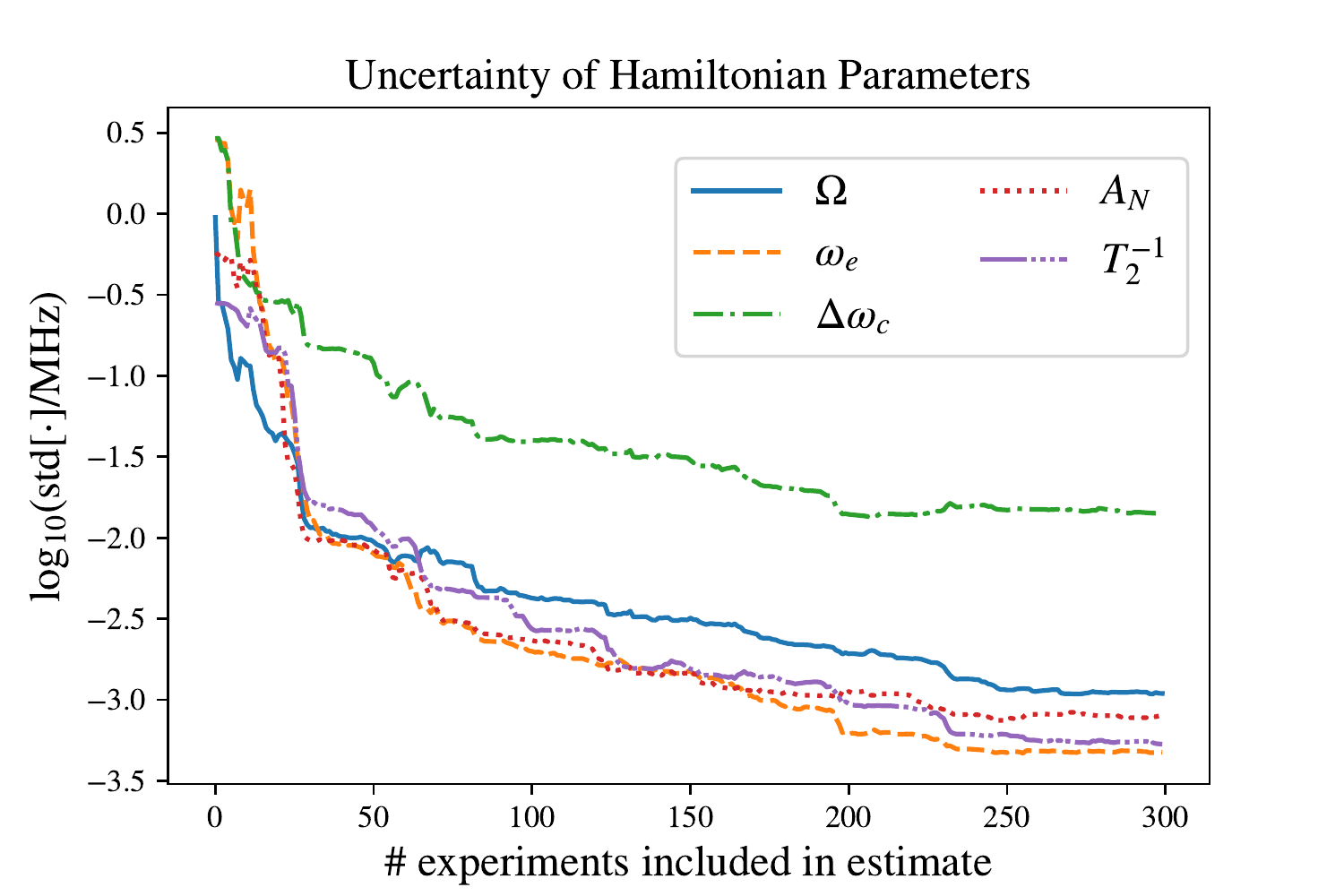}} \\
    \sidesubfloat[]{\includegraphics[width=0.45\textwidth]{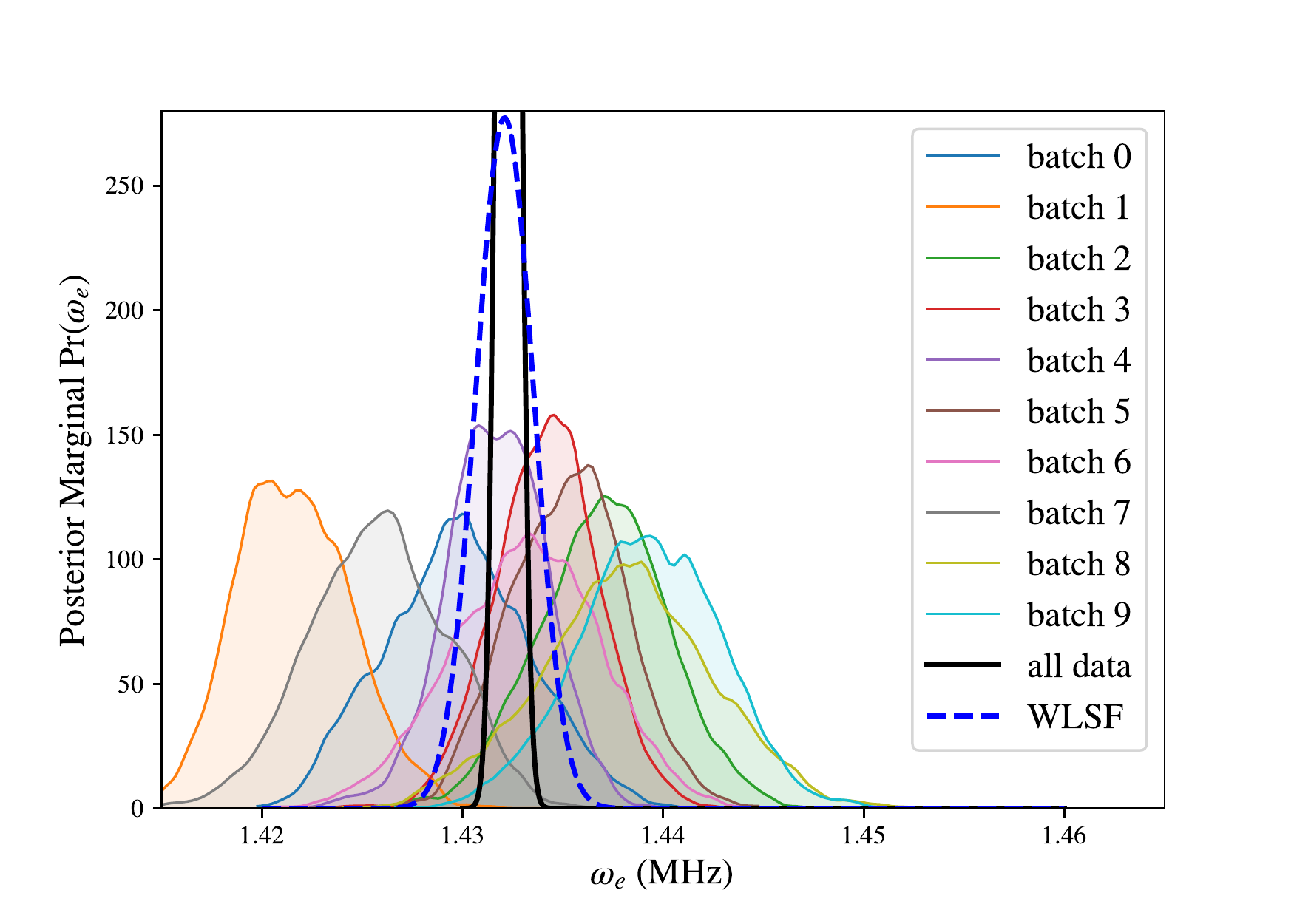}} \quad
    \sidesubfloat[]{\includegraphics[width=0.45\textwidth]{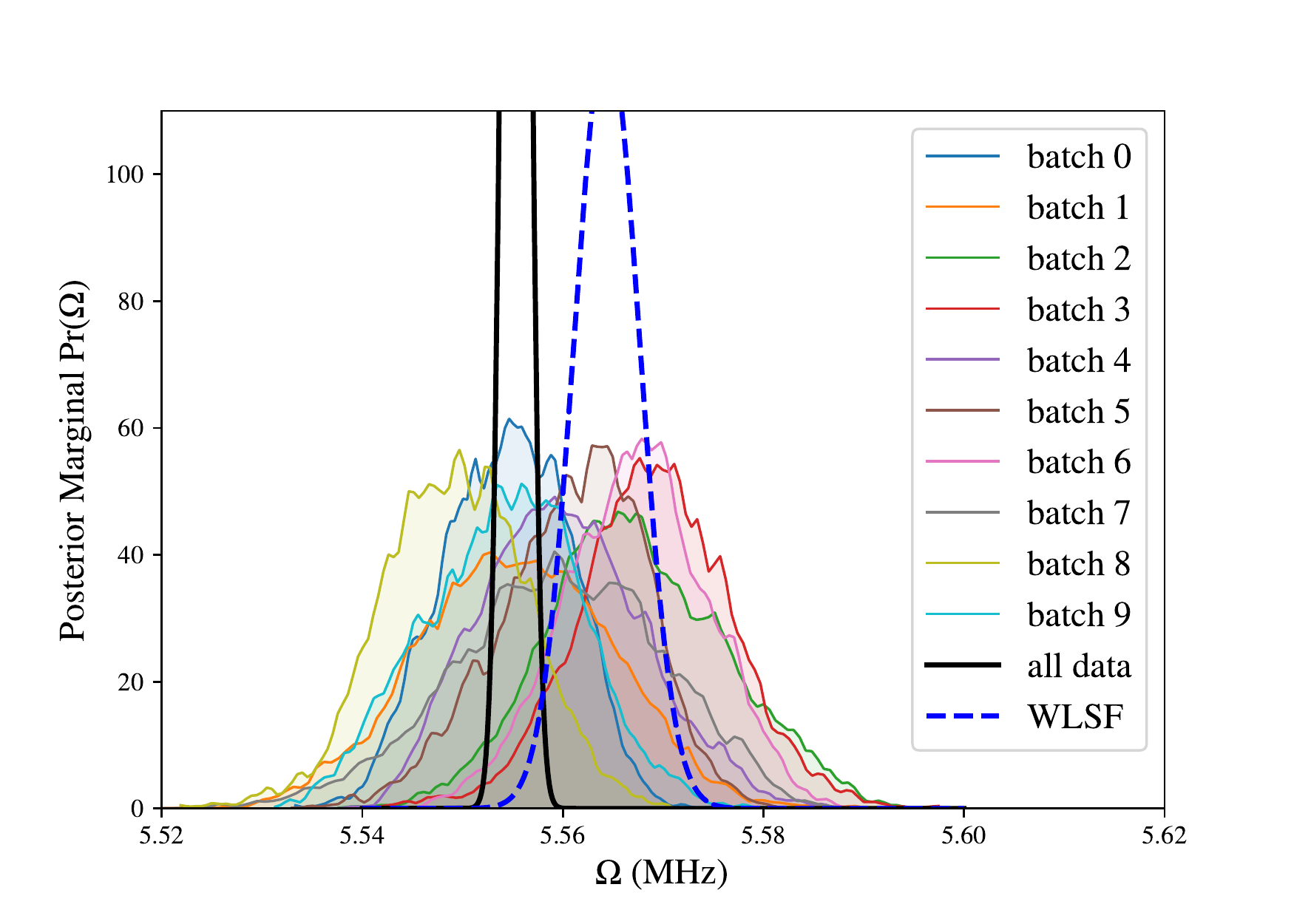}}
    \caption{Results of data processing with the SMC algorithm.
    In (a-b), the expectation value of the SMC posterior,
    $\expect_{\pi^*}[\vec{x}]=(\hat{\omega}_e,\hat{\delta\Delta},\hat{\Omega},\hat{A}_N,\hat{T}_2^{-1})$,
    is is used in a simulation of the Hamiltonian model (\autoref{sec:qhl-hamiltonian-model}),
    and shown on top of the normalized raw data.
    The raw data was normalized using the MLE in Equation~\autoref{eq:mle},
    and the $95\%$ error bars are computed with Equation~\autoref{eq:fisher-confidence} 
    for comparison.
    The posterior distribution is tight enough that simulations from randomly sampled values 
    are visually indistinguishable.
    In (c), the expectation value of the SMC posterior is shown as a function of the 
    number of Bayes' update steps in SMC. 
    It is seen that very little happens to the mean, at least at the 
    zoom level shown, after about 50 updates.
    In (d), however, we continue to see in their standard deviations decrease, 
    where the square root of the diagonal of the covariance matrix 
    is plotted on the same x-axis as (c).
    In (e) and (f) posterior marginal distributions are 
    shown for the parameters $\omega_e$ and $\Omega$, respectively.
    Each of the broad curves (coloured) correspond to results from the 
    same data-processing 
    algorithm run on completely disjoint subsets of experimental data.
    The narrow curve (black) is the result from the algorithm being run on 
    the amalgamation of these datasets.
    For comparison, the dashed curves show a Gaussian distribution 
    with mean and variance computed through a weighted least-squares 
    fit (WLSF) of the entire dataset.}
    \label{fig:qhl-results}
\end{figure}

%=============================================================================
\subsection{QHL Likelihood Function}
\label{sec:qhl-likelihood-function}

As we did for the tomography model sketched in \autoref{sec:generalized-inference-problems},
we now write down a model for our QHL problem.
We label a given experimental configuration as $\vec{c}=(t_r,t_w,t_p,k)$, where the three 
timing parameters 
were defined in ~\autoref{sec:qhl-hamiltonian-model} and $k\in\{\text{RABI},\text{RAMSEY}\}$
selects the experiment types explained in ~\autoref{sec:qhl-choices}.
Similarly, we denote a hypothetical parameter set as a vector $\vec{x}=(\omega_e,\delta\Delta,\Omega,A_N,T_2^{-1})$.
A specific pair, $(\vec{x},\vec{c})$, provides enough information to do a full 
quantum simulation of the spin-1 manifold, resulting in the probability 
of a projective $\ket{0}$ measurement given by
\begin{align}
    p_{\vec{x},\vec{c}} = \Tr \left(\ketbra{0} \mathcal{S}_{\vec{x},\vec{c}}(\ketbra{0}) \right).
    \label{eq:qhl-simulation}
\end{align}

Here, $\mathcal{S}_{\vec{x},\vec{c}}$ is the solution to the Lindblad master
equation under the Hamiltonian model described in \autoref{sec:qhl-hamiltonian-model}.
This yields the conditional model $X,Y,Z|\alpha,\beta,\vec{x};\vec{c}$ 
with $X\sim\Poisson{\alpha}$, $Y\sim\Poisson{\beta}$, and 
$Z\sim\Poisson{\beta+p_{\vec{x},\vec{c}}(\alpha-\beta)}$.
The goal of the inference problem is to deduce the true values of $\vec{x}$,
and in particular, $\omega_e$, given a dataset of photon counts.

%=============================================================================
\subsection{Bayesian Inference with Sequential Monte Carlo}
\label{sec:qhl-inference}

We begin with a prior distribution $\pi(\vec{x})$ describing 
our knowledge of the system before any measurements, given by 
the following product distribution:
\begin{align}
    \omega_e &\sim \Unif{\SI{0}{MHz},\SI{10}{MHz}} \\
    \delta\Delta &\sim \Unif{\SI{-5}{MHz},\SI{5}{MHz}} \\
    \Omega &\sim \Unif{\SI{0}{MHz},\SI{10}{MHz}} \\
    A_N &\sim \Unif{\SI{1.5}{MHz}, \SI{3.5}{MHz}} \\
    T_2^{-1} &\sim \Unif{(\SI{100}{\us})^{-1},(\SI{1}{\us})^{-1}}.
    \label{eq:qhl-prior}
\end{align}
Additionally, we empirically choose a prior for the references $\alpha$
and $\beta$
by computing the sample moments of the experimental reference count data, 
multiplying the standard deviations by $4$ to be conservative, and choosing a
product gamma distribution with these moments.
This distribution is discussed in \autoref{app:drift-prior-uncor}.
We label this distribution as
$\pi_k(\alpha,\beta)$ where $k\in\{\text{RABI},\text{RAMSEY}\}$.

The prior distribution is now sequentially updated through Bayes' law 
one triple $(x_{\vec{c}},y_{\vec{c}},z_{\vec{c}})$ at a time.
Here, $x_{\vec{c}}=\sum_{n=1,r=1}^{N,R}x_{n,{\vec{c}},r}$, 
$y_{\vec{c}}=\sum_{n=1,r=1}^{N,R}y_{n,{\vec{c}},r}$,
and $z_{\vec{c}}=\sum_{n=1,r=1}^{N,R}z_{n,{\vec{c}},r}$ are the
photon counts for a particular experiment $\vec{c}=(t_r,t_w,t_p,k)$
summed over all repetitions and averages (see \autoref{sec:combining-experiments}).
The distribution $\Pr(\alpha,\beta,\vec{x})$ 
is stored as a so-called \textit{particle approximation} consisting 
of a finite list of hypothetical values, called particles, labeled as
$\{(\alpha_i,\beta_i,\vec{x}_i)\}$ with corresponding weights $\{w_i\}$,
$\sum_i w_i=1$.
Typically on the order of $10000$ particles are used for numerical 
stability.

The distribution for the references $\alpha$ and $\beta$ is reset 
to $\pi_k(\alpha,\beta)$ before each triple of data is used.
As discussed in \autoref{sec:bayes-estimator}, we may process the
reference pair $(x_{\vec{c}},y_{\vec{c}})$ first and subsequently 
process the signal count $z_{\vec{c}}$.
The reference pair is processed by replacing the reference
prior with the analytically derived posterior, discussed in 
\autoref{app:drift-prior-uncor}.
This is done by replacing the $(\alpha,\beta)$ coordinates 
of each particle $(\alpha_i,\beta_i,\vec{x}_i)$ 
with a random variate drawn from the posterior $\pi^*(\alpha,\beta|x_{\vec{c}},y_{\vec{c}})$.
The signal is processed using the Sequential Monte Carlo (SMC) algorithm,
as implemented by the software package \qinfer~\cite{granade_qinfer_2017}.

While Bayes' update rule is agnostic to the order in which we enter data since 
each data triple is statistically independent, the order is relevant to the 
numerical implementation.
An intuitive explanation is as follows.
Suppose one is interested in determining the frequency of a cosine wave 
using amplitude data sampled at various time points, and that we assign a flat 
prior distribution over a wide range of frequencies.
If we first update our prior with the data from a late time point, the 
posterior will have many peaks because every divisor of the measurement time
will correspond to a period consistent with the observed data.
Subsequent updates will eventually inform us about which peak contains the 
true frequency.
However, if we first update our prior with data from a time point early enough
that we know (according to our prior) that less than a full period has had time to 
take place,
then the posterior will be a very broad but unimodal.
Subsequent chronological Bayes updates will tend to shift and narrow this peak.
Since the SMC algorithm --- specifically, the Liu--West 
resampler \cite{liu_combined_2001} ---
implicitly assumes unimodality, the first approach will 
usually fail and the second approach will usually succeed, assuming 
the ansatz that there is only one true value in parameter space.
Given this data processing constraint, we fed the data to the Bayes updater 
in strictly increasing times $t_w$ and $t_r$, shuffling the Rabi and 
Ramsey data together randomly.
Alternatively, an algorithm without a unimodality constraint could 
be considered~\cite{granade_structured_2016}.

One nice feature of the SMC algorithm is that it typically heralds its own 
failure through the effective sample size criterion \cite{beskos_stability_2014}.
Such failures can result from multi-modalities, as discussed above.
Another common failure path is through overly-informative data, where a single 
Bayes update causes only a handful of particles to remain relevant.
We mitigate against this partly by using a conjugate prior for the reference 
indices, as discussed in \autoref{sec:bayes-estimator}, and also by a technique
called \textit{bridging the transition}, discussed in \autoref{app:bridged-updater}.

%=============================================================================
\subsection{Results and Validation}
\label{sec:qhl-results}

Bayesian inference with the SMC algorithm was run on the entire dataset 
to obtain a posterior distribution (two-parameter marginals are plotted 
in the appendix, \autoref{app:qhl-figures}).
Recall that our entire dataset consists of 
400 averages of 30000 repetitions for each of the 300 different experimental 
configurations, corresponding to roughly 24 hours of experiment time given 
our particular optical efficiency.
This number of averages was chosen to be large to allow for more convincing 
validation of our techniques.
To this end, the 400 averages were
divided into ten disjoint and chronological batches of 40 averages each, and 
the SMC algorithm was run independently on each batch. 
Since each batch has strictly less data than the entire dataset (effectively 
lower values of $\alpha$ and $\beta$), wider posteriors are expected for these
than for the entire data set.

The main results are shown in \autoref{fig:qhl-results} 
and \autoref{tab:qhl-fits}, with supplementary 
figures found in \autoref{app:qhl-figures}.
The top two plots, \autoref{fig:qhl-results}(a-b), show that the SMC 
posterior corresponds to a sensible 
traditional fit of the data; the posterior is used to obtain a point 
estimate of each of the parameters, and these parameters are then used in
a simulation spanning the experimental configurations.
Since the posterior distribution is tight enough that 
simulations from randomly sampled values are visually indistinguishable,
these fits can be interpreted as a visual posterior predictive check,
where data simulated according to the posterior is compared with 
actual data~\cite{gelman_bayesian_2013}.
The middle two figures show convergence properties of the SMC algorithm.
Finally, the bottom two figures show that the disjoint data sets 
result in posteriors that are consistent with each other, and consistent 
with the posterior of the amalgamated data set.
Keep in mind that the parameters could be fluctuating slightly 
over long time scales. 

For comparison, we also analyzed the data using a weighted least-squares 
fit (WLSF) of of the parameters $\vec{x}=(\omega_e,\delta\Delta,\Omega,A_N,T_2^{-1})$.
This was done by using the \scipy~\cite{jones_scipy:_2001} function \lstinline+optimization.curve_fit+
to minimize the quantity
\begin{align}
    \Phi(\vec{x}) &= \sum_{\vec{c}} \left(
        \frac{
            \hat{p}_{\vec{c}} - p_{\vec{x},\vec{c}}
        }{\sigma_{\vec{c}}}
    \right)^2 \nonumber \\
    \hat{p}_{\vec{c}} & = \frac{z_{\vec{c}} - y_{\vec{c}}}{x_{\vec{c}} - y_{\vec{c}}}, \quad
    \sigma_{\vec{c}}^2
        = \frac{
            \hat{p}_{\vec{c}} (\hat{p}_{\vec{c}}+1) x_{\vec{c}} 
            +(\hat{p}_{\vec{c}}-2) (\hat{p}_{\vec{c}}-1) y_{\vec{c}} 
        }{(x_{\vec{c}} -y_{\vec{c}} )^2}
\end{align}
where the sum is taken over all experiment configurations $\vec{c}=(t_r,t_w,t_p,k)$ 
that were performed, $p_{\vec{x},\vec{c}}$ is the simulation of hypothesis $\vec{x}$
under conditions $\vec{c}$ defined in \autoref{eq:qhl-simulation}, 
$\hat{p}_{\vec{c}}$ is the \MLE~ of $p$ given the data 
$(x_{\vec{c}},y_{\vec{c}},z_{\vec{c}})$,
and the formula for the estimated variance $\sigma_{\vec{c}}^2$ is derived from the 
Cram\'er--Rao bound \autoref{eq:cramer-rao-bound}.
For simplicity, the initial guess of the WLSF function was taken to be the SMC 
point estimate.
The WLSQ fit is shown in \autoref{fig:qhl-results}(e-f) on top of the SMC 
marginal posteriors.
\autoref{tab:qhl-fits} provides a 
more comprehensive comparison, where WLSQ fits are also performed on 
each of the ten batches.
For the smaller batch sizes, the WLSQ confidence intervals are more comparable 
in size to the SMC credible regions.
We suspect this is because SMC did not have enough
data to significantly reduce posterior correlations between parameters, especially
between $\delta\Delta$ and $\omega_e$.

\begin{table}
    \renewcommand{\arraystretch}{1.5}
    \centering
    \begin{tabularx}{0.85\textwidth}{llXXXXXXXXXX}
    ~ & ~ & \multicolumn{2}{c}{$\omega_e$} & \multicolumn{2}{c}{$\delta\Delta$} & \multicolumn{2}{c}{$\Omega$} & \multicolumn{2}{c}{$A_N$} & \multicolumn{2}{c}{$T_2^{-1}$} \\ 
    ~ & ~ & $\mathbb{E}$ & $\sigma$ & $\mathbb{E}$ & $\sigma$ & $\mathbb{E}$ & $\sigma$ & $\mathbb{E}$ & $\sigma$ & $\mathbb{E}$ & $\sigma$ \\ 
    \cline{3-12}
    \multirow{2}{*}{all data\quad} & SMC~~~~ & 1432 & 0.5 & 597 & 13.9 & 5555 & 1.1 & 2171 & 0.8 & 35 & 0.5 \\ 
    & WLSF~~~~ & 1432 & 1.4 & 158 & 54.7 & 5564 & 3.3 & 2169 & 1.8 & 43 & 1.7 \\ 
    \cline{3-12}
    \multirow{2}{*}{batch 0\quad} & SMC~~~~ & 1430 & 3.6 & 639 & 115.0 & 5554 & 6.2 & 2181 & 4.7 & 44 & 2.8 \\ 
    & WLSF~~~~ & 1424 & 4.0 & 773 & 127.8 & 5551 & 9.5 & 2176 & 5.2 & 49 & 4.8 \\ 
    \cline{3-12}
    \multirow{2}{*}{batch 1\quad} & SMC~~~~ & 1422 & 2.8 & -463 & 211.0 & 5555 & 9.6 & 2172 & 3.1 & 40 & 3.6 \\ 
    & WLSF~~~~ & 1425 & 3.5 & -222 & 172.7 & 5562 & 8.4 & 2171 & 4.6 & 41 & 4.0 \\ 
    \cline{3-12}
    \multirow{2}{*}{batch 2\quad} & SMC~~~~ & 1437 & 3.2 & -323 & 279.6 & 5567 & 8.9 & 2163 & 4.9 & 48 & 4.2 \\ 
    & WLSF~~~~ & 1433 & 4.0 & -188 & 153.3 & 5566 & 8.6 & 2164 & 4.9 & 49 & 4.5 \\ 
    \cline{3-12}
    \multirow{2}{*}{batch 3\quad} & SMC~~~~ & 1434 & 2.6 & 30 & 247.2 & 5569 & 7.6 & 2166 & 3.3 & 31 & 2.0 \\ 
    & WLSF~~~~ & 1433 & 3.3 & 153 & 140.4 & 5569 & 8.3 & 2164 & 4.3 & 36 & 3.9 \\ 
    \cline{3-12}
    \multirow{2}{*}{batch 4\quad} & SMC~~~~ & 1432 & 2.5 & -120 & 233.3 & 5560 & 8.0 & 2173 & 4.2 & 40 & 2.3 \\ 
    & WLSF~~~~ & 1431 & 3.6 & -132 & 133.3 & 5563 & 7.8 & 2175 & 4.8 & 46 & 4.3 \\ 
    \cline{3-12}
    \multirow{2}{*}{batch 5\quad} & SMC~~~~ & 1435 & 2.9 & 125 & 147.8 & 5561 & 7.3 & 2177 & 4.2 & 40 & 2.7 \\ 
    & WLSF~~~~ & 1435 & 3.7 & 163 & 140.1 & 5566 & 8.4 & 2178 & 4.8 & 45 & 4.2 \\ 
    \cline{3-12}
    \multirow{2}{*}{batch 6\quad} & SMC~~~~ & 1433 & 3.7 & -41 & 248.9 & 5567 & 7.1 & 2169 & 4.5 & 44 & 3.7 \\ 
    & WLSF~~~~ & 1433 & 3.5 & 0 & 6962.6 & 5571 & 8.1 & 2166 & 4.5 & 44 & 4.0 \\ 
    \cline{3-12}
    \multirow{2}{*}{batch 7\quad} & SMC~~~~ & 1426 & 3.4 & 390 & 499.5 & 5559 & 10.4 & 2168 & 4.3 & 40 & 3.3 \\ 
    & WLSF~~~~ & 1429 & 3.3 & 410 & 127.2 & 5562 & 7.8 & 2165 & 4.1 & 40 & 3.8 \\ 
    \cline{3-12}
    \multirow{2}{*}{batch 8\quad} & SMC~~~~ & 1438 & 4.3 & 203 & 348.1 & 5549 & 7.2 & 2168 & 3.8 & 41 & 2.9 \\ 
    & WLSF~~~~ & 1440 & 3.7 & 511 & 144.9 & 5550 & 8.9 & 2168 & 4.6 & 43 & 4.3 \\ 
    \cline{3-12}
    \multirow{2}{*}{batch 9\quad} & SMC~~~~ & 1439 & 3.5 & 226 & 650.5 & 5554 & 7.8 & 2167 & 3.6 & 45 & 2.9 \\ 
    & WLSF~~~~ & 1439 & 3.9 & 538 & 140.1 & 5554 & 8.8 & 2169 & 4.8 & 46 & 4.3 \\ 
    \cline{3-12}
\end{tabularx}
    \caption{Results of the SMC and WLSF fits are shown for for the entire 
    data sat, as well as for each batch. All units are kHz.
    Both the point estimate ($\mathbb{E}$) and the marginal standard deviation 
    ($\sigma$) are displayed for each of the five fit parameters, 
    $(\Omega, \omega_e, \delta\Delta, A_N, T_2^{-1})$.}
    \label{tab:qhl-fits}
\end{table}

%=============================================================================
\section{Conclusions}
\label{sec:conclusion}
%=============================================================================

We started with a standard phenomenological model of \NV optical dynamics 
and derived random variables corresponding to photon emission during 
measurement.
Noise, limited visibility, and imperfect state preparation were considered 
in detail and added to the model, which did not affect it very much; one 
still ends up with three conditionally independent Poisson random variables,
just with lower count rates and contrast.

This material will have all been review for those familiar with this 
quantum system.
However, we deemed it important to include it explicitly so that our subsequent 
statement of the \NV measurement process as a formal statistical inference 
problem would be entirely self-contained and justified.
We were able to give a straight forward argument for 
why summing up all of the photon counts for repeated experiments of the 
same type is almost always the best thing to do, regardless of data-processing 
techniques, and despite our discussion about the headaches involved in
drifting reference counts.

We derived both frequentist and Bayesian estimators for this inference problem.
We found that the most commonly used frequentist estimator 
is exactly the maximum likelihood estimator, and its 
mean-squared-error risk is well approximated by a 
simple equation derived from the Cram\'er-Rao bound.
This fact produced a useful formula, \autoref{eq:fisher-confidence-requirement},
for estimating the amount of data required to reduce the error bars 
of $p=\Tr(\rho \ketbra{0})$ to a specified value $\Delta p$.
Quantitatively comparing the estimators revealed that the Bayesian 
estimator for $p$ is superior to the \MLE~ with respect to 
mean-squared-error risk, especially near the boundaries of $[0,1]$.

Finally, in order to convince the reader that a Bayesian approach is tractable 
in practice, we perform quantum Hamiltonian learning on experimental 
data in a fully Bayesian setting using the sequential Monte Carlo 
inference algorithm.
In addition to giving a good fit even in the case of a very wide prior, 
cross validation gives convincing
evidence that the posterior distribution over Hamiltonian parameters 
accurately reports its confidence in their values. 
Credible regions reported by SMC are generally better than confidence 
intervals due to weighted least squares fitting, but in some cases, 
not by much.

\acknowledgments{
    IH is grateful to Toeno van der Sar and Thomas Alexander 
    for helpful conversations.
    This research was undertaken thanks in part to funding from:
    The Canada First Research Excellence Fund (CFREF)
    Canadian Excellence Research Chairs (CERC),
    the Natural Sciences and Engineering Research Council of Canada (NSERC),
    the Canadian Institute for Advanced Research,
    the Province of Ontario, and Industry Canada.
}

\bibliographystyle{apsrev4-1}
\bibliography{nvmeas}

%=============================================================================
\appendix
%=============================================================================

%=============================================================================
\section{Data and Code for this Paper}
\label{app:stoch-moments-code}
%=============================================================================

This paper used Wolfram Mathematica 11 for several calculations, both numerical 
and symbolic, along with the corresponding plots found in the figures.
Quantum Hamiltonian learning was performed in Python 2.7 using a custom model
written for the \qinfer~ software package \cite{granade_qinfer_2017}.
All code and data necessary to reproduce the results of this paper
and its appendices are openly hosted in a GitHub repository; follow the link
of Reference~\cite{code_and_data}.

%=============================================================================
\section{Stochastic Moments}
\label{app:stoch-moments-code}
%=============================================================================

Mathematica version 10.0.2.0 was used to execute the code snippets in \autoref{lst:stoc-moms}
and \autoref{lst:stoc-moms2}.

\begin{lstlisting}[style=mathematica,label={lst:stoc-moms},caption={Mathematica code for computing the moments of the stochastic model defined in Equation~\autoref{eq:stoch-model} conditional on the value of $\alpha_1$.}]
    p[expr_] := TransformedProcess[
        expr,
        { 
            \[Nu] \[Distributed] OrnsteinUhlenbeckProcess[0, \[Sigma]\[Nu], \[Theta]\[Nu], \[Alpha]0 - \[CapitalGamma]],
            \[Kappa] \[Distributed] OrnsteinUhlenbeckProcess[\[Kappa]0, \[Sigma]\[Kappa], \[Theta]\[Kappa]]
        },
        t
    ];

    \[Alpha]mean1 = p[\[CapitalGamma]+\[Nu][t]][t]//Mean
    \[Beta]mean1 = p[\[CapitalGamma]+\[Kappa][t]\[Nu][t]][t]//Mean
    \[Alpha]var1 = p[\[CapitalGamma]+\[Nu][t]][t]//Variance
    \[Beta]var1 = p[\[CapitalGamma]+\[Kappa][t]\[Nu][t]][t]//Variance//Expand
    (* We need to compute the covariance manually using Cov[a,b]=Mean[a*b]-Mean[a]Mean[b] *)
    \[Alpha]\[Beta]product =(\[CapitalGamma]+\[Nu][t])(\[CapitalGamma]+\[Kappa][t]\[Nu][t])//Expand;
    \[Alpha]\[Beta]cov1 = Sum[Mean[p[expr][t]],{expr,List@@\[Alpha]\[Beta]product}] - \[Alpha]mean*\[Beta]mean//Expand
\end{lstlisting}

\begin{lstlisting}[style=mathematica,label={lst:stoc-moms2},caption={Mathematica code for using the laws of total expectation, variance, and covariance to derive the moments of the stochastic model defined in Equation~\autoref{eq:stoch-model}. The results are shown in Equation~\autoref{eq:stoch-moments}.}]
    totalExp[mean_] := 
        Mean[TransformedDistribution[mean, \[Alpha]1\[Distributed]NormalDistribution[\[Alpha]0,\[Sigma]\[Alpha]]]]
    totalVar[mean_,var_] := 
        Variance[TransformedDistribution[mean, \[Alpha]1\[Distributed]NormalDistribution[\[Alpha]0,\[Sigma]\[Alpha]]]]
        + Mean[TransformedDistribution[var, \[Alpha]1\[Distributed]NormalDistribution[\[Alpha]0,\[Sigma]\[Alpha]]]]
    totalCov[mean1_,mean2_,cov_] := 
        Covariance[TransformedDistribution[{mean1,mean2}, \[Alpha]1\[Distributed]NormalDistribution[\[Alpha]0,\[Sigma]\[Alpha]]],1,2]
        + Mean[TransformedDistribution[cov, \[Alpha]1\[Distributed]NormalDistribution[\[Alpha]0,\[Sigma]\[Alpha]]]]
        
    \[Alpha]mean = totalExp[\[Alpha]mean1]
    \[Beta]mean = totalExp[\[Beta]mean1]
    \[Alpha]var = totalVar[\[Alpha]mean1,\[Alpha]var1]
    \[Beta]var = totalVar[\[Beta]mean1,\[Beta]var1]//Expand
    \[Alpha]\[Beta]cov = totalCov[\[Alpha]mean1,\[Beta]mean1,\[Alpha]\[Beta]cov1]//Expand
\end{lstlisting}

%=============================================================================
\section{Cumulants of the Conditional Likelihood Function}
\label{app:likelihood-cumulants}
%=============================================================================

We wish to compute
\begin{align}
    I(p,\alpha,\beta)
        &= -\expect_{x,y,z}\left[\left(
            \frac{\partial}{\partial \theta_i}\frac{\partial \log \Lhood}{\partial \theta_j}
            \right)_{i,j=1}^3\right] 
        = \sum_{x=0}^\infty \sum_{y=0}^\infty \sum_{z=0}^\infty 
            \Lhood(x,y,z|p,\alpha,\beta)
            \left(
            \frac{\partial}{\partial \theta_i}\frac{\partial \log \Lhood}{\partial \theta_j}
            \right)_{i,j=1}^3 ,	\\
    K_{i,j,k}
        &= \expect_{x,y,z}\left[  
            \frac{\partial}{\partial \theta_i}
            \frac{\partial}{\partial \theta_j}
            \frac{\partial \log \Lhood}{\partial \theta_k}		
        \right],\quad\text{ and} \\
    J_{j;i,k}
        &= \expect_{x,y,z}\left[  
            \frac{\partial \log \Lhood}{\partial \theta_j}
            \cdot
            \frac{\partial}{\partial \theta_i}
            \frac{\partial \log \Lhood}{\partial \theta_k}		
        \right]
\end{align}
where $\theta_1=p$, $\theta_2=\alpha$, $\theta_3=\beta$, and from 
Equation~\autoref{eq:cond-likelihood} we have
\begin{align}
    \Lhood(x,y,z|p,\alpha,\beta) 
        &= \frac{\alpha^x\e^{-\alpha}}{x!}
            \cdot \frac{\beta^y\e^{-\beta}}{y!}
            \cdot \frac{(p\alpha+(1-p)\beta)^z\e^{-(p\alpha+(1-p)\beta)}}{z!}.
\end{align}
\autoref{lst:likelihood-cumulants}.
The results of this code for the Fisher information matrix are
\begin{align}
    I(p,\alpha,\beta)
        &= -\expect_{x,y,z}\left[\left(
            \frac{\partial}{\partial \theta_i}\frac{\partial \log \Lhood}{\partial \theta_j}
            \right)_{i,j=1}^3\right] \nonumber \\
        &= 
        \begin{pmatrix}
            \frac{(\alpha -\beta )^2}{p (\alpha -\beta )+\beta } & 
            \frac{p (\alpha -\beta )}{p (\alpha -\beta )+\beta } & 
            \frac{\alpha }{\beta +\alpha  p-\beta  p}-1 \\
            \frac{p (\alpha -\beta )}{p (\alpha -\beta )+\beta } & 
            \frac{p^2}{p \alpha -p \beta +\beta }+\frac{1}{\alpha } & 
            -\frac{(p-1) p}{p (\alpha -\beta )+\beta } \\
            \frac{\alpha }{p \alpha -p \beta +\beta }-1 & 
            -\frac{(p-1) p}{p (\alpha -\beta )+\beta } & 
            \frac{p \alpha +(p-2) (p-1) \beta }{\beta  (p (\alpha -\beta )+\beta )} \\
        \end{pmatrix}
\end{align}
with inverse
\begin{align}
    I(p,\alpha,\beta)^{-1}
        &=
        \begin{pmatrix}
            \frac{p (p+1) \alpha +(p-2) (p-1) \beta }{(\alpha -\beta )^2} & 
            \frac{p \alpha }{\beta -\alpha } & 
            \frac{(p-1) \beta }{\alpha -\beta } \\
            \frac{p \alpha }{\beta -\alpha } & 
            \alpha  & 
            0 \\
            \frac{(p-1) \beta }{\alpha -\beta } & 
            0 & 
            \beta
        \end{pmatrix}.
\end{align}
Further, the higher order cumulants turn out to be 
\begin{align}
    K_{1,\cdot,\cdot} &=
        \begin{pmatrix}
            \frac{2 (\alpha -\beta )^3}{(p (\alpha -\beta )+\beta )^2} & 
            \frac{2 \beta  (\beta -\alpha )}{(p (\alpha -\beta )+\beta )^2} & 
            \frac{2 \alpha  (\alpha -\beta )}{(p (\alpha -\beta )+\beta )^2} \\
            \frac{2 \beta  (\beta -\alpha )}{(p (\alpha -\beta )+\beta )^2} & 
            -\frac{2 p \beta }{(p (\alpha -\beta )+\beta )^2} & 
            \frac{p (\alpha +\beta )-\beta }{(p (\alpha -\beta )+\beta )^2} \\ 
            \frac{2 \alpha  (\alpha -\beta )}{(p (\alpha -\beta )+\beta )^2} & 
            \frac{p (\alpha +\beta )-\beta }{(p (\alpha -\beta )+\beta )^2} & 
            -\frac{2 (p-1) \alpha }{(p (\alpha -\beta )+\beta )^2} \\
        \end{pmatrix} \nonumber \\
    K_{2,\cdot,\cdot} &=
        \begin{pmatrix}
            \frac{2 \beta  (\beta -\alpha )}{(p (\alpha -\beta )+\beta )^2} & 
            -\frac{2 p \beta }{(p (\alpha -\beta )+\beta )^2} & 
            \frac{p (\alpha +\beta )-\beta }{(p (\alpha -\beta )+\beta )^2} \\ 
            -\frac{2 p \beta }{(p (\alpha -\beta )+\beta )^2} & 
            2 \left(\frac{p^3}{(p (\alpha -\beta )+\beta )^2}+\frac{1}{\alpha ^2}\right) & 
            -\frac{2 (p-1) p^2}{(p (\alpha -\beta )+\beta )^2} \\ 
            \frac{p (\alpha +\beta )-\beta }{(p (\alpha -\beta )+\beta )^2} & 
            -\frac{2 (p-1) p^2}{(p (\alpha -\beta )+\beta )^2} & 
            \frac{2 (p-1)^2 p}{(p (\alpha -\beta )+\beta )^2} \\
        \end{pmatrix} \nonumber \\
    K_{3,\cdot,\cdot} &=
        \begin{pmatrix}
            \frac{2 \alpha  (\alpha -\beta )}{(p (\alpha -\beta )+\beta )^2} & 
            \frac{p (\alpha +\beta )-\beta }{(p (\alpha -\beta )+\beta )^2} & 
            -\frac{2 (p-1) \alpha }{(p (\alpha -\beta )+\beta )^2} \\ 
            \frac{p (\alpha +\beta )-\beta }{(p (\alpha -\beta )+\beta )^2} & 
            -\frac{2 (p-1) p^2}{(p (\alpha -\beta )+\beta )^2} & 
            \frac{2 (p-1)^2 p}{(p (\alpha -\beta )+\beta )^2} \\ 
            -\frac{2 (p-1) \alpha }{(p (\alpha -\beta )+\beta )^2} & 
            \frac{2 (p-1)^2 p}{(p (\alpha -\beta )+\beta )^2} & 
            \frac{2}{\beta ^2}-\frac{2 (p-1)^3}{(p (\alpha -\beta )+\beta )^2} \\
        \end{pmatrix}
\end{align}
and
\begin{align}
    J_{1;\cdot,\cdot} &=
        \begin{pmatrix}
             \frac{(\beta -\alpha )^3}{(p (\alpha -\beta )+\beta )^2} & 
             \frac{(\alpha -\beta ) \beta }{(p (\alpha -\beta )+\beta )^2} & 
             \frac{\alpha  (\beta -\alpha )}{(p (\alpha -\beta )+\beta )^2} \\ 
             \frac{(\alpha -\beta ) \beta }{(p (\alpha -\beta )+\beta )^2} & 
             \frac{p^2 (\beta -\alpha )}{(p (\alpha -\beta )+\beta )^2} & 
             \frac{(p-1) p (\alpha -\beta )}{(p (\alpha -\beta )+\beta )^2} \\
             \frac{\alpha  (\beta -\alpha )}{(p (\alpha -\beta )+\beta )^2} & 
             \frac{(p-1) p (\alpha -\beta )}{(p (\alpha -\beta )+\beta )^2} &  
             -\frac{(p-1)^2 (\alpha -\beta )}{(p (\alpha -\beta )+\beta )^2} \\
        \end{pmatrix} \nonumber \\
    J_{2;\cdot,\cdot} &=
        \begin{pmatrix}
            -\frac{p (\alpha -\beta )^2}{(p (\alpha -\beta )+\beta )^2} & 
            \frac{p \beta }{(p (\alpha -\beta )+\beta )^2} & 
            -\frac{p \alpha }{(p (\alpha -\beta )+\beta )^2} \\ 
            \frac{p \beta }{(p (\alpha -\beta )+\beta )^2} & 
            -\frac{p^3}{(p (\alpha -\beta )+\beta )^2}-\frac{1}{\alpha ^2} & 
            \frac{(p-1) p^2}{(p (\alpha -\beta )+\beta )^2} \\ 
            -\frac{p \alpha }{(p (\alpha -\beta )+\beta )^2} & 
            \frac{(p-1) p^2}{(p (\alpha -\beta )+\beta )^2} & 
            -\frac{(p-1)^2 p}{(p (\alpha -\beta )+\beta )^2} \\
        \end{pmatrix} \nonumber \\
    J_{3;\cdot,\cdot} &=
        \begin{pmatrix}
            \frac{(p-1) (\alpha -\beta )^2}{(p (\alpha -\beta )+\beta )^2} & 
            -\frac{(p-1) \beta }{(p (\alpha -\beta )+\beta )^2} & 
            \frac{(p-1) \alpha }{(p (\alpha -\beta )+\beta )^2} \\ 
            -\frac{(p-1) \beta }{(p (\alpha -\beta )+\beta )^2} & 
            \frac{(p-1) p^2}{(p (\alpha -\beta )+\beta )^2} & 
            -\frac{(p-1)^2 p}{(p (\alpha -\beta )+\beta )^2} \\ 
            \frac{(p-1) \alpha }{(p (\alpha -\beta )+\beta )^2} & 
            -\frac{(p-1)^2 p}{(p (\alpha -\beta )+\beta )^2} & 
            \frac{(p-1)^3}{(p (\alpha -\beta )+\beta )^2}-\frac{1}{\beta ^2} \\
        \end{pmatrix}.
\end{align}

\begin{lstlisting}[style=mathematica,label={lst:likelihood-cumulants},caption={Mathematica code to find the
fisher information matrix, as well as two higher order cumulants.}]
    (* Define assumptions on variables *)
    $Assumptions = 0<\[Beta]<\[Gamma]<\[Alpha] && 0<=p<=1 && x>0 && y>0 && z>0;
    (* Define the log-likelihood *)
    L = (\[Alpha]^x Exp[-\[Alpha]])/Factorial[x]
        * (\[Beta]^y Exp[-\[Beta]])/Factorial[y]
        * ((p \[Alpha]+(1-p)\[Beta])^z Exp[-(p \[Alpha]+(1-p)\[Beta])])/Factorial[z];
    LL = Log[L] // FullSimplify;
    (* Verify that L is normalized as a sanity check *)
    Sum[L,{x,0,\[Infinity]},{y,0,\[Infinity]},{z,0,\[Infinity]}] // Simplify

    (* Compute Fisher matrix and its inverse *) 
    Ifisher = -Sum[Evaluate[Simplify[
                Outer[D[LL,#1,#2]&,{p,\[Alpha],\[Beta]},{p,\[Alpha],\[Beta]}]L
            ]],
            {x,0,\[Infinity]}, {y,0,\[Infinity]}, {z,0,\[Infinity]}
        ]  //  FullSimplify;
    Iinv = Ifisher // Inverse // FullSimplify;

    (* Compute K and J (takes a while, especially J) *)
    K = With[
            {summand=FullSimplify[L*Outer[D[LL,#1,#2,#3]&,{p,\[Alpha],\[Beta]},{p,\[Alpha],\[Beta]},{p,\[Alpha],\[Beta]}]]},
            Sum[summand, {x,0,\[Infinity]}, {y,0,\[Infinity]}, {z,0,\[Infinity]}]
        ] // FullSimplify;
    J = With[
            {summand=L*FullSimplify[Outer[D[LL,#2,#3]*D[LL,#1]&,{p,\[Alpha],\[Beta]},{p,\[Alpha],\[Beta]},{p,\[Alpha],\[Beta]}]]},
            Sum[summand, {x,0,\[Infinity]}, {y,0,\[Infinity]}, {z,0,\[Infinity]}]
        ] // FullSimplify;
\end{lstlisting}

%=============================================================================
\section{Maximum Likelihood Estimator}
%=============================================================================

\subsection{Derivation}
\label{app:mle-derivation}

The log-likelihood of the basic conditional model is given by
\begin{align}
    \log L
        &= 
        (x \log \alpha - \alpha - \log x!) 
        + (y \log \beta - \beta - \log y!)
        + (z \log (p\alpha + (1-p)\beta) - (p\alpha + (1-p)\beta) - \log z!).
\end{align}
Having fixed a some particular values of $x$, $y$, and $z$, the goal is 
to maximize the function $\log L(p,\alpha,\beta)$.
The value which maximizes this function is then the MLE.
The easiest method is to consider the equivalent problem maximizing the function 
\begin{align}
    (x \log \alpha - \alpha - \log x!) 
    + (y \log \beta - \beta - \log y!)
    + (z \log \gamma - (p\alpha + (1-p)\beta) - \log z!
\end{align}
subject to the constraint $\gamma=p\alpha + (1-p)\beta$.
Using the Lagrange multiplier $\lambda$, this is encoded in a Lagrangian as
\begin{align}
    \Phi &= [(x \log \alpha - \alpha - \log x!) 
        + (y \log \beta - \beta - \log y!)
        + (z \log \gamma - (p\alpha + (1-p)\beta) - \log z!)] \nonumber \\
        &\quad\quad-\lambda [\gamma-(p\alpha + (1-p)\beta)]
    \label{eq:mle-lagrangian}
\end{align}
yielding a simple set of equations
\begin{align}
    \{
        \frac{\partial \Phi}{\partial\alpha}=0,
        \frac{\partial \Phi}{\partial\beta}=0,
        \frac{\partial \Phi}{\partial\gamma}=0,
        \frac{\partial \Phi}{\partial p}=0,
        \frac{\partial \Phi}{\partial\lambda}=0
    \}
\end{align}
with no more logarithms.
These equations can be solved for $\alpha$, $\beta$,
and $p$ as a function of $x$, $y$, and $z$.
This was done in Mathematica 10.0.2.0 using the snippet found 
in \autoref{lst:lagrange-solve}.

This calculation can also be done directly by taking partial derivatives 
of $\log L$ and setting them to zero, although the algebra is 
significantly more demanding, and the order in which the equations are 
solved for affects the difficulty of subsequent steps.

\begin{lstlisting}[style=mathematica,label={lst:lagrange-solve},caption={Mathematica code to solve the Lagrange problem stated in Equation~\autoref{eq:mle-lagrangian}}]
    (* Define assumptions on variables *)
    $Assumptions = 0<\[Beta]<\[Gamma]<\[Alpha] && 0<=p<=1 && x>0 && y>0 && z>0 && \[Lambda]\[Element]Reals;
    (* Define Lagrangian *)
    \[CapitalPhi] = Log[
            \[Alpha]^x Exp[-\[Alpha]]/Factorial[x]
            *\[Beta]^y Exp[-\[Beta]]/Factorial[y]
            *(p \[Alpha]+(1-p)\[Beta])^z Exp[-\[Gamma]]/Factorial[z]
        ] - \[Lambda] (\[Gamma]-(p \[Alpha]+(1-p)\[Beta])); 
    (* Take partial derivatives to get system of equations *)
    lagrangeEquations = D[\[CapitalPhi],#]==0 & /@ {\[Alpha],\[Beta],\[Gamma],\[Lambda],p} // FullSimplify;
    (* Remove \[Lambda] from the set of equations*)
    lagrangeEquations = Rest[lagrangeEquations] /. First@Solve[First@lagrangeEquations,\[Lambda]] // FullSimplify;
    (* Remove \[Gamma] from the set of equations*)
    lagrangeEquations = lagrangeEquations[[{1,2,4}]] /. First@Solve[lagrangeEquations[[3]],\[Gamma]];
    (* Solve for the remaning variables, giving the desired result. *)
    soln = Solve[lagrangeEquations, {\[Alpha], \[Beta], p}] // FullSimplify
\end{lstlisting}

%=============================================================================
\subsection{Bias}
\label{app:mle-bias}

Throughout this section, if an integral is performed or a series is 
summed without explanation, the reader can assume it was done using
Mathematica.
For a variate $(x,y,z)$ of $(X,Y,Z)|\alpha,\beta$ consider the estimator 
\begin{align}
    \hat{p}_{\MLE,\epsilon} = \frac{z-y}{x-y+\epsilon}
\end{align}
for $p=\frac{\gamma-\beta}{\alpha-\beta}$ where
$\epsilon$ is any non-integer.
We are ultimately interested in the limiting case as $\epsilon\rightarrow 0$ 
since this gives the maximum likelihood estimator.
We wish to compute the bias of $\hat{p}_{\MLE,\epsilon}$, which is given by
\begin{align}
    \Bias[\hat{p}_{\MLE,\epsilon}]
        &= \expect_{x,y,z}[\hat{p}_{\MLE,\epsilon} - p] \nonumber \\
        &= \sum_{x=0}^\infty \sum_{y=0}^\infty \sum_{z=0}^\infty
            \frac{z-y}{x-y+\epsilon} 
            \frac{\alpha^x \e^{-\alpha}}{x!}
            \cdot \frac{\beta^x \e^{-\beta}}{y!}
            \cdot \frac{\gamma^x \e^{-\gamma}}{z!}
        -p.
\end{align}
This triple sum is not straight forward to compute.
Our strategy is to first sum over $z$ and $x$ resulting in
\begin{align}
    \expect_{x,y,z}[\hat{p}_{\MLE,\epsilon}]
        &= \sum_{y=0}^\infty
            \left[
\frac{\gamma  e^{-\alpha -\beta } \beta ^y (-\alpha )^{y-\epsilon } \Gamma (\epsilon -y)}{\Gamma (y+1)}
-\frac{e^{-\alpha -\beta } \beta ^y (-\alpha )^{y-\epsilon } \Gamma (\epsilon -y)}{\Gamma (y)} 
\right. \nonumber \\
& \quad\quad
\left.+\frac{e^{-\alpha -\beta } \beta ^y (-\alpha )^{y-\epsilon } \Gamma (\epsilon -y,-\alpha )}{\Gamma (y)}
-\frac{\gamma  e^{-\alpha -\beta } \beta ^y (-\alpha )^{y-\epsilon } \Gamma (\epsilon -y,-\alpha )}{\Gamma (y+1)} \right]
\end{align}
where $\Gamma(x)=\int_0^\infty t^{x-1}e^{-t}\dd t$ is the 
gamma function, and $\Gamma(s,x)=\int_x^\infty t^{s-1} e^{-t}\dd t$ 
is the upper incomplete gamma function.
Note that the non-integer value of $\epsilon$ allows us to avoid 
poles of the gamma function.
The first two terms of the sum are seen to be purely imaginary.
Since the expectation value is known to be real, we may ignore them.

To proceed, we make use of the complex integral 
form of the incomplete gamma function,
\begin{align}
    \Gamma(\epsilon-y,-\alpha)
        &= \int_{-\alpha}^\infty t^{\epsilon-y-1}e^{-t} \dd t 
        = \lim_{R \rightarrow \infty} \int_{-1}^R (\alpha t)^{\epsilon-y-1}e^{-\alpha t} \alpha \dd t,
\end{align}
which holds for any integration path in $\Complex$ from $-1$ to $R$ which 
does not cross the negative real axis.
For the third term we get
\begin{align}
    \sum_{y=0}^\infty
        \frac{e^{-\alpha -\beta } \beta ^y (-\alpha )^{y-\epsilon } \Gamma (\epsilon -y,-\alpha )}
            {\Gamma (y)}
        &= \int_{-1}^\infty \sum_{y=0}^\infty
            \frac{e^{-\alpha -\beta }\beta ^y (-1)^{y-\epsilon }t^{\epsilon-y-1}e^{-\alpha t}}
                {\Gamma(y)} \dd t\nonumber \\
        &= \int_{-1}^\infty
            \beta  (-1)^{1-\epsilon } t^{\epsilon -2} e^{-\alpha(1+t)-\beta(1+1/t)}\dd t
\end{align}
and similarly
\begin{align}
    \sum_{y=0}^\infty
        -\frac{\gamma e^{-\alpha-\beta } \beta ^y (-\alpha )^{y-\epsilon }\Gamma(\epsilon -y,-\alpha )}
            {\Gamma (y+1)}
        &= \int_{-1}^\infty \sum_{y=0}^\infty
            -\frac{\gamma e^{-\alpha-\beta } \beta ^y (-1)^{y-\epsilon }t^{\epsilon-y-1}e^{-\alpha t}}
                {\Gamma (y+1)} \dd t \nonumber \\
        &= \int_{-1}^\infty
            \gamma  (-1)^{1-\epsilon } t^{\epsilon -1} e^{-\alpha(1+t)-\beta(1+1/t)}\dd t
\end{align}
for the last term.
Therefore
\begin{align}
    \expect_{x,y,z}[\hat{p}_{\MLE,\epsilon}]
        &= \int_{-1}^\infty
            (-1)^{1-\epsilon } e^{-\alpha(1+t)-\beta(1+1/t)}
            (\gamma t^{\epsilon-1}+\beta t^{\epsilon-2})
            \dd t
\end{align}
Consider the integration path which is a 
semi-circle from $-1$ to $1$, and
then a straight line to $R$ along the positive real axis, 
avoiding the singularity at $t=0$.
Note that along the positive real axis, the integrand is 
purely imaginary.
We are therefore only interested in the real part of the 
integral along the unit semi-circle.
Making the change of variables $t=e^{i\phi}$ we are left with
\begin{align}
    \expect_{x,y,z}[\hat{p}_{\MLE,\epsilon}]
    &= \re\int_0^\pi
            \ii (-1)^{-\epsilon } e^{-\alpha(1+e^{\ii\phi})-\beta(1+e^{-\ii\phi})}
            (\gamma e^{\ii\phi\epsilon}+\beta e^{\ii\phi(\epsilon-1)})
            \dd \phi
\end{align}
from which we conclude
\begin{align}
    \expect_{x,y,z}[\hat{p}_{\MLE}]
    &= \lim_{\epsilon\rightarrow 0} \expect_{x,y,z}[\hat{p}_{\MLE,\epsilon}] \nonumber \\
    &= \re\int_0^\pi
            \ii e^{-\alpha(1+e^{\ii\phi})-\beta(1+e^{-\ii\phi})}
            (\gamma +\beta e^{-\ii\phi})
            \dd \phi \nonumber \\
    &= \int_0^\pi
        e^{-(\alpha+\beta)(1+\cos\phi)}\left[
            (\gamma+\beta\cos\phi)\sin((\alpha-\beta)\sin\phi)+(\beta\sin\phi)\cos((\alpha-\beta)\sin\phi)
        \right] \dd\phi.
\end{align}
From this expression we can see that $\expect_{x,y,z}[\hat{p}_{\MLE}]$ 
is exactly linear with respect to $\gamma$, and therefore the 
bias will be exactly linear with respect to $p$.
This integral is best done numerically.
What follows is a closed form approximation which will
usually be more practical.

We see that the integrand falls off exponentially 
with rate $\alpha+\beta$ as $\phi$ decreases from $\pi$.
Since it will usually hold that $\alpha+\beta \gg 2$, the 
integrand will have most of its support 
in a region close to $\pi$.
This justifies a low order series expansion of trigonometric 
functions about $\phi=\pi$, giving
\begin{align}
    \expect_{x,y,z}[\hat{p}_{\MLE}]
    &\approx \int_0^\pi
        e^{-(\alpha+\beta)(\phi-\pi)^2/2}\left[
            -(\gamma-\beta)\sin((\alpha-\beta)(\phi-\pi))-(\phi-\pi)\beta\cos((\alpha-\beta)(\phi-\pi))
        \right] \dd\phi \nonumber \\
    &= \int_{-\pi}^0
        e^{-(\alpha+\beta)\phi^2/2}\left[
            -(\gamma-\beta)\sin((\alpha-\beta)\phi)-\phi\beta\cos((\alpha-\beta)\phi)
        \right] \dd\phi \nonumber \\
    &\approx \int_{-\infty}^0
        e^{-(\alpha+\beta)\phi^2/2}\left[
            -(\gamma-\beta)\sin((\alpha-\beta)\phi)-\phi\beta\cos((\alpha-\beta)\phi)
        \right] \dd\phi \nonumber \\
    &= \frac{\gamma-\beta}{\alpha-\beta} +
        \left(
            \frac{\gamma-\beta}{\alpha-\beta} 
            - \frac{\beta}{\alpha+\beta}
        \right)
        \left(
            \sqrt{2}\frac{\alpha-\beta}{\sqrt{\alpha+\beta}}
            F\left( \frac{\alpha-\beta}{\sqrt{2}\sqrt{\alpha+\beta}} \right)
            -1
        \right) \nonumber \\
    &= p + \left(
            p - \frac{\beta}{\alpha+\beta}
        \right)
        f\left( \frac{\alpha-\beta}{\sqrt{\alpha+\beta}} \right)
\end{align}
where $F(x)=e^{-x^2}\int_0^x e^{t^2}\dd t$ is the Dawson function and
$f(x)=\sqrt{2}xF(x/\sqrt{2})-1$.
Numerics show that these approximations are accurate to 
$\order{(\alpha+\beta)^{-2}}$ for $\alpha+\beta\gtrsim 700$.
For large $x$ we have the series
$f(x)=\frac{1}{x^2}+\frac{3}{x^4}+\order{x^{-6}}$.
It follows that 
\begin{align}
    \Bias[\hat{p}_\MLE]
        &\approx
        \left(
            p - \frac{\beta}{\alpha+\beta}
        \right)
        \frac{\alpha+\beta}{(\alpha-\beta)^2}
        + \order{(\alpha+\beta)^{-2}}
\end{align}
We see that the estimator is unbiased at the single point
$p=\frac{\beta}{\alpha+\beta}$ and that worst bias 
happens at $p=0$ or $p=1$ and scales as 
$\order{\frac{1}{\alpha+\beta}}$.
Note that we have assumed that the contrast
$\frac{\alpha-\beta}{\alpha+\beta}$ stays fixed to 
make the above asymptotic arguments.

%=============================================================================
\section{Bias Corrected Estimator}
\label{app:bias-corr-estimator}
%=============================================================================

We can use Equation~\autoref{eq:mle-bias} to attempt to derive an estimator which
is less biased than the MLE by subtracting off an estimate of the 
bias.
This gives
\begin{align}
    \hat{p}_\BCE
        &= \hat{p}_\MLE + 
            \left(
                \hat{p}_\MLE - \frac{\hat\beta}{\hat\alpha+\hat\beta}
            \right)
            \frac{\hat\alpha+\hat\beta}{(\hat\alpha-\hat\beta)^2} \nonumber \\
        &= \frac{z((x-y)^2-x-y)-y((x-y)^2-2x-2y)}{(x-y)^3}
\end{align}

However, recall from \autoref{sec:cond-fisher-info} that the Cram\'er--Rao
bound gives us the lower bound  
\begin{align}
    \Var[\hat{p}_\MLE]
        &\geq 
        \frac{p (p+1) \alpha +(p-2) (p-1) \beta }{(\alpha -\beta )^2}
\end{align}
on the variance of our estimator.
If we take the average of this bound at $p=0$ and $p=1$ we get 
$\frac{\alpha+\beta}{(\alpha-\beta)^2}$ which happens to be equal to
the worst-case bias derived above.
We conclude that when $\alpha+\beta\gtrsim 300$, the bias of the $\MLE$ is 
negligible since it will be of $\order{(\alpha+\beta)^{-1}}$, well contained 
within a single standard deviation of $\hat{p}_\MLE$, 
$\order{(\alpha+\beta)^{-\frac{1}{2}}}$.

Moreover, the estimate of the bias 
$\left(\hat{p}_\MLE-\frac{\hat\beta}{\hat\alpha+\hat\beta}\right)\frac{\hat\alpha+\hat\beta}{(\hat\alpha-\hat\beta)^2}$
used above is itself likely to have a variance and bias which outweigh 
that which it is trying to correct.
Therefore, this estimator is not useful in practice.

%=============================================================================
\section{Conjugate Priors for Drift Parameters}
\label{app:conjugate-priors}
%=============================================================================

Given a likelihood function, its \textit{conjugate prior} is a special 
family of distributions such that the posterior is also in the same family
of distributions.
Therefore conjugate priors, when they exist, are very useful at reducing 
the complexity of applying Bayesian inference.

%=============================================================================
\subsection{Uncorrelated Conjugate Prior}
\label{app:drift-prior-uncor}

Since the likelihood function $L(\alpha,\beta|x,y)$ for the drift parameters
is separable into two Poisson distributions,
\begin{align}
    L(\alpha,\beta|x,y)
        &= \frac{\alpha^x\e^{-\alpha}}{x!}
            \cdot \frac{\beta^y\e^{-\beta}}{y!},
\end{align}
the product of Gamma distributions will be a conjugate prior.
Indeed, with the prior
\begin{align}
    \pi(\alpha,\beta)
        &= \pi(\alpha)\pi(\beta)
        = \pdf{Gamma}{\alpha;a_\alpha,b_\alpha}\cdot\pdf{Gamma}{\beta;a_\beta,b_\beta}
\end{align}
where $\pdf{Gamma}{\xi;a,b}=\frac{b^a\xi^{a-1}\e^{-\xi b}}{\Gamma(a)}$ and $\Gamma$ 
is the gamma function, and given the variate $(x,y)$ of $(X,Y)|\alpha,\beta$,
the posterior distribution takes the analytic form
\begin{align}
    \pi^*(\alpha,\beta)
        = \Pr[\alpha,\beta|x,y]
        &= \frac{\Pr[x,y|\alpha\beta]\pi(\alpha,\beta)}{\int\Pr[x,y|\alpha\beta]\pi(\alpha,\beta)\dd \alpha\dd\beta} \nonumber \\
        &= \pdf{Gamma}{\alpha;a_\alpha+x,b_\alpha+1}\cdot\pdf{Gamma}{\beta;a_\beta+y,b_\beta+1}.
\end{align}
This convenient fact means that if we describe our knowledge of the references
$\alpha$ and $\beta$ by the hyperparameters $(a_\alpha,b_\alpha,a_\beta,b_\beta)$, 
then the hyperparameters describing the posterior are 
$(a_\alpha+x,b_\alpha+1,a_\beta+y,b_\beta+1)$.
Note that the mean and variance of the gamma distribution $\Gam{a}{b}$ 
are given by $\mu=\frac{a}{b}$ and $\sigma^2=\frac{a}{b^2}$, respectively.
These equations can be uniquely inverted as $a=\frac{\mu^2}{\sigma^2}$ 
and $b=\frac{\mu}{\sigma^2}$.
Therefore we can equivalently, but more intuitively, describe our prior 
with the hyperparameters $(\mu_\alpha,\sigma_\alpha^2,\mu_\beta,\sigma_\beta^2)$ 
which give posterior hyperparameters 
\begin{align}
    \left(
        a_\alpha^* = x+\frac{\mu_\alpha^2}{\sigma_\alpha^2},
        b_\alpha^* = 1+\frac{\mu_\alpha}{\sigma_\alpha^2},
        a_\beta^* = y+\frac{\mu_\beta^2}{\sigma_\beta^2},
        b_\beta^* = 1+\frac{\mu_\beta}{\sigma_\beta^2},
    \right)
\end{align}
or, in terms of mean and variance,
\begin{align}
    \left(
        \mu_\alpha^* = \frac{\mu_\alpha ^2+\sigma_\alpha ^2 x}{\mu_\alpha +\sigma_\alpha ^2},
        (\sigma_\alpha^*)^2 = \frac{\sigma_\alpha ^2 \left(\mu_\alpha ^2+\sigma_\alpha ^2 x\right)}{\left(\mu_\alpha +\sigma_\alpha ^2\right)^2}
        \mu_\beta^* = \frac{\mu\beta ^2+\sigma\beta ^2 y}{\mu\beta +\sigma\beta ^2},
        (\sigma_\beta^*)^2 = \frac{\sigma\beta ^2 \left(\mu\beta ^2+\sigma\beta ^2 y\right)}{\left(\mu\beta +\sigma\beta ^2\right)^2}
    \right).
\end{align}

%=============================================================================
\subsection{Correlated Conjugate Prior}
\label{app:drift-prior-cor}

The prior introduced in the previous subsection assumes that the parameters 
$\alpha$ and $\beta$ are uncorrelated. 
This will rarely if ever be true in practice; a large positive 
correlation is expected.
Therefore, in addition to the hyperparameters 
$(\mu_\alpha,\sigma_\alpha^2,\mu_\beta,\sigma_\beta^2)$ 
we would like to add a fifth hyperparameter, $\sigma_{\alpha,\beta}$,
which describes the covariance of $\alpha$ and $\beta$.

To this end we consider a bivariate Poisson 
model inspired by Equation~\autoref{eq:drift-params-breakdown}.
In that equation it is clear that variations in $\delta$ will 
cause correlations between $\alpha$ and $\beta$.
A bivariate Poisson random variable is defined as 
$(A,B)\sim\BP{\theta_0,\theta_1,\theta_2}$ where 
$A=C_0+C_1$ and $B=C_0+C_2$ with 
$C_i\sim\Poisson{\theta_0}$, $i=0,1,2$.
This produces the probability density
\begin{align}
    \pdf{BP}{x,y;\theta_0,\theta_1,\theta_2}
        &= \frac{\e^{-(\theta_0+\theta_1+\theta_2)}\theta_1^x\theta_2^y}{x!y!}
            \sum_{i=0}^{\min(x,y)}\matrixtwobyone{x}{i}\matrixtwobyone{y}{i}i!\left( \frac{\theta_0}{\theta_1\theta_2} \right)^i.
\end{align}
This distribution has marginal distributions $A\sim\Poisson{\theta_0+\theta_1}$
and $B\sim\Poisson{\theta_0+\theta_1}$, and covariance $\Cov(A,B)=\theta_0$.
As discovered by Karlis and Tsiamyrtzis~\cite{karlis_exact_2008}, it
has an exact family of conjugate priors given by the mixture distributions
\begin{align}
    \sum_{j=0}^r w_j 
        G(\theta_0;a_0+j,b_0)
        \cdot G(\theta_1;a_1-j,b_1)
        \cdot G(\theta_2;a_2-j,b_2)
    \label{eq:bivar-poiss-gen-prior}
\end{align}
where $G$ is the probability density of the gamma distribution,
$a_i,b_i>0$, $r\in\{0,1,2,...\}$, $0\leq w_j\leq 1$, 
and $\sum_{j=0}^r w_j=1$.

For the prior that interests us, fixing $r=0$ and $b_0=1$ 
will suffice; this leaves us with five hyperparameters 
$(a_0,a_1,a_2,b_0,b_1)$ which we will bijectively map onto
the more intuitive hyperparameters 
$(\mu_\alpha,\sigma_\alpha^2,\mu_\beta,\sigma_\beta^2,\sigma_{\alpha,\beta})$.

Indeed, make the change of variables $\alpha=\theta_0+\theta_1$ and 
$\beta=\theta_0+\theta_2$ and consider the prior
\begin{align}
    \pi(\alpha,\beta,\theta_0)
        &= G(\alpha-\theta_0;a_1,b_1)
            \cdot G(\beta-\theta_0;a_2,b_2)
            \cdot G(\theta_0;a_0,1)
\end{align}
where
\begin{align}
    a_1 &= \frac{(\mu_\alpha-\sigma_{\alpha,\beta})^2}{\sigma_\alpha^2-\sigma_{\alpha,\beta}} &
    b_1 &= \frac{\mu_\alpha-\sigma_{\alpha,\beta}}{\sigma_\alpha^2-\sigma_{\alpha,\beta}} \nonumber \\
    a_2 &= \frac{(\mu_\beta-\sigma_{\alpha,\beta})^2}{\sigma_\beta^2-\sigma_{\alpha,\beta}} &
    b_2 &= \frac{\mu_\beta-\sigma_{\alpha,\beta}}{\sigma_\beta^2-\sigma_{\alpha,\beta}} \nonumber \\
    a_0 &= \sigma_{\alpha,\beta}. &
\end{align}
It then holds that, for example,
\begin{align}
    \expect[\alpha^2]
        &= \int_0^\infty\dd\theta_0\int_{\theta_0}^\infty\dd\alpha\int_{\theta_0}^\infty\dd\beta
            \alpha^2\pi(\alpha,\beta,\theta_0) \nonumber \\
        &= \int_0^\infty\dd\theta_0\int_0^\infty\dd\theta_1
            (\theta_0^2+\theta_1^2+2\theta_1\theta_3)
            G(\theta_1;a_1,b_1) \cdot G(\theta_0;a_0,1) \nonumber \\
        &= 	  \left( \frac{a_0}{1^2} + \frac{a_0^2}{1^2} \right)
            + \left( \frac{a_1}{b_1^2} + \frac{a_1^2}{b_1^2} \right)
            + 2\frac{a_1}{b_1}a_0 \nonumber \\
        &= \sigma_\alpha^2 + \mu_\alpha^2
\end{align}
so that $\Var[\alpha]=\sigma_\alpha^2$.
Similarly we get
$\expect[\alpha]=\mu_\alpha$, $\expect[\beta]=\mu_\beta$,
$\Var[\alpha]=\sigma_\alpha^2$, $\Var[\beta]=\sigma_\beta^2$, and 
$\Cov[\alpha,\beta]=\sigma_{\alpha,\beta}$.

\begin{figure}
    \begin{center}
        \includegraphics[width=0.5\textwidth]{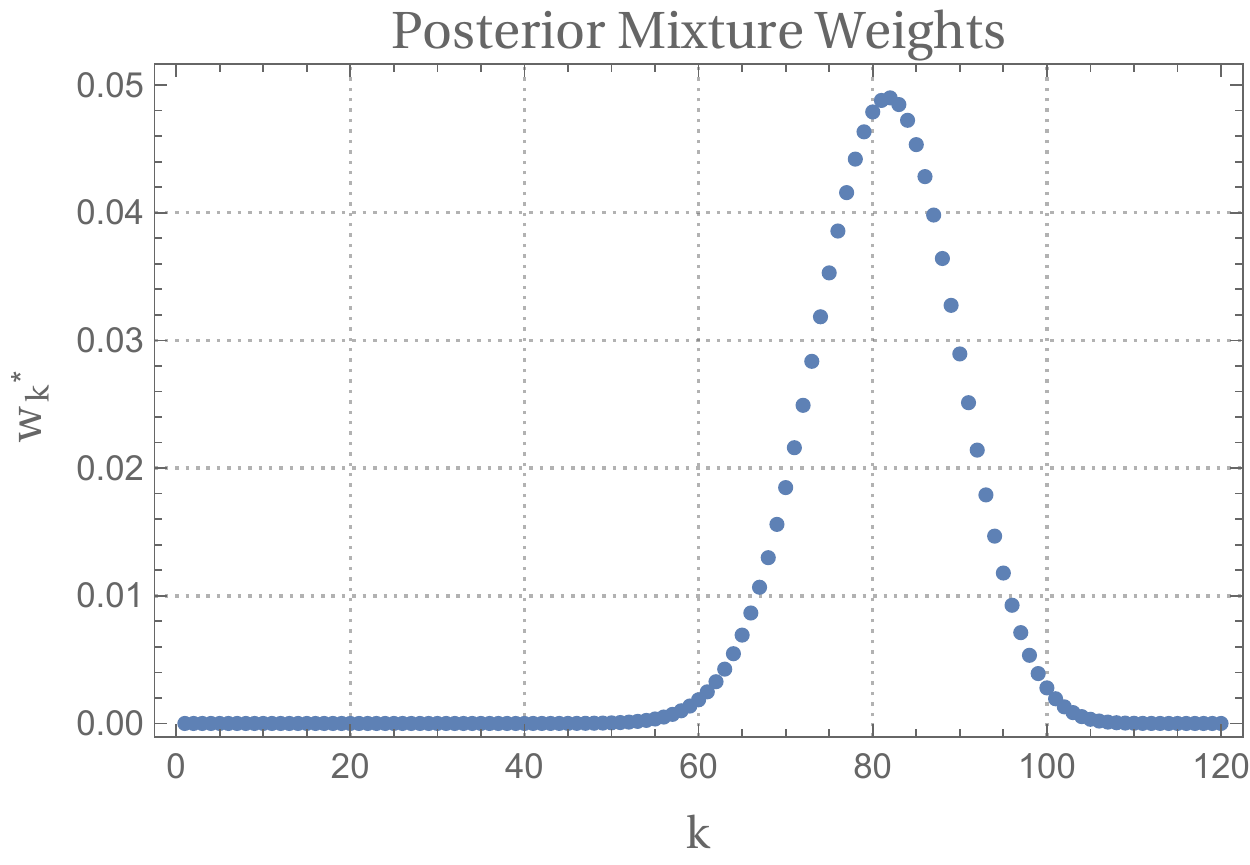}
    \end{center}
    \caption{An example of posterior mixture weights $w_k^*$ for
    $k=0,...,S_1$. Parameters used were $\mu_\alpha=200$, 
    $\mu_\beta=140$, $\sigma_\alpha=40$, $\sigma_\beta=15$, 
    $\sigma_{\alpha,\beta}=90$.
    The prior was updated with a single sample $x=220$ and $y=120$.}
    \label{fig:bivar-posterior-mixture-weights}
\end{figure}

Since this prior is conjugate to the likelihood function 
of $(X,Y)\sim\BP{\theta_0,\alpha-\theta_0,\beta-\theta_1}$, if we receive 
the iid data $(x_1,y_1),...,(x_n,y_n)$ sampled from this distribution, the
posterior will have the form of Equation~\autoref{eq:bivar-poiss-gen-prior}
with updated parameters.
Let $x=\sum_{k=1}^nx_k$ and $y=\sum_{k=1}^ny_k$.
First we have the simple updated posterior parameters 
$a_0^*=a_0$, $a_1^*=a_1+x$, $a_2^*=a_2+y$, $b_i^*=b_i+n$ for $i=0,1,2$,
which is very similar to the uncorrelated case of the previous section.
The new weights $w_j^*$ of the posterior (recall we had a single weight 
$w_0=1$ in the prior) also has a closed form, but it is cumbersome to write down.
Defining $s_i=\min(x_i,y_i)$ and $S_n=\sum_{i=1}^n s_i$, then for each 
$0\leq k\leq S_n$ we have 
$w_k=\bar{p}_k/\sum_{m=0}^{S_n} \bar{p}_k$ where
\begin{align}
    \bar{p}_k
        &= c_k^{(n)} \frac{b_1^{a_1}b_2^{a_2}}{\Gamma(a_1)\Gamma(a_2)\Gamma(a_0)}
        \Gamma(a_1-k+x)\Gamma(a_2-k+y)\Gamma(a_0+k)\left( \frac{(n+b_1)(n+b_2)}{n+b_0} \right)^k.
\end{align}
and the quantities $c_k^{(n)}$ are defined recursively as
\begin{align}
    c_k^{(n)}
        &= \sum_{r=\max(0,k-s_n^*)}^{\min(k,s_n^*)} v_r^{(n)}c_{k-r}^{(n-1)}
\end{align}
where $v_r^{(m)}=\left( (x_m-r)!(y_m-r)!r! \right)^{-1}$, $c_k^{(1)}=v_k^{(1)}$, 
and $s_m^*=\min(s_m,S_{m-1})$.
For generality, we have provided these formulas given 
$n$ random samples, when in practice, because of 
the drift discussed in the main body of this article, 
only one random sample will usually be taken.
With $n=1$, the recursive definition is unnecessary and the
weights are given by
\begin{align}
    w_k^*
        &= \frac{
            x! y! \sin (\pi  (\alpha_1+x)) \sin (\pi  (\alpha_2+y)) 
            \Gamma (k+\alpha_0) \Gamma (x-k+\alpha_1) \Gamma (y-k+\alpha_2)
        }{
            \pi^2 k! (x-k)! (y-k)! \, \Gamma (\alpha_0) \cdot
            _3\tilde{F}_2\left(\{-x,-y,\alpha_0\},\{1-x-\alpha_1,1-y-\alpha_2\},\frac{1}{2}(\beta_1+1)(\beta_2+1)\right)
        }
        \left( \frac{(1+b_1)(1+b_2)}{2} \right)^k
    \label{eq:corr-posterior-weights}
\end{align}
where $_p\tilde{F}_q(\{c_1,...,c_p\},\{d_1,...,d_q\},z)=\frac{_pF_q(\{c_1,...,c_p\},\{d_1,...,d_q\},z)}{\Gamma(d_1)\cdots\Gamma(d_q)}$
is the regularized generalized hypergeometric function and 
$_pF_q(\{c_1,...,c_p\},\{d_1,...,d_q\},z)$
is the generalized hypergeometric function.
However, the only factors in \autoref{eq:corr-posterior-weights} which are are 
relevant to numerical implementations are those which involve $k$.
Everything else can be implicitly calculated 
by demanding the normalization $\sum_{k=0}^{S_n}w_k^*=1$.
An example of posterior mixture weights is shown in 
\autoref{fig:bivar-posterior-mixture-weights}.

With these definitions, the posterior is given by the exact distribution
\begin{align}
    \pi^*(\alpha,\beta,\theta_0) 
        &= \sum_{j=0}^{S_n} w_j^* 
        G(\theta_0;a_0+j,b_0+n)
        \cdot G(\alpha-\theta_0;a_1+x-j,b_1+n)
        \cdot G(\beta-\theta_0;a_2+y-j,b_2+n).
\end{align}

Depending on the size of $S_n$, the posterior may be expensive to 
compute the value of at a given coordinate. 
However, once the posterior weights $\omega_k^*$ have been computed,
drawing a sample is essentially the same cost as drawing a sample from three 
Gamma distributions.

\begin{figure}
    \centering
    \subfloat[$\mu_\alpha=200$, 
    $\mu_\beta=120$, $\sigma_\alpha=20$, $\sigma_\beta=15$, 
    $\sigma_{\alpha,\beta}=100$, $n=1$, $x=170$, $y=100$]{\includegraphics[width=0.45\textwidth]{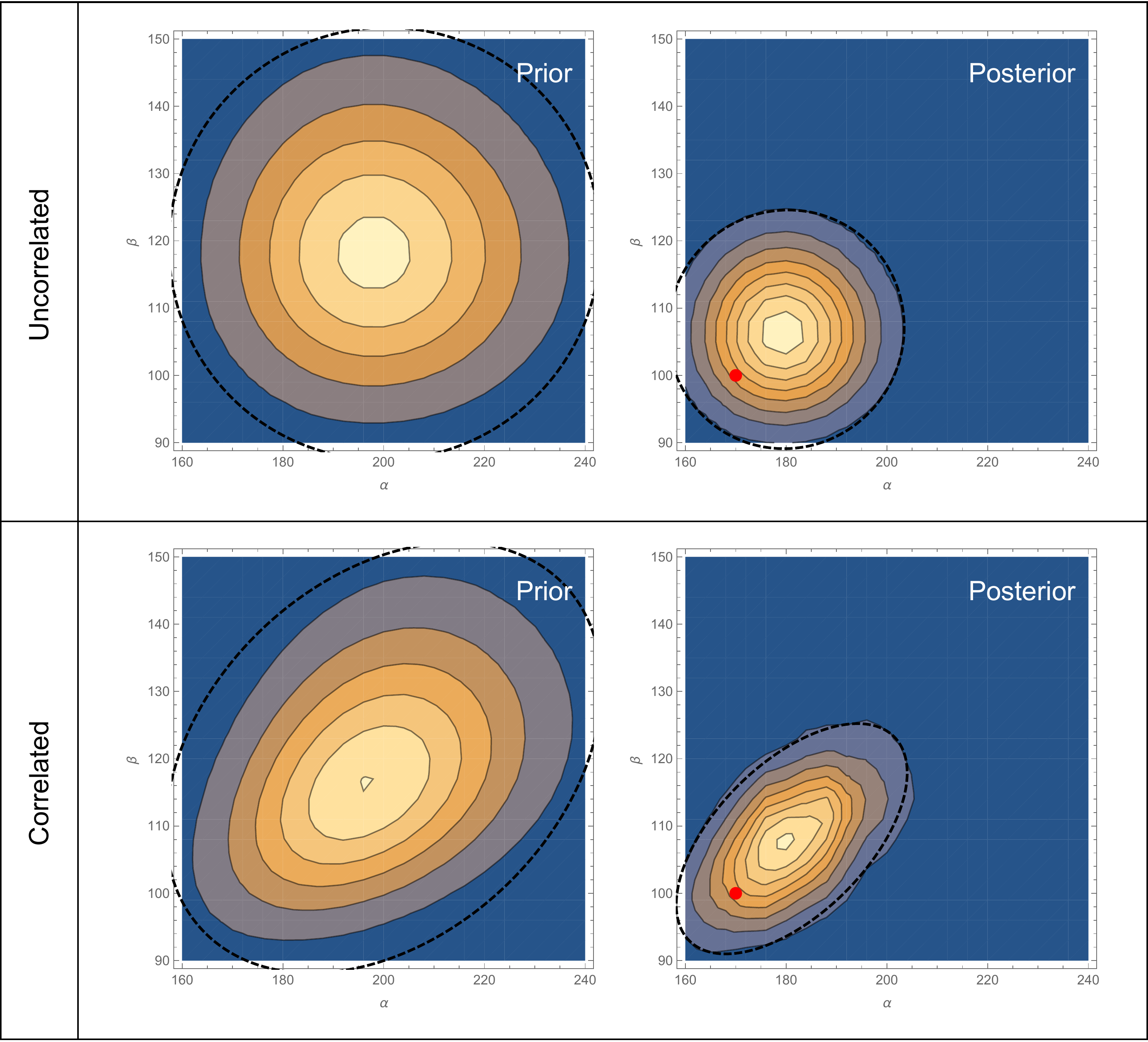}} 
    \subfloat[$\mu_\alpha=200$, 
    $\mu_\beta=120$, $\sigma_\alpha=20$, $\sigma_\beta=15$, 
    $\sigma_{\alpha,\beta}=100$, $n=1$, $x=230$, $y=100$]{\includegraphics[width=0.45\textwidth]{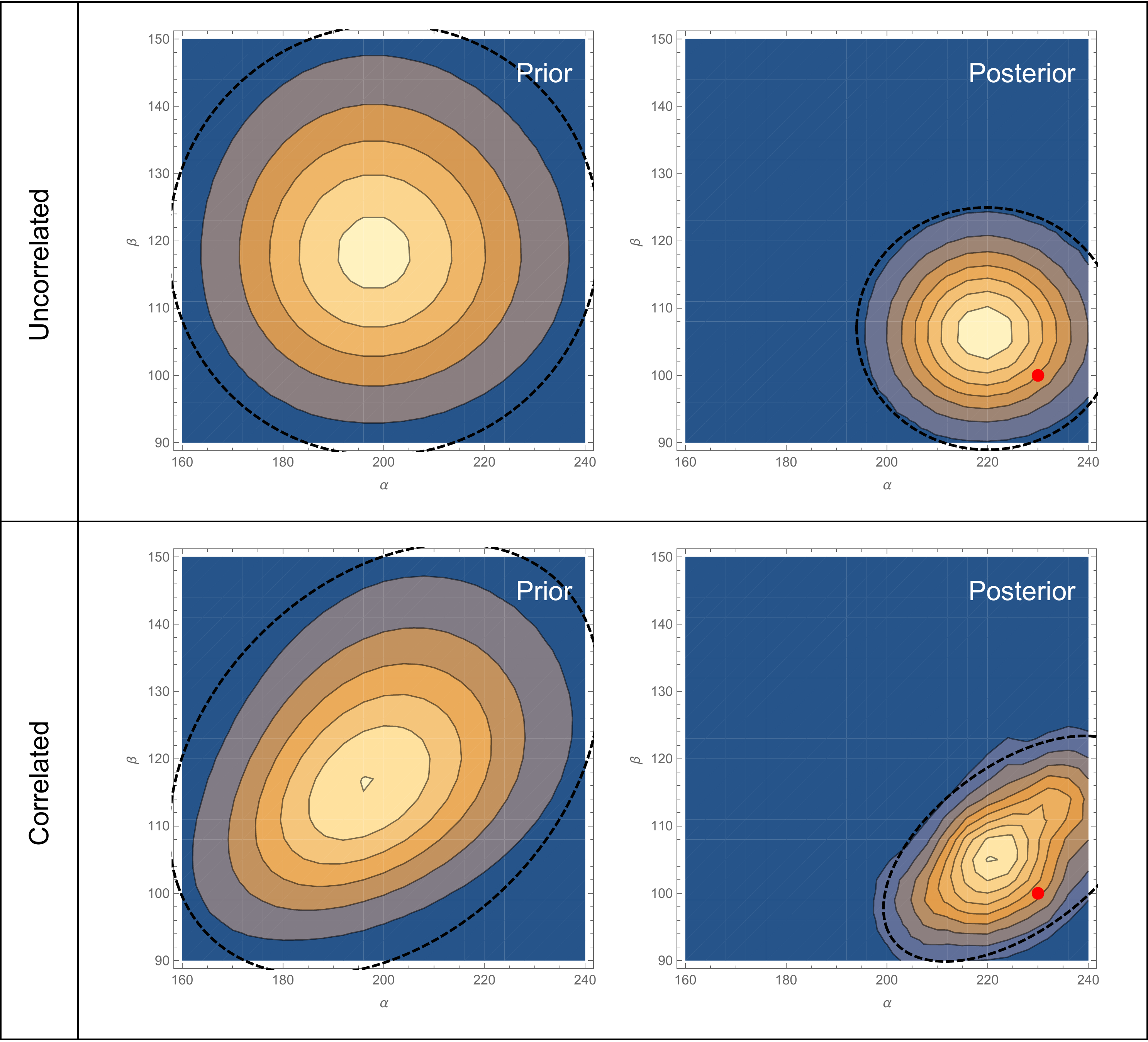}} \\
    \subfloat[$\mu_\alpha=2000$, 
    $\mu_\beta=1200$, $\sigma_\alpha=200$, $\sigma_\beta=150$, 
    $\sigma_{\alpha,\beta}=1000$, $n=1$, $x=1700$, $y=1000$]{\includegraphics[width=0.45\textwidth]{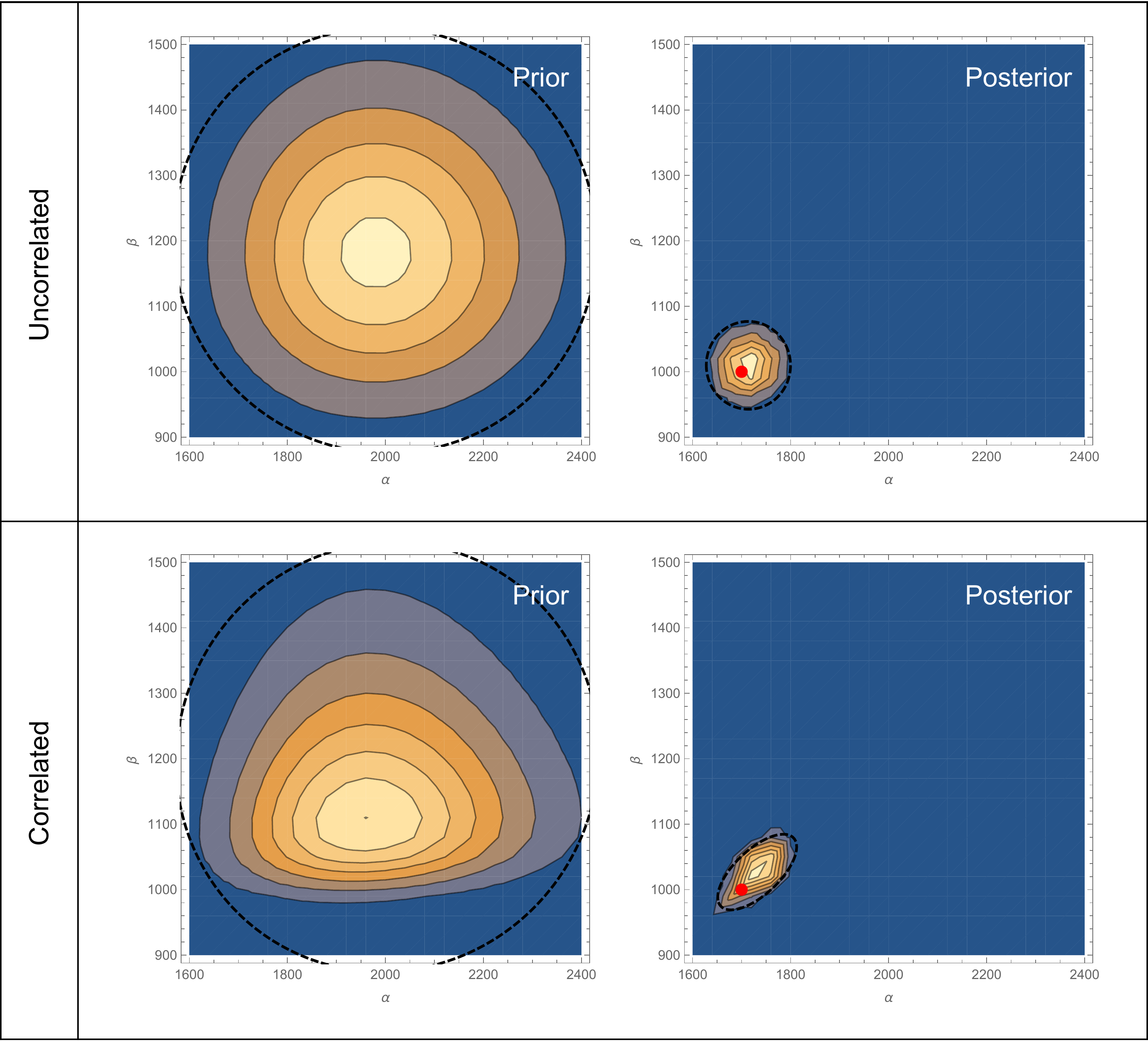}} 
    \subfloat[$\mu_\alpha=2000$, 
    $\mu_\beta=1200$, $\sigma_\alpha=200$, $\sigma_\beta=150$, 
    $\sigma_{\alpha,\beta}=1000$, $n=1$, $x=2300$, $y=1000$]{\includegraphics[width=0.45\textwidth]{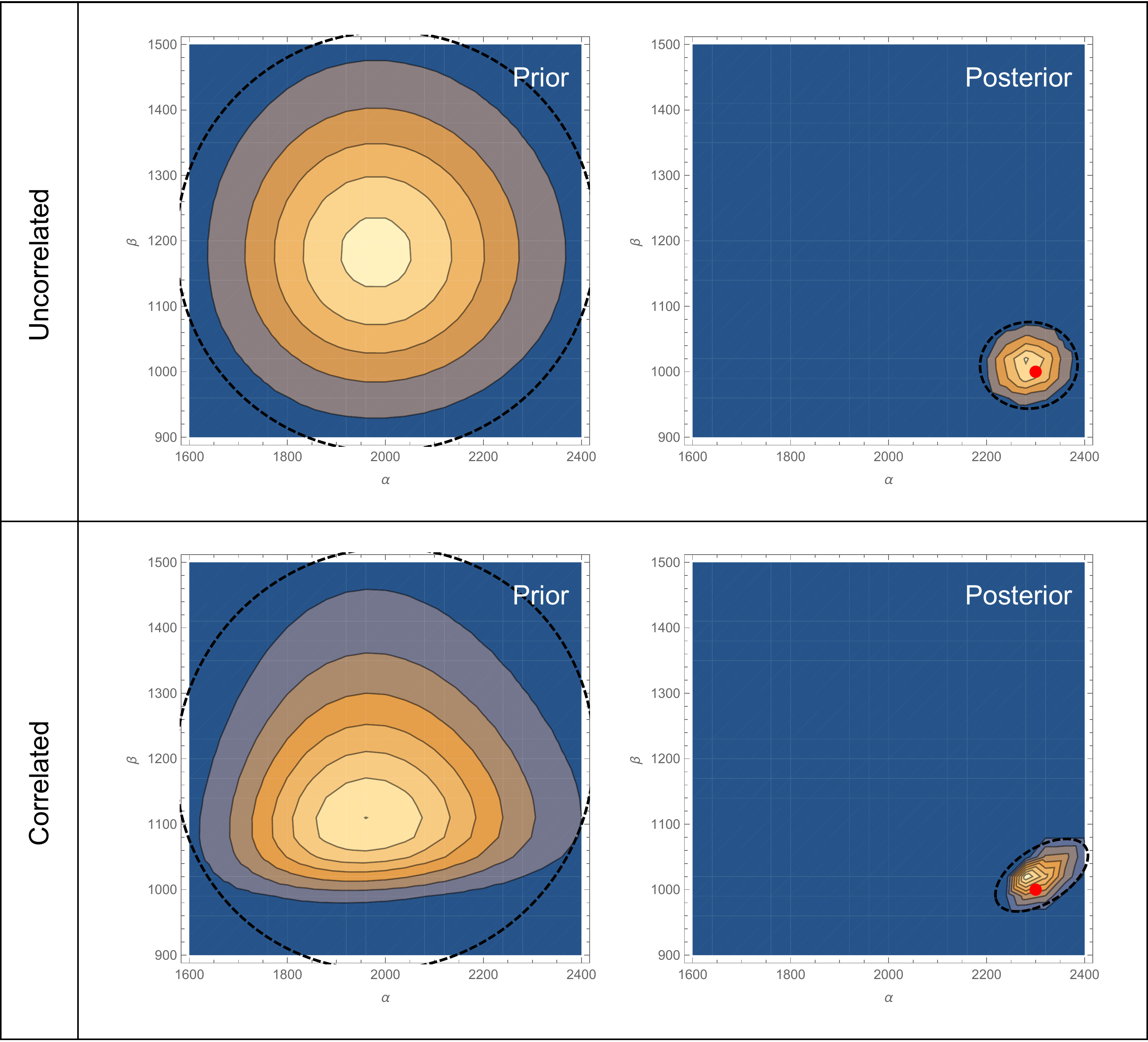}}
    \caption{Examples of Bayesian updates showing contours of the prior and posterior probability distributions,
    where the update data is depicted by red dots.
    In each of the four cases, the same data is given to both the correlated and uncorrelated priors
    described in \autoref{app:drift-prior-uncor} and \autoref{app:drift-prior-cor}, respectively.
    Black dashed ellipses represent 90\% confidence regions; their centers are at the mean
    of the distribution, and their eccentricity matrix is equal to 4.6 times the covariance matrix 
    of the distribution.
    Cases (a) and (b) represent a low data scenario, whereas cases (c) and (d) represent a high data scenario.
    Cases (a) and (c) represent correlated measurement data, whereas cases (b) and (d) represent anti-correlated measurement data.}
    \label{fig:bivar-update-examples}
\end{figure}

%=============================================================================
\section{Bridged Updater for the Referenced Poisson Model in SMC}
\label{app:bridged-updater}
%=============================================================================

For us, there are two main mechanisms that can cause the SMC algorithm 
to become unreliable.
Both have to do with the finite particle approximation and its reliance on
having enough effective particles near the true model parameter values, known 
as importance sampling.
The first mechanism is that periodicities in the model have the
effect of creating temporary multimodalities in the posterior as we 
analyze the data.
If certain early data happen to cause disproportionate support on an incorrect 
mode, we lose particles where we need them, and this might lead to a runaway 
effect where all particle weights become zero.
We mitigate against this by processing the data in ascending order, as discussed in 
\autoref{sec:qhl-inference}.

The second cause of instability is that some data points are overly informative.
From a learning perspective, informative data are great.
However, from SMC's perspective, very informative data have the tendency to 
drastically reduce the weight of most particles, causing the effective 
particle count,
\begin{align}
    n_{\text{eff}} = 1/\sum_{n=1}^{N} w_n^2,
    \label{eq:eff-particles}
\end{align}
to become a tiny fraction of the actual particle count, $N$.
For example, it is not unusual (nor is it common) for data, with our QHL model,
to reduce the effective particle count to below 100 while there are 16000 
actual particles.
Storing a multi-dimensional distribution (seven dimensions in our QHL example) 
on 100 particles is a bad idea, and causes the remainder of the 
inference to become suspect.
A costly solution to this problem is to, for example, increase to 160000 particles,
so that the effective particle count does not dip much below 1000.
The purpose of this section is to discuss a less costly solution.

The idea is as follows. 
It is easy to detect when an update \textit{would} cause the effective particle count to 
drop below some threshold, say 1000; one simply need compute 
\autoref{eq:eff-particles} on the posterior weights before overwriting the
current weights.
If this flag is raised, instead of performing the update 
(i.e. overwriting the current weights), we instead perform a sequence 
of less informative updates that do not correspond to actual observations, but
that do result in the same posterior.
This technique is called \textit{bridging the transition}
\cite{andrieu_particle_2010}, and it gives a chance
for the updater to resample the particles ``mid-update'' so that particles can 
be relocated to where they are actually needed.
This technique is only amenable to certain likelihood functions, and thankfully,
owing to some nice properties of the Poisson distribution, ours is one of them.

To see this, consider the generic \NV model at some particular step of SMC:
\begin{subequations}
\begin{align}
    s &\sim \pi(s) \\
    \alpha,\beta &\sim \pi(\alpha,\beta) \\
    X | \alpha &\sim \Poisson{\alpha} \\
    Y | \beta &\sim \Poisson{\beta} \\
    Z | s,\alpha,\beta &\sim \Poisson{p(s)\alpha + (1-p(s))\beta}.
\end{align}
\end{subequations}
Here, $s$ is a set of parameters we wish to learn, 
$\pi(s)$ is our current particle distribution describing these parameters,
$\pi(\alpha,\beta)$ is our prior on the next data point's 
references, $p(s)$ is the function that takes the parameters of interest 
and returns $p=\Tr \rho P_0$, and $(X,Y,Z)$ is the next data point triplet.
On obtaining the variate $(x,y,z)$, the next SMC update step 
should result in the posterior distribution 
$\Pr(s,\alpha,\beta|x,y,z)$, which, using Bayes' rule, is proportional to
\begin{subequations}
\begin{align}
    \Pr(s,\alpha,\beta|x,y,z)
        &\propto \Pr(x,y,z|s,\alpha,\beta)\pi(\alpha,\beta)\pi(s) \\
        &= \Pr(z|s,\alpha,\beta)\Pr(x|\alpha)\Pr(y|\beta)\pi(\alpha,\beta)\pi(s) \\
        &\propto \Pr(z|s,\alpha,\beta)\pi_{x,y}^*(\alpha,\beta)\pi(s).
\end{align}
\end{subequations}
As noted in \autoref{sec:bayes-estimator}, this formula shows that our 
update consists of first updating $\alpha$ and $\beta$ analytically 
using a conjugate prior, and subsequently updating $s,\alpha,\beta$ using 
$\pi_{x,y}^*(\alpha,\beta)\pi(s)$ as a prior.
The proportionalities allow us to neglect those factors which depend only
on $x$, $y$, and/or $z$ but not any of $s$, $\alpha$, or $\beta$; 
all particles see the same values of $x$, $y$, and $z$, so any such factors
will be canceled out when enforcing the normalization condition of the particle weights.

\begin{figure}
    \begin{center}
        \includegraphics[width=0.8\textwidth]{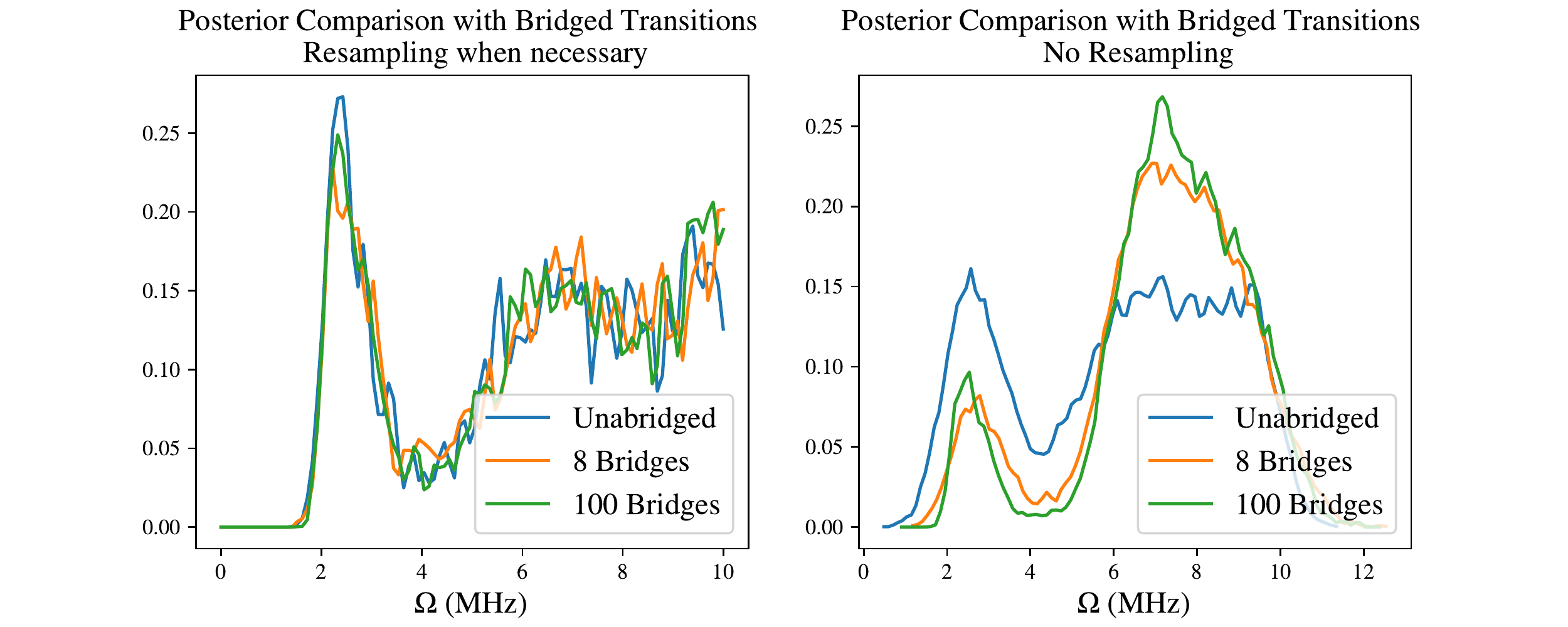}
    \end{center}
    \caption{A particle distribution was initialized to the prior of 
    \autoref{eq:qhl-prior} with 16000 particles, 
    and separately updated with the data from a single point of a Rabi experiment in 
    six different ways.
    We show a slice through the posterior for each case.
    On the left are bridged and un-bridged updates with no resamples allowed,
    The final effective particle count was about 1800 for all three of these updates.
    This demonstrates the bridging technique works in practice.
    On the right are bridged and un-bridged updates with resamples taken 
    whenever the distribution was detected to have fewer than 8000 effective particles.
    These two bridge cases maintained at least 8000 effective particles at all 
    times. Since the posterior is far from normal, we can expect the resampler 
    to introduce distortions.}
    \label{fig:qhl-bridge-transitions}
\end{figure}

To bridge this transition, we start with the particles in the state 
$\pi_{x,y}^*(\alpha,\beta)\pi(s)$ and notice that
\begin{subequations}
\begin{align}
    \Pr(z|s,\alpha,\beta)
        &\propto (p(s)\alpha + (1-p(s))\beta)^z e^{-p(s)\alpha - (1-p(s))\beta} \\
        &= \left(
                p(s)\frac{\alpha}{m} + (1-p(s))\frac{\beta}{m}
            \right)^{m \frac{z}{m}}
            m^z
            e^{-m p(s)\frac{\alpha}{m} - m(1-p(s))\frac{\beta}{m}} \\
        &\propto \pdf{\text{Pois}}{\frac{z}{m};
            p(s)\frac{\alpha}{m}+(1-p(s))\frac{\beta}{m}}^m
\end{align}
\end{subequations}
so that the update can be achieved instead with $m$ Poisson updates with reduced 
count data $z/m$ and reference particle locations for $\alpha$ and 
$\beta$ also reduced by $m$.
This is illustrated in \autoref{fig:qhl-bridge-transitions}.
Note that our fictional data can have fractional photon counts; this is 
not a problem, the Poisson mass function is well defined for non-integer data,
assuming the factorial is implemented using the gamma function, which it will 
be in any modern programming language.

%=============================================================================
\section{QHL Supplemental Figures}
\label{app:qhl-figures}
%=============================================================================

\begin{figure}[h]
    \sidesubfloat[]{\includegraphics[width=0.45\textwidth]{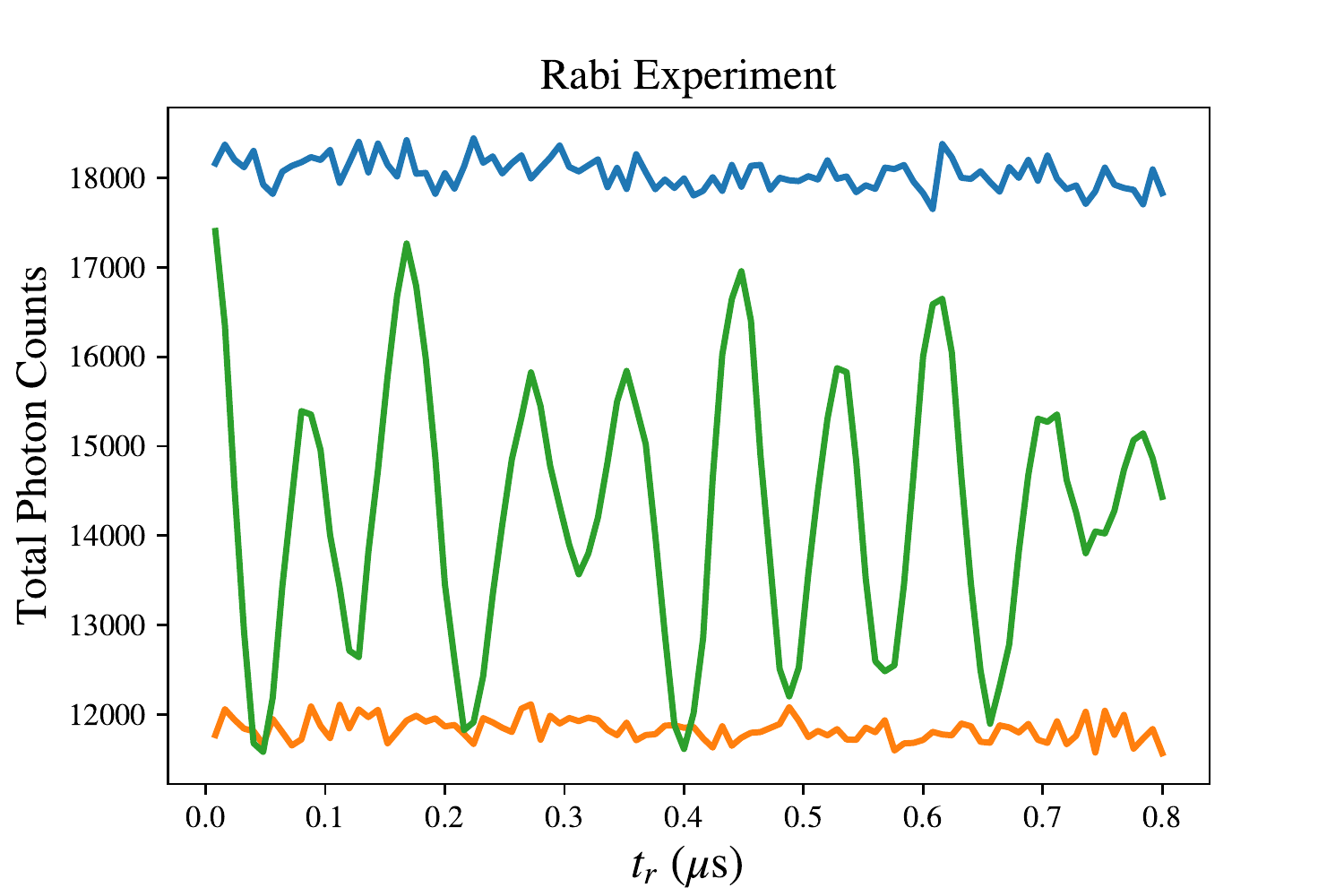}} \quad
    \sidesubfloat[]{\includegraphics[width=0.45\textwidth]{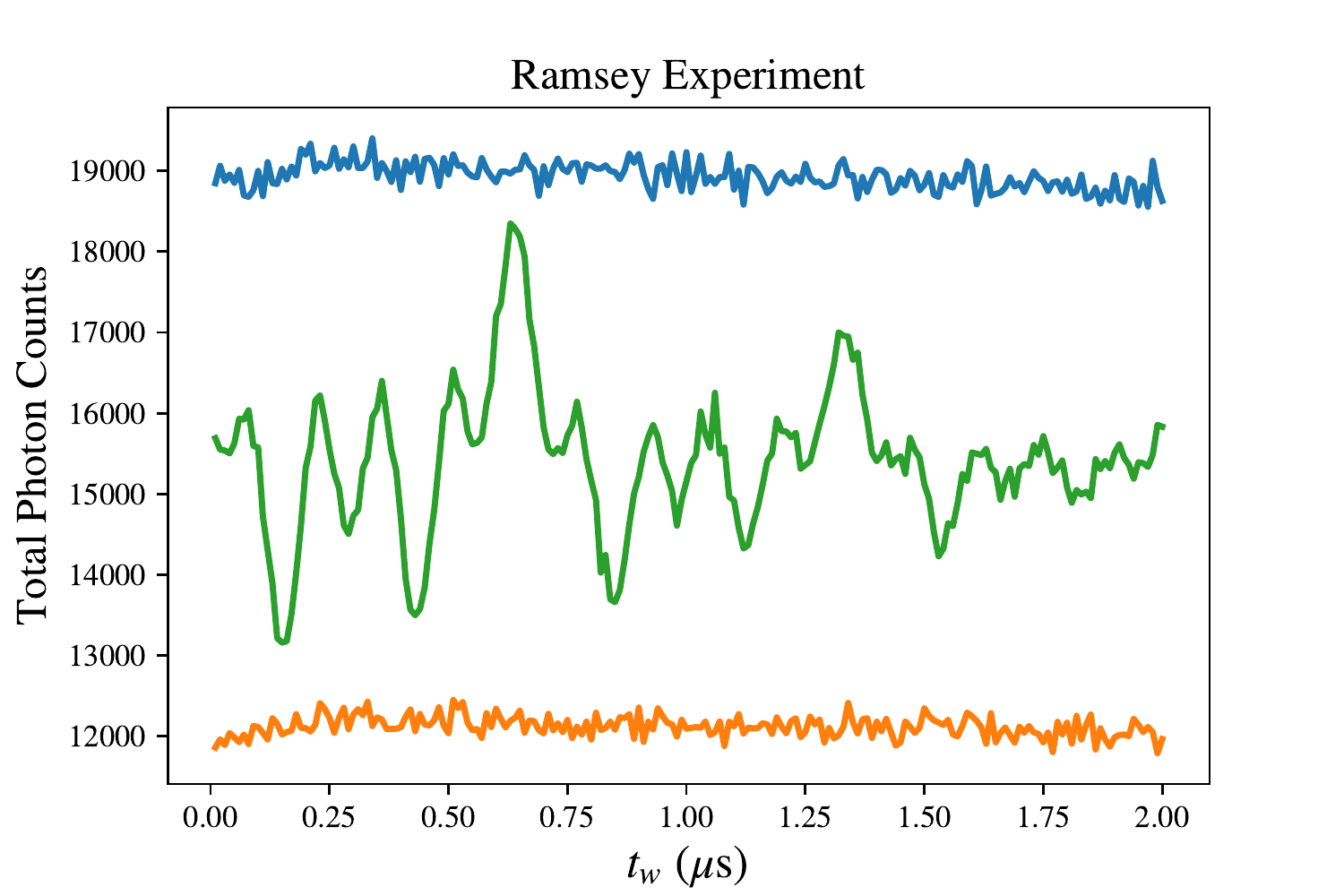}} \\
    \sidesubfloat[]{\includegraphics[width=0.45\textwidth]{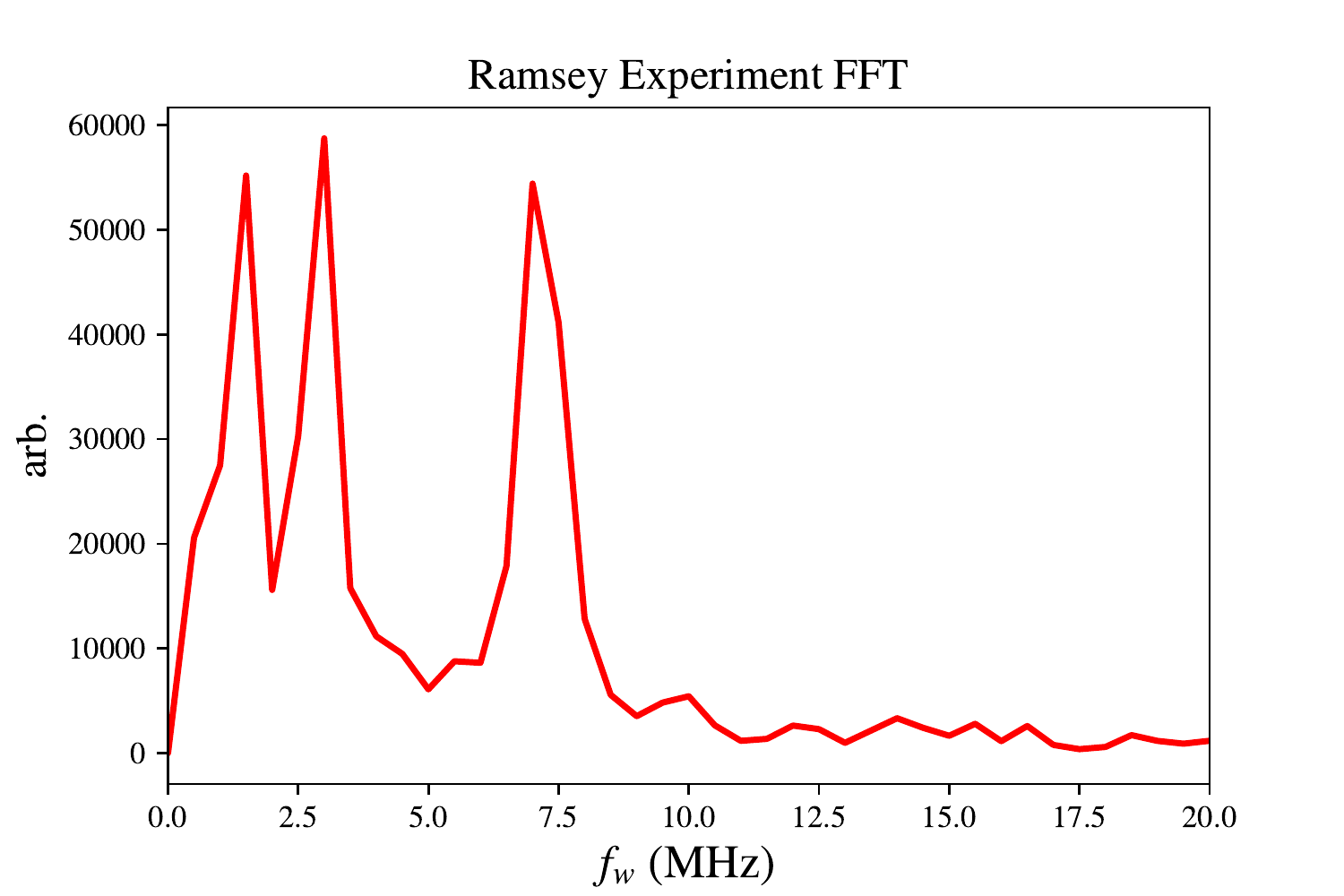}} \quad
    \sidesubfloat[]{\includegraphics[width=0.45\textwidth]{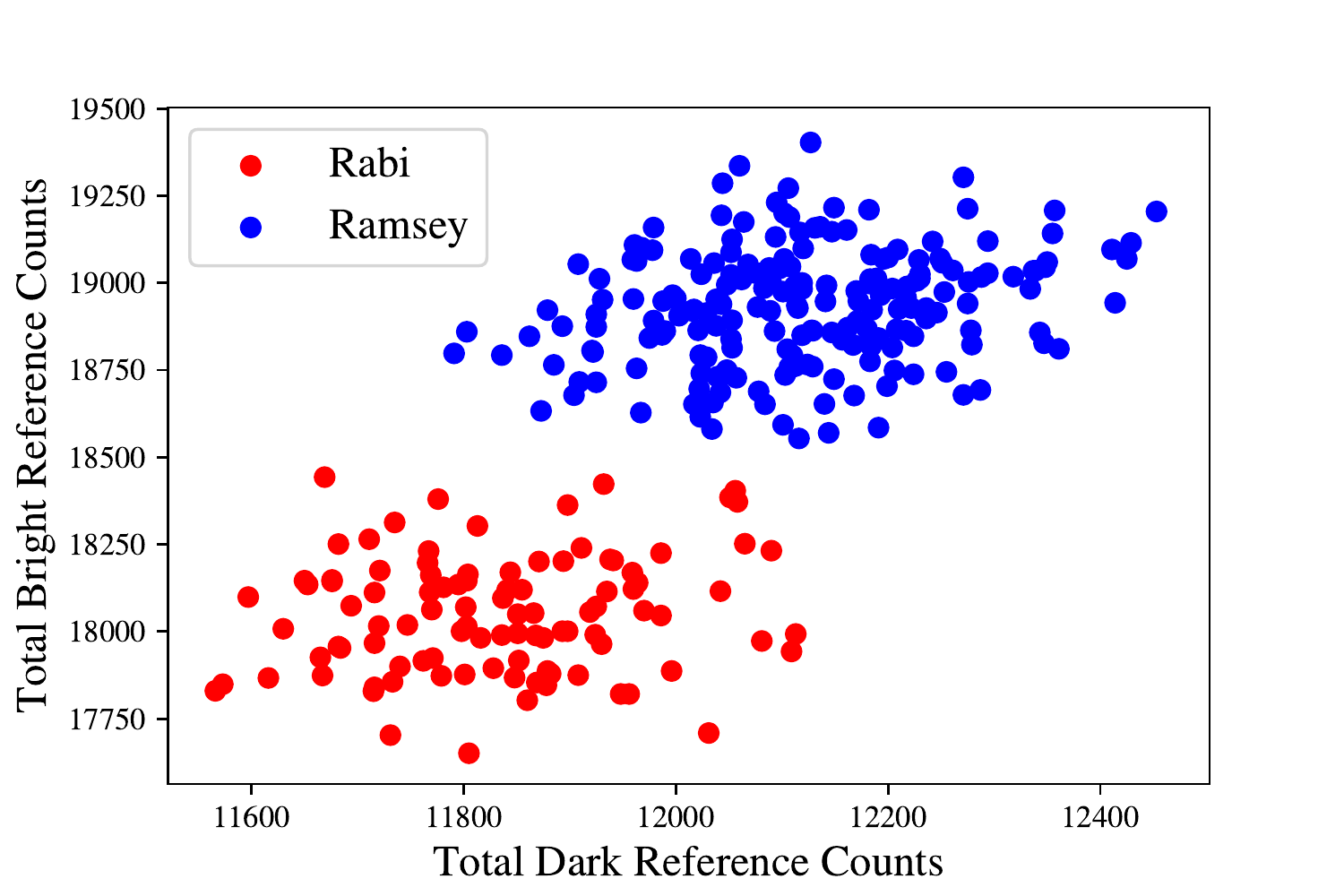}}
    \caption{(a),(b) Time domain data from Rabi and Ramsey experiments. Photon 
    counts are summed over all $400\times 30000$ repetitions at each experiment parameter
    on the x-axis. Bright and dark references are shown in addition to the 
    signal of interest. (c) The discrete Fourier transform of the Ramsey experiment. 
    (d) Scatter plot of the summed reference counts for both experiments.
    Each point represents a different experiment configuration, the discrepancy 
    between distributions is due to performing the experiments on different days of the week.}
    \label{fig:raw-data}
\end{figure}

\begin{figure}[h]
    % here we explicitly include the .png because the .pdf is 25MB and 
    % singlehandedly exceeds the arXiv quota
    \includegraphics[width=0.95\textwidth]{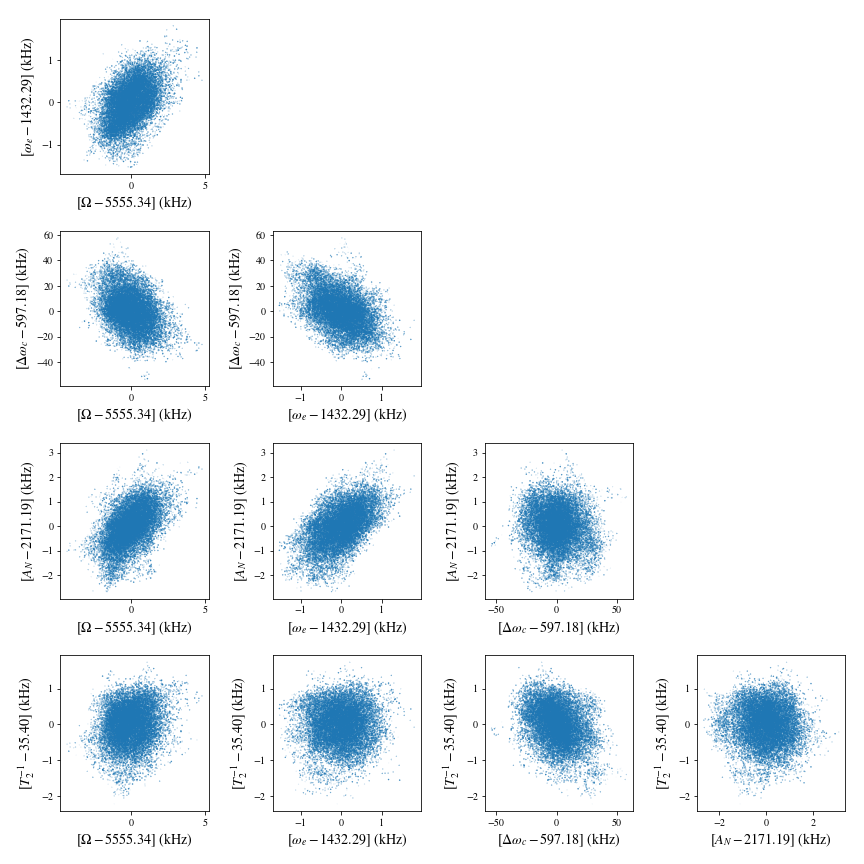}
    \caption{Two-parameter marginals of the QHL posterior
    distribution, where each dot is a member of the particle approximation
    projected onto the corresponding axes.
    The plots have been centered around the mean value of the distribution,
    the components of which are specified in the axis labels.}
    \label{fig:qhl-two-param-marginals}
\end{figure}

\begin{figure}[h]
    \includegraphics[width=0.951\textwidth]{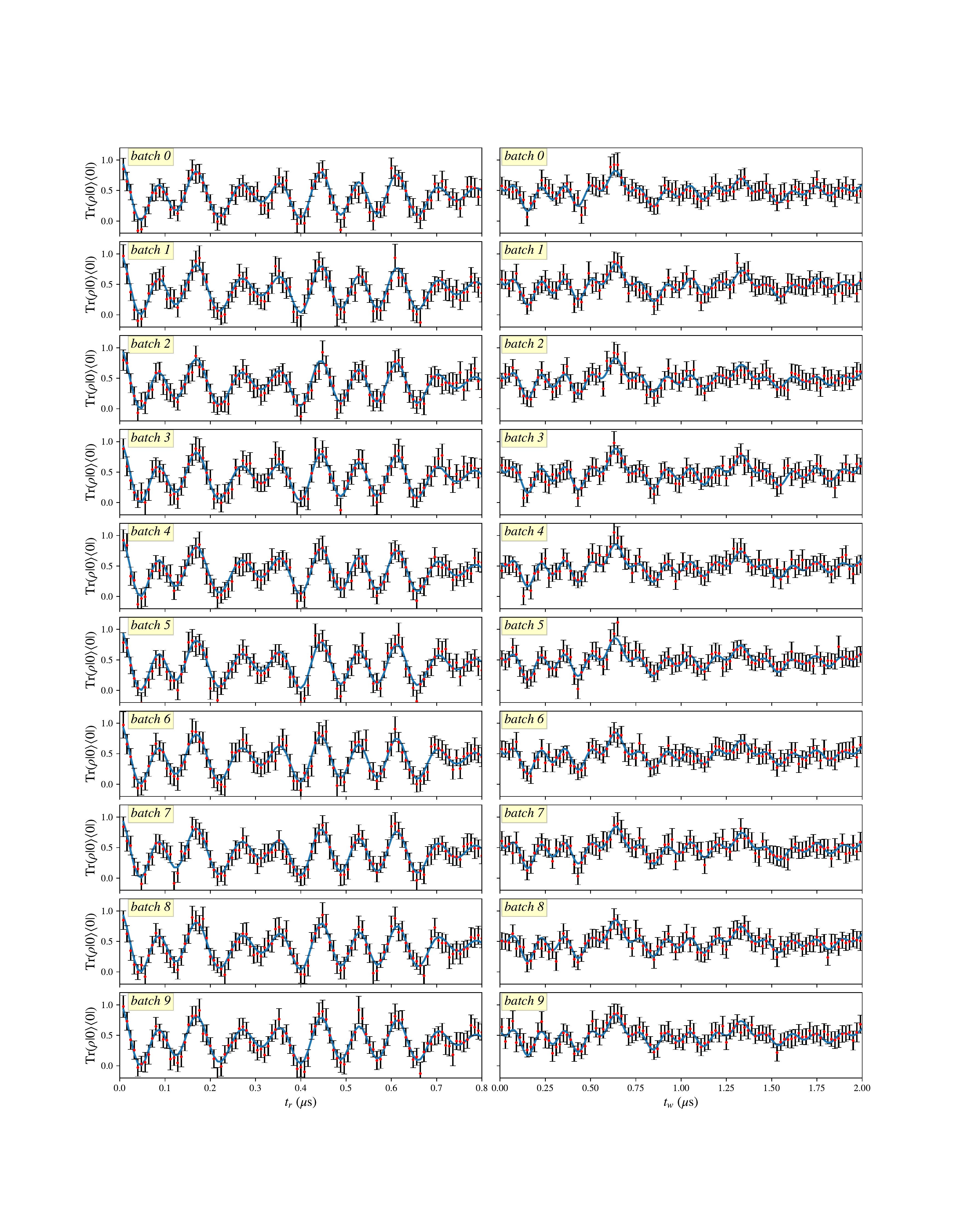}
    \caption{Fits to the data from each of the 10 batches of 40 averages.
    The left column contains the Rabi experiments, and the right column
    contains the Ramsey experiments.
    The points are the normalized data used in the corresponding SMC algorithm,
    with error bars calculated using \autoref{eq:cramer-rao-bound}.}
    \label{fig:qhl-batch-fits}
\end{figure}

In this appendix section a few extra plots are included related 
to \autoref{sec:qhl-example} of the main body.
\autoref{fig:raw-data} shows the raw (summed) data.
\autoref{fig:qhl-two-param-marginals} shows all two-parameter 
marginals of the parameter posterior distribution using the entire 
data-set.
\autoref{fig:qhl-batch-fits} shows simulation results due to SMC fits 
for each of the 10 batches.

\floatsetup{subcapbesideposition=top}

\floatsetup{subcapbesideposition=top}

\end{document}